\documentclass[twocolumn,preprintnumbers,floatfix,superscriptaddress,nofootinbib]{revtex4}

\usepackage{graphicx}
\usepackage{color,xcolor}
\usepackage{hepunits}
\usepackage[colorlinks=true,urlcolor=brown,linkcolor=blue,citecolor=magenta]{hyperref}
\usepackage{amsmath}
\usepackage{parskip}
\usepackage[english]{babel}

\newcommand{\abs}[1]{\left\lvert #1 \right\rvert}
\newcommand{\angu}[1]{\left\langle #1 \right\rangle}
\newcommand{\bxd}[1]{\lbrack #1 \rbrack}
\newcommand{\ov}[1]{\overline{#1}}

\newcommand{\Sec}[1]{Sec.~\ref{#1}}

\newcommand{\App}[1]{Appendix~\ref{#1}}
\newcommand{\Apps}[2]{Appendices~\ref{#1} and \ref{#2}}

\newcommand{\Fig}[1]{Fig.~\ref{#1}}
\newcommand{\Figs}[2]{Figs.~\ref{#1} and \ref{#2}}
\newcommand{\Eq}[1]{Eq.~(\ref{#1})}
\newcommand{\Eqs}[2]{Eqs.~(\ref{#1}) and (\ref{#2})}
\newcommand{\Eqst}[2]{Eqs.~(\ref{#1})-(\ref{#2})}

\newcommand{\be}{\begin{equation}}
\newcommand{\ee}{\end{equation}}
\newcommand{\ba}{\begin{eqnarray}}
\newcommand{\ea}{\end{eqnarray}}
\newcommand{\bl}{\left}
\newcommand{\br}{\right}

\newcommand{\gsim}{\lower.7ex\hbox{$\;\stackrel{\textstyle>}{\sim}\;$}}
\newcommand{\lsim}{\lower.7ex\hbox{$\;\stackrel{\textstyle<}{\sim}\;$}}

\newcommand{\Phis}{\angu{\Phi}_s}
\newcommand{\Phib}{\angu{\Phi}_b}
\newcommand{\dd}{\mathrm{d}}

\newcommand{\GV}{\frac{\Gamma}{\mathcal{V}}}
\newcommand{\GVh}{\Gamma/\mathcal{V}}
\newcommand{\GVa}[1]{\frac{\Gamma #1}{\mathcal{V}}}

\newcommand{\mpl}{m_{\rm Pl}}
\newcommand{\ie}{{\it i.e.}}
\newcommand{\eg}{{\it e.g.}}
\newcommand{\pt}{\mathrm{PT}}
\newcommand{\hpt}{\mathrm{hPT}}
\newcommand{\cpt}{\mathrm{cPT}}
\newcommand{\eq}{\mathrm{eq}}
\newcommand{\crit}{\mathrm{crit}}
\newcommand{\mx}{\mathrm{max}}
\newcommand{\mn}{\mathrm{min}}
\newcommand{\tot}{\mathrm{tot}}
\newcommand{\eff}{\mathrm{eff}}
\newcommand{\bub}{\mathrm{bub}}

\newcommand{\fv}{\mathrm{fv}}
\newcommand{\tv}{\mathrm{tv}}
\newcommand{\gw}{\mathrm{gw}}

\newcommand{\bc}{\mathrm{bc}}
\newcommand{\sw}{\mathrm{sw}}
\newcommand{\lbar}{\overline{\lambda}}

\newcommand{\chii}{{\chi, i}}
\newcommand{\rc}{{r, c}}
\newcommand{\vs}{{\rm vs}}
\newcommand{\rd}{{\rm rd}}
\newcommand{\rmx}{{r, \mathrm{max}}}
\newcommand{\rrd}{{r, \mathrm{rd}}}
\newcommand{\rhpt}{{r, \mathrm{hPT}}}
\newcommand{\rcpt}{{r, \mathrm{cPT}}}

\newcommand{\chiR}{\mathrm{\chi R}}

\newcommand{\vepsrh}{\varepsilon_{\rm rh}}
\newcommand{\ard}{a_\rd}
\newcommand{\Ard}{A_\rd}
\newcommand{\Drd}{D_\rd}

\newcommand{\eV}{\mathrm{eV}}

\newcommand{\GeV}{\mathrm{GeV}}
\newcommand{\TeV}{\mathrm{TeV}}

\newcommand{\Kel}{\mathrm{K}}

\newcommand{\ignore}[1]{}

\newcommand\eea{\end{aligned} \end{equation}}
\newcommand\bea{\begin{equation} \begin{aligned}}

\begin{document}
\title{Gravitational-wave signatures from reheating}

\author{Manuel A. Buen-Abad}
\email{buenabad@umd.edu}
\affiliation{Maryland Center for Fundamental Physics, Department of Physics, University of Maryland, College Park, Maryland 20742, USA}
\affiliation{Dual CP Institute of High Energy Physics, C.P. 28045, Colima, M\'{e}xico}

\author{Jae Hyeok Chang}
\email{jaechang@umd.edu}
\affiliation{Maryland Center for Fundamental Physics, Department of Physics, University of Maryland, College Park, Maryland 20742, USA}
\affiliation{Department of Physics and Astronomy, Johns Hopkins University, Baltimore, MD 21218, USA}

\author{Anson Hook}
\email{hook@umd.edu}
\affiliation{Maryland Center for Fundamental Physics, Department of Physics, University of Maryland, College Park, Maryland 20742, USA}

\begin{abstract}
We initiate a study of the gravitational-wave signatures of a phase transition that occurs as the Universe's temperature increases during reheating.  
The gravitational-wave signatures of such a heating phase transition are different from those of a cooling phase transition, and their detection could allow us to probe reheating. In the lucky case that the gravitational-wave signatures from both the heating and cooling phase transitions were to be observed, information about reheating could in principle be obtained utilizing the correlations between the two transitions. Frictional effects, leading to a constant bubble-wall speed in one case, will instead behave as an ``antifriction'' force in the other and accelerate the bubble wall. This antifriction will often take the bubble into a runaway regime, significantly enhancing the amplitude of the heating phase transition gravitational-wave signal.  The efficiency, strength, and duration of the phase transitions will be similarly correlated in a reheating-dependent way.

\end{abstract}

\maketitle

\section{Introduction}

The remarkable transparency of the Universe to light allows us to look far back in time and learn about the early Universe.  Using this fact, we can observe the clumping of matter as a function of redshift, as well as infer early Universe properties from the cosmic microwave background (CMB) power spectrum.  At around a redshift of $z \sim 1100$, however, this treasure trove of information ceases as the Universe becomes opaque to light.

Gravitational waves (GWs) offer a unique opportunity to look even further back in time.  Unlike light, the Universe is never opaque to GWs, and thus they allow us to observe the Universe at its very youngest.  This unique opportunity comes at the cost of it being much harder to observe GWs than it is to observe light.  It is only recently that they have been observed for the first time by the LIGO-Virgo Collaboration~\cite{LIGOScientific:2016aoc}.  With future detectors such as LISA~\cite{LISA:2017pwj,Baker:2019nia,Caprini:2015zlo,Caprini:2019egz}, BBO~\cite{Crowder:2005nr,Corbin:2005ny,Harry:2006fi}, and DECIGO~\cite{Seto:2001qf,Kawamura:2011zz,Yagi:2011wg,Isoyama:2018rjb} on the horizon, the prospects of GW detection are bright.

One of the main mechanisms by which GWs can teach us about the early Universe comes in the form of a stochastic GW background (SGWB), the CMB of GWs. Stochastic GWs can result from any number of well-motivated early Universe phenomena such as inflation~\cite{Grishchuk:1974ny,Starobinsky:1979ty,Rubakov:1982df,Guzzetti:2016mkm}, reheating/preheating~\cite{Khlebnikov:1997di,Easther:2006gt,Easther:2006vd,GarciaBellido:2007dg,GarciaBellido:2007af,Garcia-Bellido:2007nns,Garcia-Bellido:2007fiu,Dufaux:2007pt,Dufaux:2008dn,Dufaux:2010cf,Figueroa:2017vfa,Adshead:2018doq,Adshead:2019lbr}, phase transitions~\cite{Kosowsky:1991ua,Kosowsky:1992vn,Kamionkowski:1993fg,Grojean:2006bp,Huber:2007vva,Kahniashvili:2008pf,Huber:2008hg,Caprini:2015zlo,Caprini:2019egz,Hindmarsh:2020hop,Athron:2023xlk}, topological defects~\cite{Caprini:2018mtu,Christensen:2018iqi}, and second-order scalar perturbations~\cite{Acquaviva:2002ud,Mollerach:2003nq,Baumann:2007zm,Espinosa:2018eve,Kohri:2018awv}.
There exists a massive literature on stochastic GWs and what can be learned from them; see \eg, Refs.~\cite{Grojean:2006bp,Schwaller:2015tja,Chang:2019mza,Gouttenoire:2019rtn,Cui:2019kkd,Buchmuller:2019gfy,Dror:2019syi,Dunsky:2019upk,Blasi:2020wpy,Machado:2019xuc,Geller:2018mwu,Hook:2020phx,Brzeminski:2022haa,Bodas:2022zca,Bodas:2022urf}. The GW source that we focus on in this article is a first-order phase transition (PT). In PTs the Universe evolves from a metastable or false vacuum state to a stable or true vacuum state through the nucleation and subsequent expansion of ``bubbles'' of the new phase. The complicated dynamics of bubble collisions can generate GWs.

One of the most interesting early Universe events is the process called {\it reheating}.  Inflation cools the Universe to a temperature $T$ of zero due to its exponential expansion.  Meanwhile the late-time Universe is well described by the standard model of cosmology, where the Universe is a hot thermal bath cooling due to the expansion of the Universe.  Clearly sometime in between these two events the Universe must have gone from $T=0$ to $T > 0$, a process referred to as reheating (RH).  This makes RH a particularly special era in the history of the Universe, since in almost all models the temperature only ever decreases afterwards.  Subsequent events such as entropy dumps only serve to cool the Universe more slowly rather than cause new RH events. Because RH likely only occurred once in our Universe, it is a unique and interesting event to study, and it is the target that we aim to elucidate.

In this article, we wish to probe how RH can be tested experimentally. Because RH is the only time when the temperature of the Universe increases, we are led to study signatures that arise from a period of increasing temperature. We therefore study the GW signature resulting from a heating phase transition (hPT) as opposed to the more commonly studied cooling phase transition (cPT).\footnote{This abbreviation is not to be confused, of course, with CPT symmetry.}

The GWs of a cPT (which we denote as cGWs) are chiefly generated by the colliding bubble walls, plasma sound waves, and plasma turbulence.  The resulting signature is commonly characterized by the efficiencies ($\kappa_\cpt$), the strength of the phase transition ($\alpha_\cpt$), the velocity of the bubble wall ($v_{w,\cpt}$), the duration ($\beta_\cpt^{-1}$), and the temperature of the phase transition ($T_\cpt$).  The same quantities, {\it mutatis mutandis}, characterize the GW signature of an hPT (hGW for short), with the addition of two more that parametrize reheating. In this work we take RH to begin at a Hubble scale $H_i$ corresponding to a time when the reheaton, the particle whose decay reheats the Universe, starts decaying with a rate $\Gamma_\chi$.  The hPT parameters, \eg, $\beta_\hpt$, are related to their corresponding cPT parameters, \eg, $\beta_\cpt$, and they can be expressed in terms of each other up to $\mathcal{O}(1)$ numbers and RH parameters.

In general, the difference between an hPT and a cPT will be more than just a change in the parameters, as the dependence of the GW signal on both the PT parameters and frequency will change.  For example, in a cPT plasma sound waves last for an entire Hubble time before the expansion of the Universe damps them.  The result of this prolonged emission time is that the power in GWs is enhanced by a factor of $\beta_\cpt/H_\cpt$.  In a heating PT, sound waves will instead be damped by the heating process, which adds total energy to the plasma damping the sound waves and giving a smaller enhancement of $\beta_\hpt/\Gamma_\chi$.  While in this paper we focus more on the amplitude of the GWs, these same processes could also change the frequency dependence of the GW signal in interesting ways. In addition, novel plasma effects related to the restoration of symmetry that accompanies an hPT can enhance the amplitude of the GW signal coming from bubble collisions.

In this article, we study how an hPT percolates and describe how hPT and cPT parameters are related in a particularly simple model. Additionally, we study the special case when the heating and cooling PTs are both observable at future GW detectors.  In these lucky scenarios, information about reheating can likely be obtained.  Even if only an hPT were observed, it is possible that much could be learned from its frequency distribution.

In Sec.~\ref{Sec: toy PT} we present a toy model with a phase transition, the object of our study.  In Sec.~\ref{Sec: PT} we investigate the details of how a heating phase transition completes, and review the cooling case.  Section~\ref{Sec: GWs} describes how GWs are generated by a heating and cooling phase transition, highlighting their differences.  Section~\ref{Sec: prospects} details what is needed for an hPT to be found at future GW detectors.  Finally, we conclude in Sec.~\ref{Sec: conclusion}, and supplement our results with four appendices. For the reader's convenience, in Table~\ref{Tab:notation} we provide a list of our notation and some of the parameters most commonly used in the literature.

\begin{widetext}
\begin{center}
\begin{table}[ht]
\centering
  \begin{tabular}{|| c | l ||}
  \hline
    Variable & Meaning\\
    \hline
    $H_i$ & Hubble expansion rate when reheating starts. \\
    $\Gamma_\chi$ & Decay rate of the reheaton. \\
    $\rho_\chi$, $\rho_r$, $T$ & Energy densities in the reheaton and radiation, and the temperature of the latter. \\
    $\Phi$ & Higgs-like scalar field, whose spontaneous-symmetry-breaking potential drives the phase transition. \\
    $\Phib$, $\Phis$ & Broken and symmetric minima of the temperature-dependent $\Phi$ potential. \\
    $\{ \mu, A, \lambda \}$ & Coefficients of the quadratic, cubic, and quartic terms in the $\Phi$ potential. \\
    $0 < \Delta < 1$ & The useful combination $\Delta = 4 A^2/(3 \lambda \mu^2)$, which controls much of the physics of the phase transition. \\
    $T_\mx(t_\mx)$ & Maximum temperature during reheating, and the time at which it is reached. \\
    $T_c (t_c)$ & Critical temperature and time: when the broken and symmetric phases are degenerate. \\
    $T_0 (t_0) $ & Binodal temperature and time: when the symmetric phase becomes a maximum of the $\Phi$ potential. \\
    $T_1 (t_1)$ & Spinodal temperature and time:  when the broken phase no longer exists. \\
    $T_n(t_n)$ & Nucleation temperature and time: when one bubble per Hubble patch is formed. \\
    $T_\cpt(t_\cpt)$ & Percolation temperature and time: when the {\it cooling} phase transition completes. \\
    $T_\hpt(t_\hpt)$ & The same as above, but for a {\it heating} phase transition. \\
    $\GVh$ & Bubble nucleation rate per unit volume. \\
    $S$ & Euclidean bounce action of the bubble nucleation rate. \\
    $h(t)$ & Metastable volume fraction: the fraction of the volume of the Universe in the false vacuum. \\
    $n_b(t)$ & Bubble number density. \\
    $\ov{R}_\pt$ & Average distance between bubbles at percolation time. \\
    $\beta$ & Inverse duration of the phase transition.\\
    \hline
  \end{tabular}
  \caption{Summary of the notation used in this paper to denote various quantities of interest.}
  \label{Tab:notation}
\end{table}
\end{center}
\end{widetext}

\section{A Toy Model With a Phase Transition} \label{Sec: toy PT}

There are two main ingredients in the toy model that we consider. The first one deals with how RH takes place. While there is a plethora of ways to achieve RH, as a representative toy model we consider a Universe whose energy density is entirely contained in a reheaton field $\chi$, with RH proceeding via $\chi$ decays. For simplicity we assume that the daughter particles constitute an interacting dark sector (DS) with $g_*$ degrees of freedom, which quickly form a thermal bath.\footnote{For the $\mathcal{O}(\TeV)$ temperatures we consider, an interaction rate $\Gamma_{\rm th} \sim g^2 T$ is efficient enough to thermalize the DS plasma.} Eventually this DS plasma reheats the visible sector (VS) containing the Standard Model (SM) via some portal interactions, whose form is irrelevant to our purposes and we thus leave unspecified. This scenario is characterized by two parameters: the Hubble scale $H_i$ at which $\chi$ starts to decay (which depends on the initial energy density $\rho_\chii$ of the reheaton) and the decay rate $\Gamma_\chi$.\footnote{The reheaton may itself be the inflaton in simple models such as $m^2 \chi^2$ inflation. In this case we expect $\Gamma_\chi \sim H_i \sim m_\chi$ for $\mathcal{O}(1)$ couplings, since inflation terminates when $\chi \sim \mpl$, where $\mpl$ is the Planck mass~\cite{Kofman:1997yn}. In other models such as hybrid inflation~\cite{Linde:1993cn}, the reheaton and inflaton are separate particles, and $H_i$ and $\Gamma_\chi$ are not necessarily related.} Roughly speaking, the reheaton-dominated era lasts for a time $\Delta t_\chi \sim 1/\Gamma_\chi$ equal to its lifetime, after the onset of its decay.

The other ingredient of our toy model is the field $\Phi$ responsible for the first-order phase transition. $\Phi$ is a component of a thermal bath after reheating, and will eventually generate GWs. As is standard, we base our model of $\Phi$ on the Higgs boson, where thermal corrections give rise to a cubic term in its potential and, consequently, to a first-order phase transition.  We consider a finite temperature $T$ potential \cite{Linde:1981zj,Enqvist:1991xw,Quiros:1999jp}
\be\label{eq:pot}
    V = \frac{\mu^2}{2} (T^2 - T_0^2) \Phi^2 - \frac{A}{3} T \Phi^3 + \frac{\lambda}{4!} \Phi^4 \ .
\ee
The parameters $A$, $\mu$, $\lambda$, and  $T_0$ have model-dependent values but we allow them to vary freely, in order to ensure the wide applicability of our results. These parameters come about from the coupling of $\Phi$ to other particles and the subsequent mass difference between the two phases $\Phis \equiv \angu{\Phi}_{\rm symmetric} = 0$ and $\Phib \equiv \angu{\Phi}_{\rm broken} \ne 0$. The particles and their interactions with $\Phi$ also play a crucial role in the dynamics driving bubble expansion. Note that the term $-\mu^2 T_0^2 \Phi^2 / 2$ clearly corresponds to a tachyonic tree-level mass for $\Phi$, which leads to the usual spontaneous-symmetry-breaking mechanism at zero temperature, with $\angu{\Phi}_0 \equiv \angu{\Phi}_{b,T=0} = \sqrt{3/\lambda} \mu T_0$ and $V_0 \equiv V(\angu{\Phi}_0) = - 3 \mu^4 T_0^4 / (8 \lambda)$.

\begin{figure}
  \includegraphics[width=\linewidth]{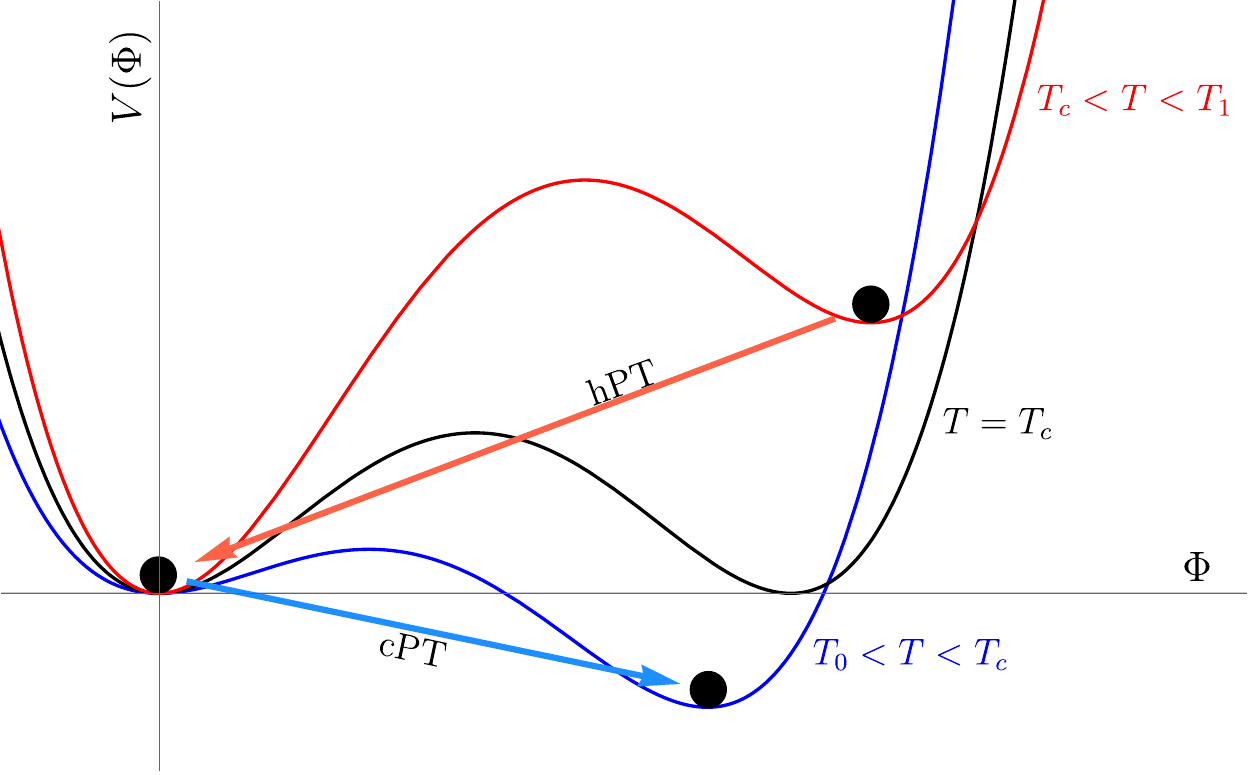}
  \caption{Finite-temperature potential $V(\Phi)$, with its symmetric ($\Phis = 0$) and broken ($\Phib \ne 0$) minima. For $T < T_0$ ($T > T_1$) the minimum corresponding to the symmetric (broken) phase disappears. At the critical temperature $T_c$ both minima are degenerate. A cooling phase transition ({\bf blue arrow}) from the metastable symmetric phase to the stable broken phase can take place for subcritical temperatures $T < T_c$, whereas a heating phase transition ({\bf red arrow}) from the metastable broken phase to the stable symmetric phase can occur for supercritical temperatures $T > T_c$.}
  \label{fig:pot}
\end{figure}

There are several temperature values that are important in our model, as illustrated in \Fig{fig:pot}. These temperatures are determined by the parameter combination $\Delta \equiv 4 A^2/(3 \lambda \mu^2)$. The first is $T_0$, the temperature below which the symmetric phase $\Phis$ ceases to be a minimum: for $T < T_0$ only the broken phase, with $\Phib$, is a minimum.  The second temperature is the {\it critical temperature} $T_c = T_0/\sqrt{1 - \Delta}$ at which the broken and symmetric phases have equal energies. Since degenerate minima are a requirement for there to be a first-order PT, we demand our potential parameters to satisfy $\Delta < 1$, which allows for the critical temperature to exist. For subcritical temperatures ($T<T_c$) the broken phase is energetically preferred by the system, whereas for supercritical temperatures ($T>T_c$) the symmetric phase is more energetically favorable.  Finally there is the temperature $T_1 = T_0 \sqrt{8/(8 - 9 \Delta)}$ above which the broken phase ceases to exist: for $T > T_1$, the only minimum is the symmetric phase. Note that if $\Delta  \geq 8/9$ the broken phase always exists. In the literature $T_0$ and $T_1$ are sometimes called the {\it binodal} and {\it spinodal} temperatures, respectively. If, however, $A = 0$ ($\Delta = 0$ and $T_0 = T_1 = T_c$) there is no potential barrier separating the symmetric and broken phases and the phase transition is second order. Furthermore, thermal corrections to the self-energy of particles (what is commonly called ``daisy resummation'' \cite{Gross:1980br,Parwani:1991gq,Carrington:1991hz,Arnold:1992rz,Quiros:1999jp}) may lower or erase the $\Phi$ potential barrier at high temperatures, thus weakening the strength of the first-order phase transition, or even negating it altogether. This puts a more stringent lower bound on $A$ and therefore on $\Delta$. As an estimate of this bound, we demand that these corrections be smaller than 50\% at $T_c$, which means $\Delta \gsim 0.27 \lambda/\mu^2$ or $A \gsim 0.45 \lambda$. See \Fig{fig:daisy_runaway} and \App{app:daisy} for more details.

\section{Phase Transitions during Reheating} \label{Sec: PT}

In this section we detail the dynamics of phase transitions during reheating. While the cases of heating and cooling phase transitions are very similar, there are important differences. Because of this, we review some of the previous literature on first-order phase transitions, occasionally highlighting our new results. Of these, our discussion of heating phase transitions during reheating takes the center stage. Although mentioned in passing in Refs.~\cite{Jiang:2015qor,Co:2020xaf}, a detailed study of the properties of hPTs taking place during RH has not, to the best of our knowledge, been published in the literature.

\begin{figure}
  \includegraphics[width=\linewidth]{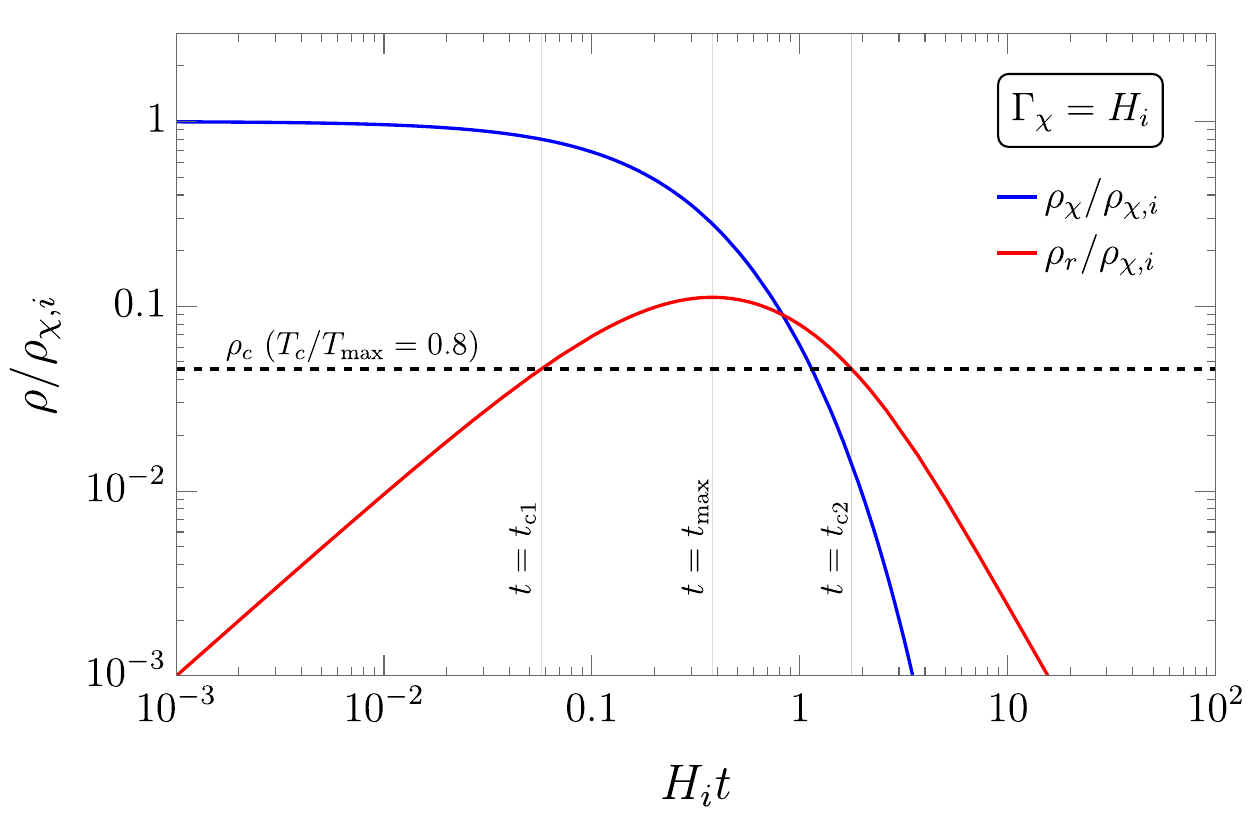}
  \caption{Thermal history of the Universe in our toy model, in which a reheaton energy density $\rho_\chi$ ({\bf blue}) reheats a radiation energy density $\rho_r$ ({\bf red}) via decays. The reheating history is entirely determined by two parameters: the reheaton decay rate $\Gamma_\chi$ and the Hubble expansion rate $H_i$ at the time $t_i = 0$ at which these decays begin to take place. The curves in the figure only depend on the ratio of the two parameters; we chose $\Gamma_\chi = H_i$ as our reference point. The {\bf dashed black line} represents the value $\rho_{r,\,c}$, where the radiation is at its critical temperature, here taken to be 80\% of the maximum temperature, $T_c=0.8~T_\mx$. $t_{c1}$ and $t_{c2}$ are the two times when $T = T_c$.}
  \label{fig:rh}
\end{figure}

The thermal history that we consider is shown in \Fig{fig:rh}.  Initially all of the energy density is in $\chi$, the temperature is zero and $\Phi$ is in the broken phase.  As $\chi$ decays, a DS thermal bath develops and the temperature of the Universe increases as a function of time. During this ``heating'' era the energy density in the radiation grows linearly with time, $\rho_r \approx \Gamma_\chi t \, \rho_\chii$. Eventually the temperature grows larger than $T_c$ (\ie~ the radiation density is $\rho_r > \rho_r(T_c) \equiv \rho_\rc$) and the broken phase in which the Universe finds itself becomes metastable. This means that a first-order hPT can now occur via the nucleation of bubbles of the stable, symmetric phase, which subsequently expand. At some point, corresponding to a temperature $T_\hpt$, these bubbles fill the entire Universe, which has now fully transitioned to the symmetric phase, and we can say that the hPT is completed.

Once the Universe reaches its maximum temperature $T_\mx$ at a time $t_\mx$, it begins to cool down due to Hubble expansion.  Famously, $T_\mx$ is larger than what is commonly known as the reheating temperature, the temperature of the plasma after the energy transfer from the reheaton \cite{Albrecht:1982mp,Dolgov:1982th,Abbott:1982hn,Scherrer:1984fd,Allahverdi:2010xz}. It depends on how much of the reheaton energy could be transformed into radiation before one Hubble time, roughly $\rho_\rmx \sim \rho_\chii \, \min[1, \Gamma_\chi/H_i]$.  After this maximum the temperature eventually falls below $T_c$, and the symmetric phase of the Universe is now metastable. The previous process repeats but in reverse, with bubbles of the broken phase forming and growing, eventually filling up the Universe at some time when it is at a temperature $T_\cpt$, at which point it can be said that the first-order cPT is finished.

There are a few necessary conditions for PTs to take place. The first one is $T_c < T_\mx$, namely that the Universe should reheat above the critical temperature of the system. Otherwise the broken phase is always stable and the Universe remains in it throughout reheating, which means no PT takes place. It is also important that the PT completes before the metastable minimum disappears, \ie, $T_\cpt > T_0$ and $T_\hpt < T_1$. If this is not satisfied, $\Phi$ will simply roll down to the stable minimum before a significant number of bubbles are formed, thereby preventing the production of a sizable amount of GWs. There are two times $t_{c1}$ and $t_{c2}$ when $T = T_c$, which take place while the plasma is heating and cooling respectively; see \Fig{fig:rh}. In addition, we denote by $t_1$ the time at which $T = T_1$ and the broken phase disappears, and by $t_0$ the time at which $T = T_0$ and the symmetric phase disappears. Therefore, the conditions stated above can be understood as follows: the hPT must complete at a time $t_\hpt$ within the interval $(t_{c1}, \min[t_{c2}, t_1])$, whereas the cPT must complete at a time $t_\cpt$ within $(t_{c2}, t_0)$.

The similarities and differences between cPTs and hPTs force us to make a brief housekeeping comment on notation: we use the generic subindex ``PT'' to denote that a quantity is evaluated at the end of a PT ($t_\pt$), for statements that apply equally to the heating or cooling cases. However, if it is imperative to specify whether the quantity in question is evaluated at the end of an hPT or a cPT ($t_\hpt$ or $t_\cpt$, respectively), we make this explicit. Throughout the rest of this section we review the dynamics of PTs, highlighting the differences between the well-studied cPTs and the novel hPTs.

The most common definition in the literature of the end of a first-order PT is the time $t_\pt$ at which the fraction $h(t_\pt)$ of the volume of the Universe found in the metastable phase ({\it metastable volume fraction} for short) has been reduced to $1/e$ \cite{Enqvist:1991xw,Hindmarsh:2020hop}. This is often called the {\it percolation time}. It can be shown \cite{Guth:1981uk,Guth:1982pn,Enqvist:1991xw,Hindmarsh:2020hop} that $h(t)$ is given by\footnote{For simplicity we ignore the expansion of the Universe in \Eqs{eq:h}{eq:nb}. We have checked that this simplification has only a negligible impact on our results for the parameter space of interest; see \App{app:bubbles}.}
\be\label{eq:h}
    h(t) = \exp \bl[ - \int\limits_{t_c}^{t} \! \dd t' ~ \GVa{(t')} \, \frac{4 \pi}{3} v_w^3 (t - t')^3 \br] \ ,
\ee
where $v_w$ is the bubble-wall velocity, and $\GVh$ is the bubble nucleation rate per unit volume \cite{Linde:1980tt,Linde:1981zj,Hindmarsh:2020hop}
\be\label{eq:GV}
    \GV \approx T^4 \bl( \frac{S}{2 \pi} \br)^{3/2} \, e^{-S} \ .
\ee
Here $S$ is the Euclidean bounce action associated with nucleating a $\Phi$ critical bubble of the lower-energy phase (see \App{app:action} for its precise definition).

\begin{figure}
  \includegraphics[width=\linewidth]{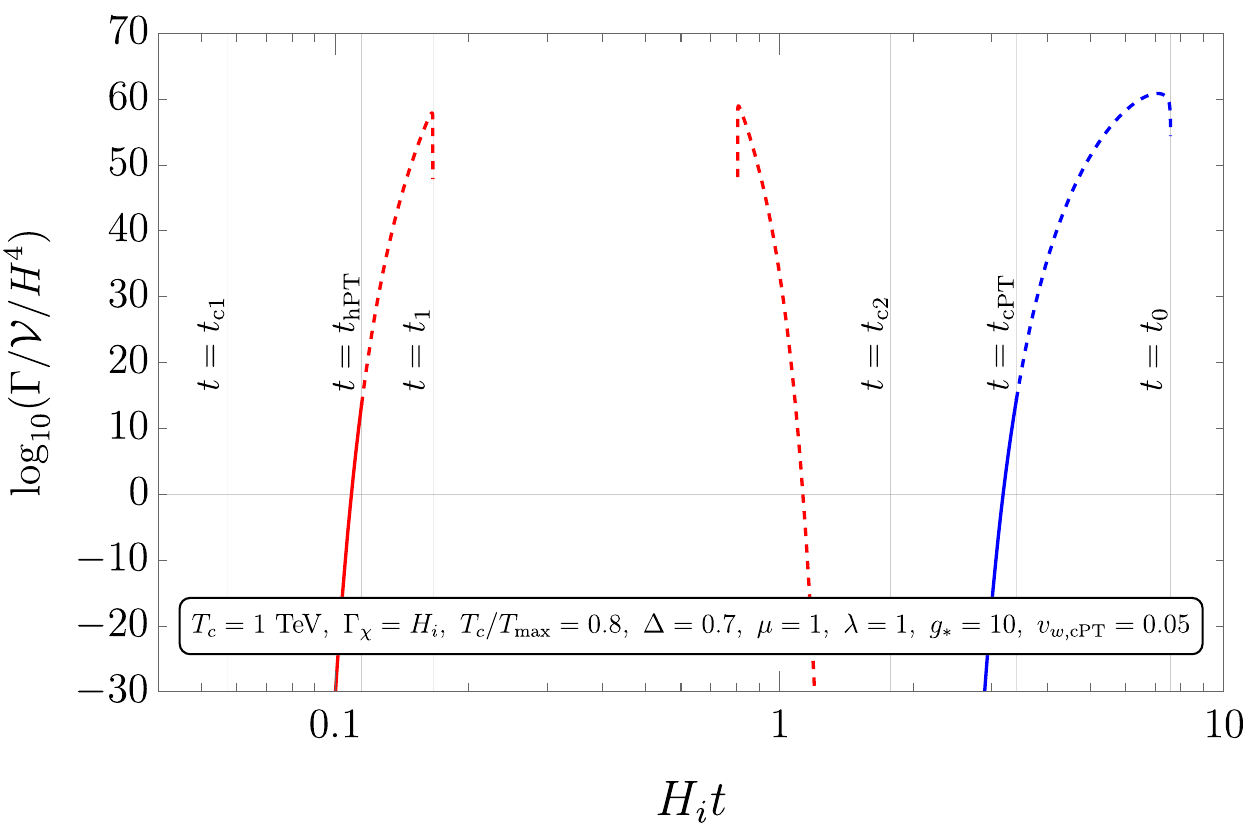}
  \caption{Example of the bubble nucleation rate $\GVh$ divided by $H^4$ as a function of time $t$. The {\bf red curve} corresponds to nucleation during an hPT ($t \in (t_{c1}, \min[t_{c2}, t_1]$), whereas the {\bf blue curve} corresponds to nucleation during a cPT ($t \in (t_{c2}, t_0$)). The {\bf dashed lines} indicate the {\it unrealized} evolution of $\GVh$, since they correspond to times after the phase transitions have actually finished. The horizontal line marks $\GVh = H^4$. The vertical lines correspond to different times of interest. For this plot we chose the parameters corresponding to the starred benchmark point in \Fig{fig:param}, namely $T_c = 1~\TeV$, $H_i = 2 \times 10^{-15}~\TeV = \Gamma_\chi$ (\ie, $T_c = 0.8~T_\mx$), and $g_* = 10$, the potential coefficients $\{ \mu, A, \lambda \} = \{ 1, 0.72, 1 \}$ ($\Delta = 0.7$), and a bubble-wall speed in the cPT of $v_{w,\cpt} = 0.05$.}
  \label{fig:nucl}
\end{figure}

Because at $T_c$ the broken and symmetric phases are degenerate, there is no free energy available to generate a phase transition from one to the other. As such, $S$ diverges and $\GVh$ vanishes. Since $S$ is a monotonically decreasing function of $\abs{T-T_c}$, the rate $\GVh$ grows exponentially from $0$ as the temperature moves away from $T_c$ in either direction, \ie, both during the hPT and the cPT; this behavior can be seen in \Fig{fig:nucl}. As long as the parameter $A$ controlling the height of the potential barrier that separates both phases is not too large, this exponential growth will generally guarantee a time at which the probability of nucleating one bubble in a Hubble space-time patch approaches 1. The time at which this happens is called the {\it nucleation time} $t_n$ \cite{Guth:1981uk,Morrissey:2012db,Caprini:2015zlo,Caprini:2019egz} and it is roughly given by
\be\label{eq:nuclTime}
    \GVa{(t_n)} \approx H(t_n)^4 \ .
\ee

The exponential sensitivity of $\GVh$ to $S$ means that in most cases the phase transition will occur in what is called the {\it exponential nucleation regime} \cite{Turner:1992tz,Kamionkowski:1993fg,Cutting:2018tjt,Hindmarsh:2019phv}, which can be observed in \Fig{fig:nucl}.\footnote{The exponential nucleation regime is the norm in most of our parameter space, for both hPTs and cPTs. Nevertheless, there is another. Because $\GVh$ vanishes at both $t_c$ (since $S(T_c) = \infty$) and $t_0$ or $t_1$ (since $S(T_0) = S(T_1) = 0$), Rolle's theorem guarantees that the function $\GVh$ has a maximum as a function of time. In the region of parameter space for which this maximum is comparable to the Hubble rate (\ie, $\GVh \lsim H^4$), the so-called {\it simultaneous nucleation regime} \cite{Cutting:2018tjt,Hindmarsh:2019phv} takes place. In this regime most bubbles of the new phase are nucleated at the time when $\GVh$ is at its largest. This occurs more easily when $T_\mx$ is reached during an hPT, which happens for large bounce actions. For more details about this regime, we direct the reader to \App{app:bubbleAppx}.}
In this regime, we can approximate the bounce action as $S \sim S_0 + S_1 t$ so that the nucleation rate is changing exponentially quickly.  As a result the times $t_c \lsim t_n \lsim t_\pt$ all take place in quick succession, and most of the bubbles of the new phase are nucleated towards the tail end of the process. In this regime, since $t_n \approx t_c$, we can write
\ba
    \GVa{(t_n)} & \approx & H(t_c)^4 \label{eq:nuclExp} \\
    \Rightarrow \bl( \GVa{(t_n)} \br)^{1/4} & \approx & 10^{-15}~\TeV ~  \sqrt{ \frac{\rho_\tot(t_c)}{1~\TeV^4} } \ , \label{eq:nuclExp2} \\
    \text{and } \ S(t_n) & \approx & 4 \ln \bl( T_c/H(t_c) \br) \ , \label{eq:SnuclExp}
\ea
where in the last equality we have solved for $S(t_n)$ from \Eq{eq:nuclExp} by keeping the dominant exponential behavior in \Eq{eq:GV} and ignoring the prefactor.

Once percolation is achieved and the PT ends at $t_\pt$, the bubbles of the new phase are large enough that they are very close to each other and begin to collide. It is these collisions and the subsequent behavior of the system that give rise to gravitational waves. The quantity of interest is the {\it mean bubble separation scale at percolation} $\ov{R}_\pt$, defined in terms of the bubble number density $n_b(t)$ as follows:
%\
\ba
    \ov{R}_\pt & \equiv & n_b(t_\pt)^{-1/3} \ , \label{eq:Rpt} \\
    \text{with} \  n_b(t) & \equiv & \int\limits_{t_c}^{t} \! \dd t' ~ \GVa{(t')} \, h(t') \ . \label{eq:nb}
\ea

From $\ov{R}_\pt$ and the bubble-wall velocity $v_w$ one can obtain the characteristic time scale $\beta^{-1}$ of the PT\footnote{Reference~\cite{Hindmarsh:2019phv} called this quantity $\beta_\eff$.}
\be\label{eq:beta_eff}
    \beta \equiv \bl( 8 \pi \br)^{1/3} \, \frac{v_w}{\ov{R}_\pt} \ ,
\ee
It can be shown that in the exponential nucleation regime \cite{Turner:1992tz,Kamionkowski:1993fg,Cutting:2018tjt,Hindmarsh:2019phv}
\be\label{eq:beta_exp}
    \beta \approx \bl. \frac{d \ln \Gamma}{d t} \br\vert_{t_\pt} \approx -S'(t_\pt) = - \bl. S \frac{d \ln S}{d \ln T} \frac{d \ln T}{d t} \br\vert_{t_\pt}  \ ,
\ee
which is the definition of $\beta$ more commonly found in the literature. However \Eq{eq:beta_eff} has a wider range of applicability and it is more closely related to the peak frequency of the GW spectrum \cite{Cutting:2018tjt,Cutting:2020nla}. Because of this, we use \Eq{eq:beta_eff} in our results, which are numerically calculated.

There is one last significant difference between cooling and heating phase transitions, regarding the manner in which their respective bubbles expand. Indeed, while during a cPT bubbles generally reach a constant subluminal velocity, in an hPT they instead typically enter a {\it runaway regime}, in which their wall velocity quickly approaches the speed of light ($v_w \rightarrow 1$). Below we justify this claim in a more or less quantitative manner, leaving a more detailed discussion and a description of the runaway parameter space to \App{app:expansion}. Finally, we would like to caution the reader that the growth of bubbles in the presence of a plasma is governed by very complex dynamics, and it is the subject of ongoing research \cite{Bodeker:2009qy,Espinosa:2010hh,Bodeker:2017cim,Dorsch:2018pat,Caprini:2019egz,Hoche:2020ysm}.

Once nucleated, the bubbles of the new stable minimum grow due to the free energy difference inside and outside of their wall. We can determine whether these bubbles run away by considering a relativistic bubble wall (where friction is maximized and antifriction is minimized) and asking if the net pressure acting on it is pushing outwards, driving the wall ever faster. Mathematically, written in terms of the total force per unit area $P_\tot$ acting on the bubble wall, this runaway condition is given by
\be
    P_\tot = \Delta V_0 + \Delta P_T > 0 \ , \label{eq:runaway_condition}
\ee
where $\Delta V_0 \equiv V_{0,\, \rm out} - V_{0,\, \rm in} = \mathrm{sign} ( T_c - T ) \abs{V_0}$ is the zero-temperature potential difference between the outside and the inside of the bubble (for cPTs (hPTs) the zero-temperature broken minimum $V_0 < 0$ is inside (outside) the bubble), and $\Delta P_T$ is the pressure difference produced by the plasma, which is the same as the leading mean-field contribution to the thermal potential \cite{Bodeker:2009qy}:
\be
    \Delta P_T \approx \frac{T^2}{24} \, \sum\limits_i c_i N_i \Delta m_i^2 = - \mathrm{sign}(T_c - T) \frac{\mu^2}{2} T^2 \Phib^2 \ . \label{eq:pressure_difference}
\ee
Here $\Delta m_i^2 \equiv m_{i, \, \mathrm{out}}^2 - m_{i, \, \mathrm{in}}^2$ is the $i$-th particle's mass difference between the outside and the inside of the bubbles, $N_i$ accounts for its degrees of freedom, and $c_i = 1 \ (1/2)$ for bosons (fermions). The second equality in \Eq{eq:pressure_difference} stems from the definition of $\mu$ in terms of the particle interactions (see \Eq{eq:muA_defs}) and from the fact that the masses of the particles depend on their Yukawa couplings to $\Phi$. The sign takes into account that $\Phi = \Phib$ inside cPT bubbles (for which  $T < T_c$) , while the opposite is true for hPTs.

\begin{figure}[tb]
	\centering
        \includegraphics[width=.8\linewidth]{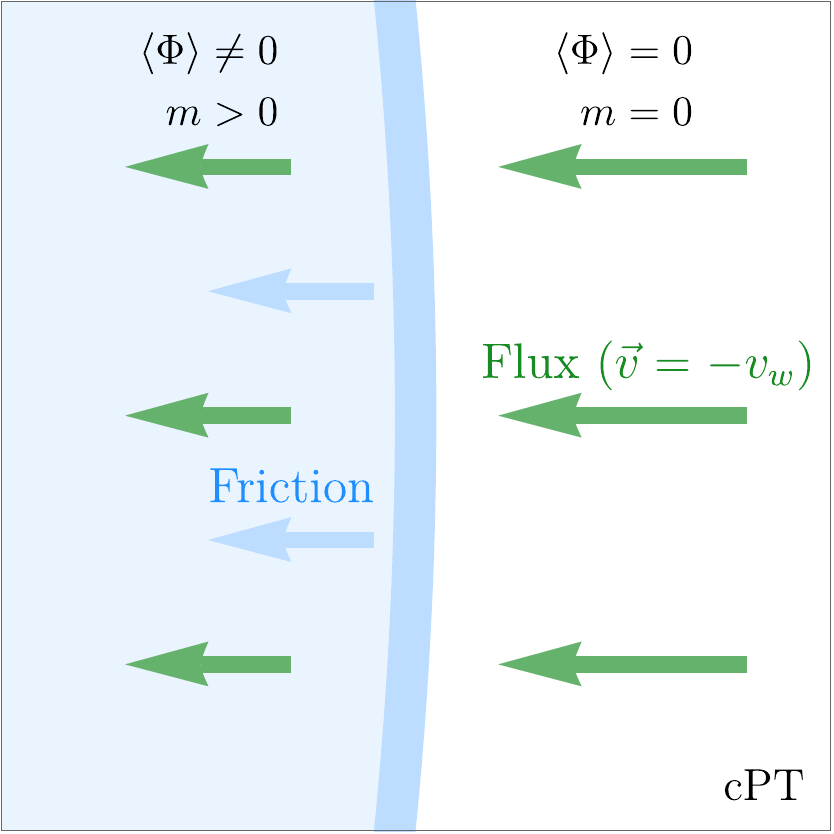}
        \includegraphics[width=.8\linewidth]{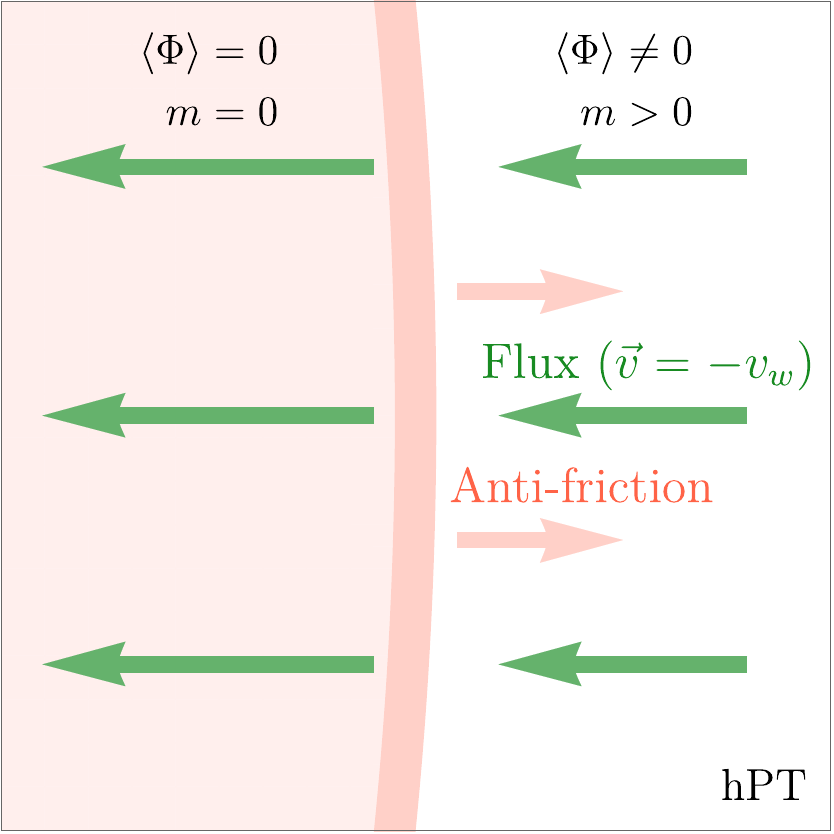}
	\caption{Dynamics of the bubble wall in its rest frame. The plasma particles of the incoming flux ({\bf green arrows} to the right of the bubble wall) move with a velocity of $-v_w$. {\bf Top:} the case of a cooling phase transition, where the plasma particles {\it gain} mass upon entering the bubble ({\bf blue region}), thereby slowing down ({\bf short green arrows} to the left of the bubble wall). Momentum conservation means the plasma exerts a friction force on the wall ({\bf blue arrows}), which moves it to the left; in the plasma rest frame the bubble wall is slowed down. {\bf Bottom:} the case of a heating phase transition, where the particles {\it lose} mass inside the bubbles ({\bf red region}) and thus accelerate ({\bf long green arrows}), thereby accelerating the bubble wall ({\bf red arrows}).}
	\label{fig:momentum}
\end{figure}

This sign of $\Delta P_T$ can be readily interpreted in terms of momentum conservation in the bubble rest frame \cite{Bodeker:2009qy,Bodeker:2017cim} with the help of \Fig{fig:momentum}. In this frame the bubble wall is static, while an incoming flux of plasma particles is moving towards it with velocity $- v_w$. For cPTs (with $T < T_c$) the massless plasma particles transmitted through the wall gain a mass $m_{i, \, \mathrm{in}}^2 > 0$, thus lowering their momentum. Momentum conservation then implies that the bubble wall needs to make up for the missing momentum by moving along the direction of the incoming flux of plasma particles, which corresponds to a slowing down of the bubble expansion. For hPTs (with $T > T_c$) the process is the opposite: the transmitted particles lose their outside mass $m_{i, \, \mathrm{out}}^2 > 0$ and thus gain momentum, which the bubble wall needs to balance out by accelerating in the direction opposite to the incoming particle flux.

As a result, this pressure difference acts as a plasma friction on the bubbles of a cPT, or as a plasma {\it anti}friction on the bubbles of an hPT. The former case has been much discussed in the previous literature, while the latter is presented in detail, to the best of our knowledge, for the first time in this work.\footnote{Phase transitions during reheating were briefly mentioned in Refs.~\cite{Jiang:2015qor,Co:2020xaf} in the contexts of both the Standard Model and inflaton models with dynamical decay rates respectively.} Intuitively, the mass{\it less} plasma particles {\it outside} the bubbles of a cPT experience a $\Phi$ potential barrier at the wall. This means that these particles can bounce off of the bubble walls or lose momentum upon entering the bubble, thereby exerting a friction on the bubbles and slowing down their growth. This friction typically ends up balancing the force driving the bubble expansion, and a constant $v_w$, often subluminal, is reached \cite{Bodeker:2009qy,Espinosa:2010hh,Bodeker:2017cim,Hoche:2020ysm}. Heating phase transitions, on the other hand, have almost exactly the opposite behavior. During an hPT the plasma particles are mass{\it less} in the {\it interior} of the bubbles, where $\Phis = 0$. As a result, it becomes energetically favorable for these particles to get inside the bubbles: the plasma is ``sucked in'' by them. As the plasma particles pass through, they transfer their energy to the bubble walls, accelerating them, and leading to a bubble runaway regime.

Owing to the fact that $T_\cpt \sim T_c \sim T_\hpt$ for generic values of the potential parameters, we can see that the bubble runaway conditions for cPTs and for hPTs described in \Eq{eq:runaway_condition} are reflections of each other: when the condition is satisfied in one case, it will typically {\it not} be satisfied in the other. This makes sense because for runaways to take place in cPTs, the friction of the plasma acting on the bubble has to be small, whereas in hPTs the antifriction has to be large; ultimately it is $\mu$, which parametrizes the strength of the interaction between the plasma particles and the $\Phi$ field, that determines the size of both friction and antifriction. Therefore, if a given $\mu$ produces enough plasma friction to cause the bubbles of a cPT to expand at a constant wall speed, it will also cause that very same plasma to exert instead an antifriction on the bubbles of the hPT, accelerating them into a runaway. Because the plasma is transferring its energy into the bubble walls in order to accelerate them, this means that the energy available in an hPT for GW production in bubble collisions can be very large.

\section{Gravitational Waves during Reheating} \label{Sec: GWs}

The frequency spectrum of a SGWB is typically described in terms of the fraction of the total energy density of the Universe found in GWs, per frequency $e$-fold: $(d \rho_\gw/d \ln f)/\rho_\tot$. Denoting with an asterisk all those quantities evaluated at the time $t_*$ at which the GWs are produced, we can find the SGWB spectrum today by accounting for the difference in $\rho_\tot$ and the redshift as follows \cite{Grojean:2006bp,Caprini:2019egz,Hindmarsh:2020hop}:
\ba
    \Omega (f) & = & F_* \, \frac{1}{\rho_{\tot,*}} \frac{d \rho_{\gw,*}}{d \ln f} \nonumber \\
    & \approx & F_* \, \bl( \frac{H_*}{\beta} \br)^2 \bl( \frac{\kappa \rho_{\rm s,*}}{\rho_{\tot,*}} \br)^2 \, N S(f) \ , \label{eq:sgwb}\\
    \text{with } \  F_* & \equiv & a_*^4 \bl( \frac{H_*}{H_0} \br)^2 \ . \label{eq:Fstar}
\ea
In the first equality the prefactor $F_*$ accounts for the radiation-like redshifting of the GWs with the scale factor $a_*$ and for the ratio of the total energy densities at $t_*$ and today,\footnote{$\rho_\crit$ here denotes the critical energy density of the Universe today, and is not to be confused with $\rho_\rc$, the energy density of radiation at $T_c$.} $\rho_{\tot,*}/\rho_\crit = H_*^2/H_0^2$. In the last equality we have written the energy density in GWs in terms of the energy density $\rho_{\rm s,*}$ of their source: $\rho_{\gw,*} \sim G \tau^2 ( \kappa \rho_{\rm s,*} )^2$. Here $G \sim H^2/\rho_\tot$ is Newton's constant, $\tau \sim \beta^{-1}$ is the typical time scale of GW production during a PT, and $\kappa$ is an efficiency factor that quantifies how much of the energy $\rho_{\rm s,*}$ in the source goes into GWs. The factors $N$ and $S(f)$ account for an overall normalization and spectrum, respectively, and they may depend on other phase transition parameters, such as the bubble-wall velocity $v_w$.

The sources of GWs during a PT are typically of three kinds: bubble collisions, sound waves, and magnetohydrodynamic turbulence. The first one involves the energy stored in the $\Phi$ bubble walls, while the last two come from the response of the plasma to the nucleation and percolation of the bubbles of the new phase. Each of them has different frequency spectra and dependences on the PT parameters. Their precise form and relative contribution to the overall GW signal is the subject of ongoing research (see Refs.~\cite{Caprini:2015zlo,Caprini:2019egz,Hindmarsh:2020hop} for reviews). In cPTs, the case most commonly studied in the literature, runaway bubbles are not typically expected and the contribution from sound waves tends to be the largest \cite{Alanne:2019bsm,Hindmarsh:2020hop}.\footnote{When runaway bubbles do occur in cPTs, however, their resulting GW spectrum can be very prominent; see Ref.~\cite{Lewicki:2022pdb}.} On the other hand, as discussed in the previous section, the same plasma exerting a friction on cPT bubbles will instead exert an antifriction on hPT bubbles, accelerating them. Because of this, most of the energy is stored in the bubble walls, and we expect that in hPTs the dominant GW contribution comes from the collisions of runaway bubbles. Since we are interested in the detectability of GWs from reheating as a proof of concept, and since the contribution from turbulence is the most uncertain \cite{Caprini:2015zlo,Caprini:2019egz,Hindmarsh:2020hop}, we will not consider it throughout the rest of this paper, and we focus instead on GWs coming from bubble collisions for hPTs and from plasma sound waves for cPTs.

\subsection{Bubble-wall collisions}

In this subsection we briefly discuss the gravitational waves generated by bubble-wall collisions. We study this signature in the context of hPTs, where it is the dominant source of GWs, whereas they are typically subdominant in cPTs \cite{Alanne:2019bsm,Hindmarsh:2020hop}.
The GW spectrum of a bubble-wall collision is typically calculated numerically with the addition of the envelope approximation \cite{Kosowsky:1991ua,Kosowsky:1992vn,Huber:2008hg,Weir:2016tov,Cutting:2018tjt,Lewicki:2019gmv,Lewicki:2020jiv,Lewicki:2020azd,Lewicki:2022pdb}, which approximates the bubbles as an expanding set of infinitely thin shells that disappear when the transition completes. 
In these numerical calculations, the Hubble expansion is typically neglected as the PTs being studied are assumed to complete very quickly.  In the hPTs that we consider, we make a similar assumption.  However, as the injection of energy is dictated by $\Gamma_\chi$, we instead assume that the phase transition completes quickly relative to both $\Gamma_\chi$ and $H_\hpt$.  Under this assumption, the numerical results apply equally well to cPTs and hPTs.

The GW spectrum is found numerically to be\footnote{In some more recent numerical studies~\cite{Cutting:2018tjt}, the scalar field oscillates after the bubble collision, giving a typical time scale longer than $1/\beta$. If these results hold, then the frequency dependence $S(f)$ may change.}
\ba
\Omega_\bc(f) & = & F_* \, \bl( \frac{\kappa_\Phi \alpha}{1 + \alpha + R_\chi} \br)^2 \bl( \frac{H_*}{\beta} \br)^2 \nonumber \label{eq:Ombc}\\
&& \times N_\bc(v_w) S_\bc(f) \ , \label{eq:Nbc}\\
N_\bc(v_w) & = & \frac{0.11 v_w^3}{0.42 + v_w^2 } \ , \\
S_\bc(f) & = & \frac{3.8 (f/f_\bc)^{2.8}}{1 + 2.8 (f/f_\bc)^{3.8}} \ , \label{eq:Sbc}\\
f_\bc & = & a_* \beta \, \bl( \frac{0.62}{1.8 - 0.1 v_w + v_w^2} \br)  \ . \label{eq:fpkbc}
\ea

Below we briefly explain the various parameters describing the GW spectrum from bubble collisions, with the aid of \Eqst{eq:1Rchi_cases_app}{eq:dlnTdt_cases_app}. For illustrative purposes we focus on the case of a short era of reheaton domination ($\Delta t_\chi \sim \Gamma_\chi^{-1} \lesssim H_i^{-1}$), for which these parameters take simple forms. For more details, as well as the case of an arbitrary duration of reheaton domination, we refer the reader to \App{subsec:analytic_rh}.\footnote{Note that in the typical case of cPTs occurring within the visible sector during radiation domination, $R_\chi = 0$, $H_*^2 = \pi^2 g_{\rm SM,*} T_{\rm SM}^4/( 90 \mpl^2 )$, and $a_* = ( g_{\rm SM,0}/g_{\rm SM,*} )^{1/3} ( T_{\gamma 0}/T_{\rm SM,*} )$, and thus \Eqst{eq:Ombc}{eq:fpkbc} reduce to those found in the literature. The same is true of \Eqst{eq:Omsw}{eq:fpksw} below.}

\paragraph*{$t_*$}: The time at which the GWs are generated. It corresponds roughly to when the bubble collisions take place, which in turn is very close to the percolation time $t_\hpt$. We therefore take $t_* \approx t_\hpt$.

\paragraph*{$T_*$}: The temperature of the plasma at the time when the GWs are generated. From the previous paragraph, $T_* \approx T_\hpt > T_c$. Except for very fine-tuned $\Phi$-potential parameters, the hPT and critical temperatures are similar in scale, $T_\hpt \sim T_c$.

\paragraph*{$H_*$}: The hPT typically finishes at a time $t_\hpt$ during the reheaton-dominated era, so $H_* \approx H_\hpt \approx H_i$.

\paragraph*{$R_\chi$}: The energy density of the reheaton over the radiation density, at $t_*$: $R_\chi \equiv \rho_{\chi,*} / \rho_{r,*} \approx \rho_\chii/\rho_\rhpt = 3 H_i^2 \mpl^2 / \rho_\rhpt$.

\paragraph*{$a_*$}: The scale factor at which the hPT takes place. If the reheaton-domination era is shorter than one Hubble time, it is given roughly by $a_* \sim 8 \times 10^{-17} ( 1~\TeV/T_\mx )$. A longer reheaton domination corrects this expression with a factor that depends on $\Gamma_\chi/H_i$. There is also a mild dependence on the degrees of freedom of the dark and visible sectors.

\paragraph*{$F_*$}: This redshift factor depends on the photon and total energy densities today, as well as the duration of the reheaton-dominated era. It is approximately given by $F_* \sim 4 \times 10^{-5}$ for short reheaton domination.

\paragraph*{$\alpha$, $\kappa_\Phi$}: The vacuum energy in the scalar $\Phi$ compared to the energy in the radiation; and the corresponding efficiency factor that quantifies how much of it goes into the bubble walls (essentially gradient energy of the $\Phi$ field), determined by \Eq{eq:runaway_condition} in the runaway bubble regime:
\ba
    \alpha & \equiv & \frac{\abs{V_0}}{\rho_r} \ , \label{eq:alpha_hpt}\\
    \kappa_\Phi & \equiv & \frac{P_\tot}{\abs{V_0}} = \frac{\alpha_\infty - \alpha}{\alpha} \ , \label{eq:kappa_hpt}\\
    \text{where } \quad \alpha_\infty & \equiv & \frac{\Delta P_T}{\rho_r} \ . \label{eq:alpha_infty}
\ea
In our work we never consider vacuum domination but only reheaton domination and radiation domination, \ie, $\alpha < 1, \, R_\chi$. Note that, up to a minus sign, \Eq{eq:kappa_hpt} is the same expression quantifying how much energy goes into the bubble wall for runaway cPTs \cite{Bodeker:2009qy,Espinosa:2010hh,Caprini:2015zlo,Bodeker:2017cim,Caprini:2019egz,Ellis:2019oqb,Ellis:2020nnr}. The sign difference, seen in \Eq{eq:runaway_condition}, stems from the fact that the direction of the PT is the opposite in hPTs than in cPTs. As discussed in the previous section this sign means that, while in most of their parameter space cPT bubbles do not run away (reaching a subluminal terminal bubble-wall velocity) and $\kappa_\Phi$ is consequently a number much smaller than 1, for hPTs we have instead $\alpha_\infty > \alpha$ for most of their parameter space, because the plasma antifriction makes the hPT bubbles run away, greatly increasing the energy stored in their walls. Note that our definition of $\alpha$ here is directly borrowed from previous literature, which has only dealt with cPTs. It is therefore {\it not} the most natural way to parametrize the energy density available in an hPT, which is not $\abs{V_0}$. As a result of this definition, $\alpha_\infty > \alpha$ in runaway hPTs means that $\kappa_\Phi$ can be much larger than $1$. See \Eq{eq:runaway1} and \Fig{fig:daisy_runaway} as well as \App{app:expansion} for a more detailed discussion on the runaway condition.

\paragraph*{$v_w$}: As discussed in the previous section, for most of the region of parameter space in which a strongly first-order phase transition takes place, the plasma exerts an antifriction on the bubble walls during an hPT, leading to a runaway regime and therefore $v_w \rightarrow 1$. The parameter space leading to a runaway hPT can be found in \Fig{fig:daisy_runaway} in \App{app:expansion}.

\paragraph*{$\beta$}: The inverse of the characteristic time scale $\tau$ of the PT, given by \Eq{eq:beta_exp} in the exponential regime. Up to $\mathcal{O}(1)$ factors $S(t_\hpt) \lsim S(t_n)$, while $d \ln T/d t \lsim \Gamma_\chi \rho_\chii/(4 \rho_\rhpt)$ from the equations governing RH (\Eq{eq:rad_cont}). From this and our discussions in \Apps{app:actionAppx}{app:bubbleAppx}, we find
\be\label{eq:betahPT}
    \beta_\hpt \sim 4 \ln \bl( T_c/H_i \br) \, \abs{\frac{d \ln S}{d \ln T}}_\hpt \, \frac{\Gamma_\chi}{4} \frac{\rho_\chii}{\rho_\rhpt} \ ,
\ee
\be\label{eq:DSDT_hPT}
    \text{where } \quad \abs{\frac{d \ln S}{d \ln T}}_\hpt \sim 91 \bl( \Delta^{-1} - 1.08 \br) \ .
\ee

\subsection{Plasma sound waves}

For the cPTs that we consider, the main source of GWs are the sound waves. The SGWB created by sound waves is found to be
\ba
\Omega_\sw(f) & = & F_* \, \bl( \frac{\kappa_\sw \alpha}{1 + \alpha + R_\chi} \br)^2 \bl( \frac{H_*}{\beta} \br) \nonumber \label{eq:Omsw}\\
&& \times N_\sw(v_w) S_\sw(f) \ , \label{eq:Nsw}\\
N_\sw(v_w) & = & 0.159 \, v_w \ , \\
S_\sw(f) & = & \bl( \frac{f}{f_\sw} \br)^3 \bl( \frac{7}{4 + 3 (f/f_\sw)^2} \br)^{7/2} \ , \label{eq:Ssw}\\
f_\sw & = & a_* \frac{2}{\sqrt{3}} \frac{\beta}{v_w}  \ . \label{eq:fpksw}
\ea
Note that, unlike the GWs sourced by bubble collisions, the spectrum from sound waves scales like one power of $H_*/\beta$ rather than two. This is because the fluid bulk motion sourcing the GWs lasts for about a Hubble time, longer than the PT duration $1/\beta$ \cite{Caprini:2015zlo,Caprini:2019egz}.\footnote{Although recently Refs.~\cite{Ellis:2018mja,Ellis:2020awk} have shown that this is not a generic feature of all models.} We continue focusing on the simple case of a short reheaton-dominated era, in which case the cPT typically occurs during radiation domination.

\paragraph*{$t_*$, $T_*$, $H_*$, $R_\chi$, $a_*$, $F_*$}: The GWs are generated at a time $t_*$ shortly after the cPT is completed at $t_\cpt$; we therefore simply assume $t_* \approx t_\cpt$. For typical potential parameters $T_* \approx T_\cpt \lesssim T_c$. The values of $H_*$ and $R_\chi$ at this time will be smaller than their hPT counterparts due to both the expansion of the Universe and the reheaton decays. If the cPT completes firmly during the post-reheating radiation-dominated era then $R_\chi \approx 0$, while $H_* = H_i \sqrt{\rho_\rcpt/\rho_\chii} = \pi^2 g_* T_\cpt^2 /(90 \mpl)$, $a_* \approx 8 \times 10^{-17} (1~\TeV/T_\cpt)$, and $F_* \approx 4 \times 10^{-5}$. However, the cPT may also take place during reheaton domination. Estimates of these quantities for both cases are listed in \Eqst{eq:1Rchi_cases_app}{eq:dlnTdt_cases_app} in \App{subsec:analytic_rh}.

\paragraph*{$v_w$}: In a typical cPT, the plasma exerts a friction that balances out the expansion force of the new phase, which leads to the bubble walls moving at a constant, typically subluminal speed. Complex model-dependent dynamics govern the expansion of bubbles in a thermal plasma, and their terminal velocity can in principle be derived from these; in this work we simply bypass the issue by taking $v_w$ in a cPT to be a free parameter.

\paragraph*{$\alpha$, $\kappa_\sw$}: These quantify the energy released during a cPT and how much of it goes into the bulk motion of the fluid. There are several ways to compute these quantities in the literature. The simplest one relies on the {\it bag model}, where the vacuum energy is used, and thus $\alpha = \abs{V_0}/\rho_r$ \cite{Espinosa:2010hh}. Recent theoretical and numerical developments advocate instead for the use of the trace of the stress-energy-momentum tensor \cite{Hindmarsh:2017gnf,Caprini:2019egz,Giese:2020rtr,Giese:2020znk}. However, in this latter model a more thorough knowledge of the plasma fluid is necessary in order to compute $\kappa_\sw$, (\eg, the plasma sound speeds of both the symmetric and broken phases), knowledge that we do not have. We simply use the bag model and the corresponding fits to $\kappa_\sw(\alpha, v_w)$, conveniently provided in Ref.~\cite{Espinosa:2010hh}. For the subluminal cPT bubble-wall velocities we consider in this work, $\kappa_\sw \approx 5 \alpha v_w^{6/5}$.

\paragraph*{$\beta$}: Also given by \Eq{eq:beta_exp} in the exponential regime. Since the temperature is cooling due to the Hubble expansion of the Universe, $d \ln T/d t = -H_*$. It can then be shown that during radiation domination
\be\label{eq:betacPT}
    \beta_\cpt \sim 4 \ln \bl( T_c/H_i \br) \, \abs{\frac{d \ln S}{d \ln T}}_\cpt \, H_* \ ,
\ee
\be
    \text{where } \quad \abs{\frac{d \ln S}{d \ln T}}_\cpt \sim 9.8 \bl( \Delta^{-1} - 0.52 \br) \ ; \label{eq:DSDT_cPT}
\ee
the estimate for the general case is shown in \Eq{eq:beta_exp_estim} (see \Apps{app:actionAppx}{app:bubbleAppx} for more details).

Finally, we say a few words about GWs from sound waves in hPTs. Since the runaway regime is common in hPTs, the plasma puts energy into accelerating the bubble walls and therefore bubble collisions dominate GW production, with sound waves most likely playing a subdominant role. It is difficult to tell, without dedicated numerical simulations, whether the GWs from sound waves in hPTs have a different frequency spectrum or amplitude than their cPT counterparts. This certainly seems possible: a first difference between both scenarios is that while in cPTs the bubble walls inject energy into the plasma, leading to radially outward fluid bulk motion and subsequently sound waves, in hPTs energy is removed from the bath and put into the bubble walls, with the sound waves moving in the opposite direction to the walls. Yet another difference is the duration of the sound waves sourcing the GWs. For a cPT, the Hubble expansion eventually damps the sound waves. For an hPT, the reheaton deposits its energy in the thermal bath at a rate $\sim \Gamma_\chi \rho_\chi / \rho_r$, thereby increasing the radiation energy density and damping the sound waves. The duration of the sound waves in hPTs is then $\sim \rho_\rhpt/(\Gamma_\chi \rho_\chi)$ As a result, we expect that the amplitude of the GWs from sound waves in an hPT will differ from that in a cPT by a factor of $\sim (H_\cpt / \Gamma_\chi) (\rho_\rhpt / \rho_\chi)$, accounting for their shorter relative duration.

\subsection{Comparison between the GW spectra from cooling and heating phase transitions}

The above discussion allows us to compare the GW spectra for hPTs and cPTs, which are dominated by collisions and sound waves, respectively. In the simplest case of a short reheaton-dominated era the ratios of both take their simplest forms:
\begin{widetext}
\ba
    \frac{\Omega_\hpt}{\Omega_\cpt} & \sim & 0.5 \, \frac{F_\hpt}{F_\cpt} \bl( \frac{\kappa_\Phi}{\kappa_\sw} \br)^2 \frac{\bl( 1 + R_\chi \br)^2_\cpt}{\bl( 1 + R_\chi \br)^2_\hpt} \frac{H_\hpt^2 \beta_\cpt}{H_\cpt \beta_\hpt^2} \, v_{w,\cpt}^{-1} \nonumber\\
    & \sim & \bl( \frac{H_i}{\Gamma_\chi} \br)^2 \bl( \frac{\rho_\rc/\rho_\chii}{0.03} \br)^4 \bl( \frac{F_\hpt}{F_\cpt} \br) \bl(\frac{0.05}{v_{w,\cpt}}\br) \bl( \frac{\kappa_\Phi}{30} \br)^2 \bl( \frac{0.01}{\kappa_\sw} \br)^2 \bl( \frac{30}{\ln \bl( T_c/H_i \br)} \br) \bl( \frac{\abs{\frac{d \ln S}{d \ln T}}_\cpt / 10}{\abs{\frac{d \ln S}{d \ln T}}_\hpt^2 / 100} \br) \ , \label{eq:Omegas_ratio} \\
    \frac{f_\hpt}{f_\cpt} & \sim & 0.2 \, v_{w,\cpt} \, \frac{a_\hpt}{a_\cpt} \frac{\beta_\hpt}{\beta_\cpt} \nonumber\\
    & \sim & 0.5 \, \bl( \frac{\Gamma_\chi}{H_i} \br) \bl( \frac{T_c}{T_\mx} \br) \bl( \frac{0.03}{\rho_\rc/\rho_\chii} \br)^{3/2} \bl( \frac{v_{w,\cpt}}{0.05} \br) \bl( \frac{\abs{\frac{d \ln S}{d \ln T}}_\hpt / 10}{\abs{\frac{d \ln S}{d \ln T}}_\cpt / 10} \br) \ , \label{eq:fpks_ratio}
\ea
\end{widetext}
where we denote the cPT bubble-wall speed by $v_{w,\cpt}$, we take $T_\pt \approx T_c$, and the numerical benchmark values we show are typical of our parameter space. We remind the reader that these expressions were derived only for a shortly lived reheaton-domination era, that they are to be taken only as heuristic, and are to be trusted only as order-of-magnitude estimates.

From the equations above it can be seen that, barring a fine-tuning of the $\Phi$ potential parameters in order to get vastly different $\kappa_\Phi$ and $\abs{d \ln S /d \ln T}$, the amplitude of the SGWB from both hPT and cPT can be of the same order of magnitude for a modest coincidence between $\rho_\rc$ and $\rho_\chii$, of about a couple of orders of magnitude (or equivalently a coincidence between $T_c$ and $T_\mx$ of a factor of a few). This coincidence has to be more severe the larger the hierarchy between $\Gamma_\chi$ and $H_i$. The peak frequencies of the spectra can also be close to each other, with $f_\hpt$ slightly smaller than $f_\cpt$ for subluminal cPT bubble-wall velocities and the aforementioned coincidence between $\rho_\rc$ and $\rho_\chii$ (since of course $T_c \lesssim T_\mx$). Increasing the $\Gamma_\chi/H_i$ ratio inverts the order of the peaks, with $f_\hpt$ eventually becoming bigger than $f_\cpt$. Furthermore, in the limit of a long reheaton-dominated era ($\Gamma_\chi \ll H_i$), the hGWs are typically quieter than the cGWs. This is both because the ratio $\rho_\rc/\rho_\chii$ is smaller (since $\rho_\rc \leq \rho_\rmx \sim \rho_\chii \Gamma_\chi/H_i$ in this limit), and because more time has elapsed between $t_\hpt$ and $t_\cpt$, which means that the relative redshift suppression ($F_\hpt / F_\cpt$) between both spectra becomes more significant.

\begin{figure}
  \includegraphics[width=\linewidth]{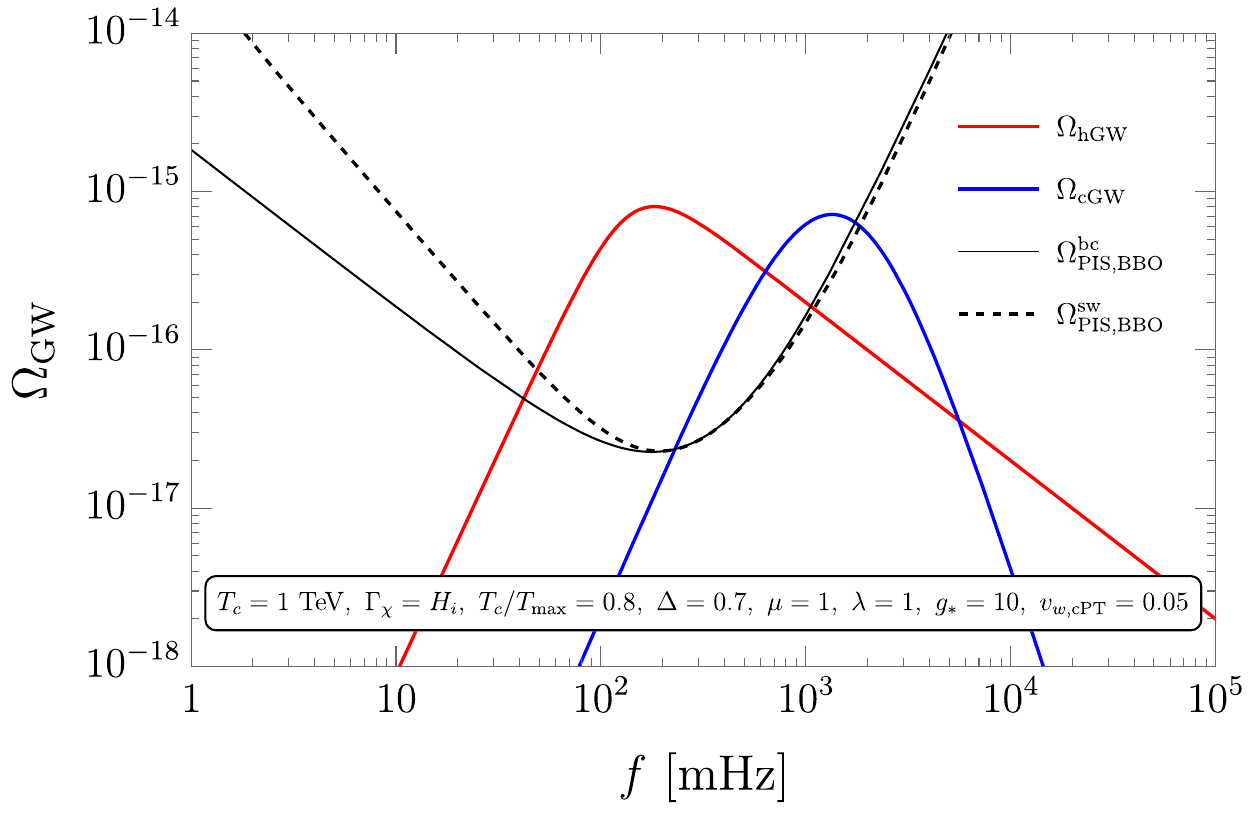}
  \caption{Stochastic GW background from the phase transitions occurring during reheating. The {\bf red curve} represents the GW spectrum arising from bubble collisions during the hPT (hGW) and the {\bf blue curve} represents the GW spectrum originating from sound waves in the cPT (cGW). Both spectra are detectable by the future GW probe BBO \cite{Crowder:2005nr,Corbin:2005ny,Harry:2006fi}, as shown by the PIS curves \cite{Schmitz:2020syl} ({\bf black lines}, {\bf solid} for GWs from bubble collisions, and {\bf dashed} for GWs from sound waves). For this plot we chose the parameters corresponding to the starred benchmark point in \Fig{fig:param}, namely $T_c = 1~\TeV$, $H_i = 2 \times 10^{-15}~\TeV = \Gamma_\chi$ (\ie, $T_c = 0.8~T_\mx$), and $g_* = 10$, the potential coefficients $\{ \mu, A, \lambda \} = \{ 1, 0.72, 1 \}$ ($\Delta = 0.7$), and a bubble-wall speed in a cPT of $v_{w,\cpt} = 0.05$.}
  \label{fig:gw}
\end{figure}

Example cPT and hPT SGWBs, for a benchmark parameter space point, are shown in \Fig{fig:gw}. The characteristics described above are easily appreciated in this figure. As this example illustrates the same detector, in this case the future BBO probe \cite{Crowder:2005nr,Corbin:2005ny,Harry:2006fi}, can observe both spectra. In the lucky case that both signals are loud and have sufficiently separated peak frequencies, so as to not have one of them buried under the other, one could then in principle distinguish between them, identifying which belongs to the cPT and which to the hPT. Then the correlations between them, of which \Eqs{eq:Omegas_ratio}{eq:fpks_ratio} are examples, would allow one to extract both the scale of reheating and the reheaton decay rate.

\section{Prospects at Future Gravitational-Wave Detectors} \label{Sec: prospects}

\begin{figure*}
  \includegraphics[width=0.8\linewidth]{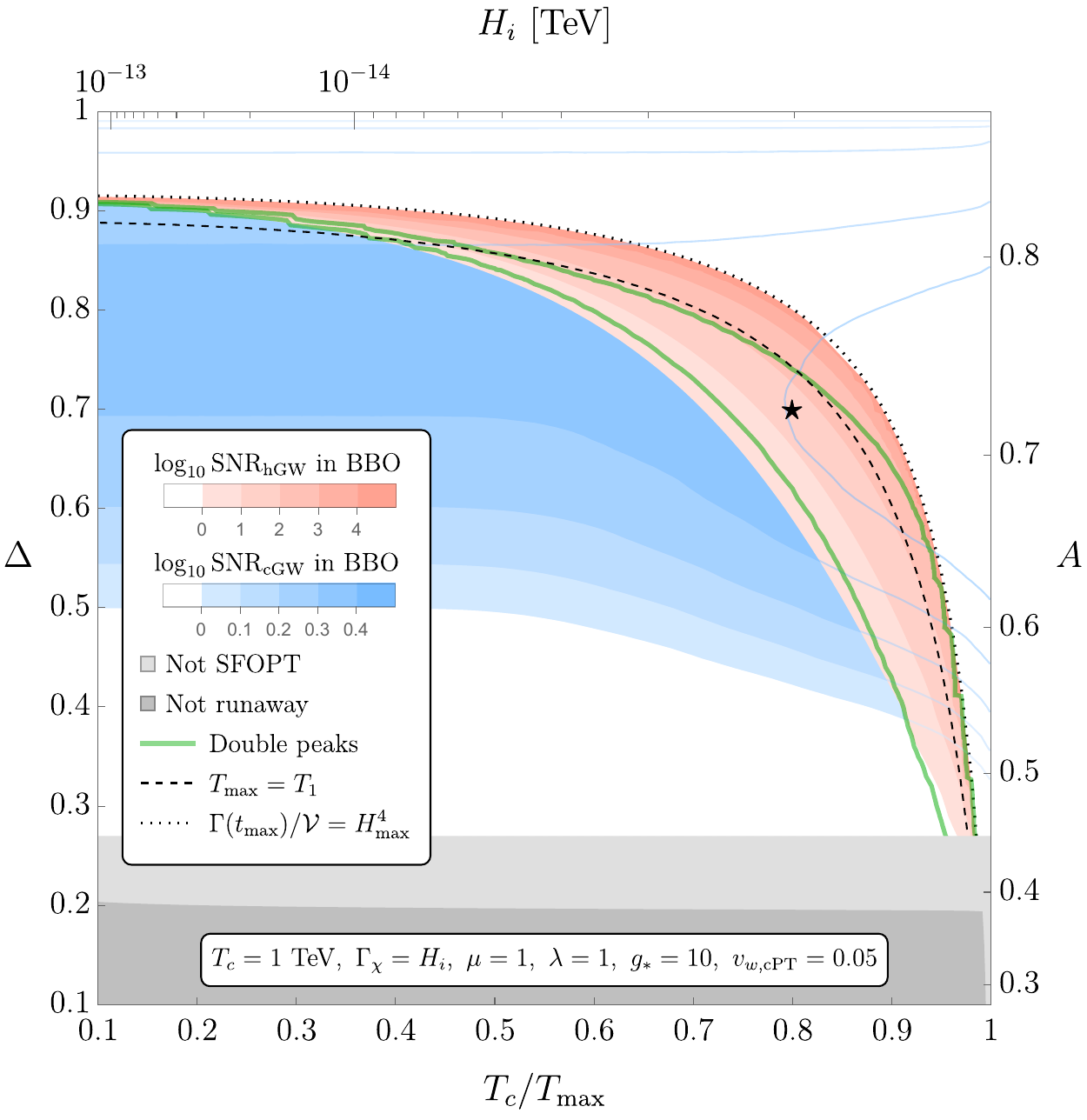}
  \caption{SNR contours for 1-year observation time of the SGWB spectra generated during reheating by hPT bubble collisions ({\bf red}) and cPT sound waves ({\bf blue}), for the upcoming BBO detector \cite{Crowder:2005nr,Corbin:2005ny,Harry:2006fi}, as a function of the ratio $T_c/T_\mx$ and the potential parameter $\Delta$. The corresponding values of $H_i$ and $A$ are also shown. The {\bf star} corresponds to the benchmark point used in the previous figures. Points within the {\bf green contour} have a total SGWB with double peaks (from both hPT and cPT). The $\Phib$ minimum never disappears in the region above the {\bf dashed line}, since $T_\mx < T_1$. In the space above the {\bf dotted line} $\GVh < H^4$ at $t_\mx$, which means that no PT takes place and no GWs are generated. In the {\bf dark grey region} there are no runaway hPT bubbles, whereas in the {\bf light grey region} the daisy contributions to the thermal potential prevent an SFOPT. For this plot we have chosen $T_c = 1~\TeV$, $\Gamma_\chi = H_i$, $g_* = 10$, $\{ \mu, \lambda \} = \{ 1, 1 \}$, and a bubble-wall speed in a cPT of $v_{w,\cpt} = 0.05$.}
  \label{fig:param}
\end{figure*}

In this section we explore the visibility of GWs generated by cooling and heating PTs in future detectors. We focus on the upcoming BBO experiment \cite{Crowder:2005nr,Corbin:2005ny,Harry:2006fi}, as it is the most relevant detector for our parameter choice. We quantify visibility in terms of the signal-to-noise ratio (SNR) of the GW spectra.

To obtain the SNR of the PT GWs we employ the peak-integrated sensitivity (PIS) curves $\Omega_\textrm{PIS}(f)$ introduced in Ref.~\cite{Schmitz:2020syl}. We compute the SNRs of GWs, originating mostly from bubble collisions in hPTs and from plasma sound waves in cPTs (abbreviated as hGWs and cGWs respectively), by simply comparing the amplitudes of the GWs at their peak with $\Omega_\textrm{PIS}(f)$:
\begin{equation}
    \textrm{SNR}=\left[ \frac{t_\textrm{obs}}{1 \textrm{ year}}\left(\frac{\Omega_\textrm{GW}^{\textrm{bc/sw}}(f_\textrm{peak})}{\Omega_\textrm{PIS}^{\textrm{bc/sw}}(f_\textrm{peak})} \right)^2 \right]^{1/2}\ ,
\end{equation}
where $t_\textrm{obs}$ is the observation time, $f_\textrm{peak}$ is the frequency at which $\Omega_\textrm{GW}$ is the peak, and ``bc" (``sw") corresponds to the GWs from bubble collision (sound waves).

The results for hGWs and cGWs are shown in Fig.~\ref{fig:param} in terms of the $T_c/T_\mx$--$\Delta$ parameter space (and the corresponding values of $H_i$ and $A$, respectively). The benchmark point used in previous figures of this paper, such as in \Fig{fig:gw}, is marked with a star in \Fig{fig:param}. The SNR contours for $t_\textrm{obs}=1$ year for hGWs (cGWs) at BBO are shown in red (blue), where the increasing opacity indicates larger SNR. We have fixed $T_c = 1~\TeV$, $\Gamma_\chi = H_i$, $g_* = 10$, $\{ \mu, \lambda \} = \{ 1, 1 \}$, and $v_{w,\cpt} = 0.05$. This specific choice of values has only a modest impact on our results, and the GW features described in this section are generic. We direct our reader to Appendix \ref{app:additional} for a more detailed study of the parameter space.

Note that $\Delta$, which controls the height of the potential barrier separating the broken and symmetric minima of the $\Phi$ potential and thus the action $S$, strongly determines the strength of the GWs. For both hPTs and cPTs a larger $\Delta$ makes $d \ln S/ d \ln T$ smaller, which increases the duration $\beta^{-1}$ of the PT. Eventually, however, sufficiently large values of $\Delta$ will kill the GW signature by making the PT impossible, as clearly seen in the region above the dotted line. Indeed, this region corresponds to those points with $\Gamma(t_\mx)/\mathcal{V} < H(t_\mx)^4$. Since at $t_\mx$ $\GVh$ is at its largest (because the temperature is at its maximum and thus the action $S$ is at its minimum), no bubbles are produced within a Hubble patch in this region. This means that the Universe remains in the broken phase throughout all of reheating, never transitioning to the symmetric phase (via an hPT), and therefore never coming back to the broken phase (via a cPT). Thus no PT takes place and therefore no appreciable GWs are produced. We nevertheless show the continuation of the cPT contours in this empty region for illustrative purposes.

The parameter $T_c/T_\mx$ (which can be turned into $\rho_\rc/\rho_\chii$ for a given $\Gamma_\chi/H_i$ ratio) also has a crucial impact on the visibility of the hGWs. A large hierarchy between $T_c$ and $T_\mx$ means that the time elapsed between $t_\hpt$ and $t_\cpt$ (which are on opposite sides of the reheating curve of \Fig{fig:rh}) is also large. As such, the GWs associated with the hPT are produced much earlier than those generated during the cPT, and are therefore more redshifted and correspondingly quieter. Thus their signal falls outside of the BBO sensitivity window. The combination of this effect and the one controlled by $\Delta$ described in the previous paragraph gives the BBO-visible hPT region its characteristic crescent shape. The cGWs have a milder dependence on $T_c/T_\mx$. A strong coincidence between these two temperatures means $\rho_r$ represents a larger share of the total energy density, which makes the GWs louder. On the other end, the more different $T_c$ and $T_\mx$ are, the later the cPT takes place. For sufficiently large hierarchies the cPT occurs squarely during radiation domination and its GWs become insensitive to reheating. This is shown in \Fig{fig:param} as the insensitivity of the cPT contours to low values of $T_c/T_\mx$.

The region enclosed by the green contour corresponds to points where the peaks from both cGWs and hGWs can be distinguished. This ``double peaks region'' is therefore defined as the parameter space where the peak amplitude of hGWs is larger than the amplitude of the cGWs at that same frequency, and vice versa. These points are potentially the best ones in terms of how much we could learn from these GWs. Indeed, observing both GW peaks in a GW detector could allow us to extract the most information about reheating by studying the correlations between cooling and heating PTs, as we have attempted to do in this paper. The region located above the dashed line has $T_\mx < T_1$, which means that the plasma never reaches $T_1$ and therefore $\Phib$ never disappears. The dark gray region is the parameter space where the plasma antifriction during the hPT is not strong enough to cause the bubble walls to enter a runaway regime. Finally, the light gray region corresponds to those points where the daisy contribution becomes so large that the cubic term in \Eq{eq:pot} is severely suppressed and there is no strongly first-order PT (SFOPT). For more details on both grey areas see \Apps{app:expansion}{app:daisy}, respectively.

\section{Conclusions} \label{Sec: conclusion}

In this article, we explored phase transitions that occurred when the Universe was heating up, a process called reheating.  Reheating is a unique and interesting cosmological event in the history of the Universe, about which we currently know nothing.  If a phase transition occurred during reheating, then the resulting GW spectrum would carry information about the reheating process to us and teach us something about this exciting era of the early Universe.

We discussed how the GW signature of heating phase transitions depends on the phase transition and reheating parameters.  In some optimistic scenarios, such as the one depicted in \Fig{fig:gw} and in the green region of \Fig{fig:param}, it would be possible to see the same phase transition in both its heating and cooling directions.  If this came to pass then, by cross correlating the observed spectra and our knowledge of the phase transition in both directions, one could learn more about the process of reheating.  For example, the time scales $\beta^{-1}$ of the two phase transitions, which appear in the peak frequency of their respective GW spectra, are related by $\Gamma_\chi/H_{\cpt}$ up to $\mathcal{O}(1)$ numbers.  Therefore, by comparing the measured values of the peaks, one would be able to obtain the decay width of the reheaton.  Many other features are correlated between heating and cooling phase transitions.  Finite bubble-wall velocities in the cooling case tend to be correlated with runaway bubbles in the heating case, and vice versa if the roles were switched.  These correlations offer an opportunity to extract information about reheating.

While not explored in our article, it would be extremely interesting if the frequency spectrum of the GWs produced during a heating phase transition were radically different from what has been studied in the literature. In this paper we argued that the main contribution to the GWs in this case comes from bubble collisions, and assumed that the envelope approximation can be used to describe its spectrum. This assumption, while justified (see \Sec{Sec: GWs}), needs to be corroborated by dedicated numerical simulations. Furthermore, it is entirely possible that plasma sound waves represent a non-negligible source of GWs during a heating phase transition, and it is reasonable to expect that they would be different at the $\mathcal{O}(1)$ level from those in the cooling case, since cooling phase transitions tend to inject energy into the plasma while heating ones tend to remove it.  Additionally, the sound waves of cooling phase transitions are damped by the Hubble expansion, whereas those from heating transitions are damped by the injection of energy from reheaton decays. If differences of this sort between both spectra were to be firmly established, it is possible that the detection of GWs generated by a heating phase transition would by itself be enough to infer detailed information about reheating.  Future research along this direction is both necessary and of great interest to anyone attempting to uncover the physics of reheating.

\acknowledgments{
The authors thank William DeRocco, Peizhi Du, Isabel Garcia Garcia, Junwu Huang, Gordan Z. Krnjaic, Patrick Meade, Filippo Sala, for helpful discussions. MBA thanks Stephanie Buen Abad for reviewing this manuscript. MBA, JHC, and AH are supported in part by the NSF grant PHY-2210361 and the Maryland Center for Fundamental Physics. JHC is also supported in part by JHU Joint Postdoc Fund.}

\bibliography{gw_pt}
\bibliographystyle{apsrev4-1}

\clearpage
\newpage
\appendix
\onecolumngrid

\section*{APPENDICES}

The following appendices contain a detailed description of our model, including the reheating history of the Universe and the phase transitions that it undergoes. We describe the numerical methods employed to obtain the dynamics of the cooling and heating phase transitions and their subsequent GW spectra. The {\tt Mathematica} code companion to our paper, which we dubbed {\tt graphare}, computes the GW signatures arising from phase transitions during reheating, and is available at {\tt \href{https://github.com/ManuelBuenAbad/graphare}{github.com/ManuelBuenAbad/graphare}}. All of the results in this paper are obtained numerically with the aid of {\tt graphare}. It is nevertheless useful to find approximate analytic expressions that can help understand our results. We include such analytic expressions, which we utilize in the main text of this paper to aid in our discussions, in these appendices.\\

\section{Reheating History}
\label{app:RH}

The portion of our numerical tools dealing with the reheating history can be found in the {\tt DSReheating.wl} package of our {\tt graphare} code, along with an explanatory {\tt Mathematica} notebook titled {\tt 01\_reheating.nb}.

\subsection{Reheating equations and radiation temperature}

We begin by assuming that the Universe is dominated by a matter-like reheaton field $\chi$, which then reheats the Universe by decaying into light particles, with a decay rate $\Gamma_\chi$. For simplicity we assume {\it i.} that the decay products are particles of a DS with $g_*$ relativistic degrees of freedom in total, within which the PTs will take place, and {\it ii.} that the particles in this DS have interactions which are significant enough to reach thermodynamic equilibrium, thereby forming a thermal bath or plasma. The DS temperature is simply related to its energy density via $\rho_r = g_* \pi^2 T^4 / 30$. The DS will eventually reheat the VS above the $\TeV$ scale through a portal interaction, the details of which are irrelevant to our story. Note that thermal equilibrium in the DS can be easily achieved even for very small couplings among its particles, since the thermalization rate, heuristically given by $\Gamma_{\rm th} \sim g^2 T$, can easily be larger than the Hubble expansion rate $H \gsim T^2/\mpl$ for the temperatures $T \sim \mathcal{O}(\TeV)$ we consider.

The equations governing the evolution of the energy densities in the reheaton and DS radiation-like fluid are
\ba
    \dot{\rho}_\chi + 3 H \rho_\chi & = & - \Gamma_\chi \rho_\chi \ , \label{eq:chi_cont}\\
    \dot{\rho}_r + 4 H \rho_r & = & \Gamma_\chi \rho_\chi \ , \label{eq:rad_cont}
\ea
where $H^2 = (\rho_\chi + \rho_r)/(3 \mpl^2)$, the dot denotes derivatives with respect to time, and we start the clock at a time $t_i = 0$ when the $\chi$ decays start. At this time the initial densities are $\rho_\chii \ne 0$ and $\rho_{r,i} = 0$. The duration of the reheaton-dominated ($\chi$D) era is mainly determined by its lifetime, $\Delta t_\chi \sim \Gamma_\chi^{-1}$. An hPT takes place when the temperature is increasing, which roughly speaking can only occur during $\chi$D and before one Hubble time has elapsed, \ie, $t \lesssim \min[ \Gamma_\chi^{-1}, H_i^{-1} ]$. This is because once the reheaton particles have decayed away the Universe is in an era dominated by radiation (RD), which cools adiabatically (see \Eq{eq:rad_cont}); and because after one Hubble time the Universe, whether $\chi$D or RD, begins to cool down. During this heating era the radiation energy density grows linearly with time: $\rho_r \approx \Gamma_\chi t \, \rho_\chii$.

\begin{figure}
  \includegraphics[width=0.5\linewidth]{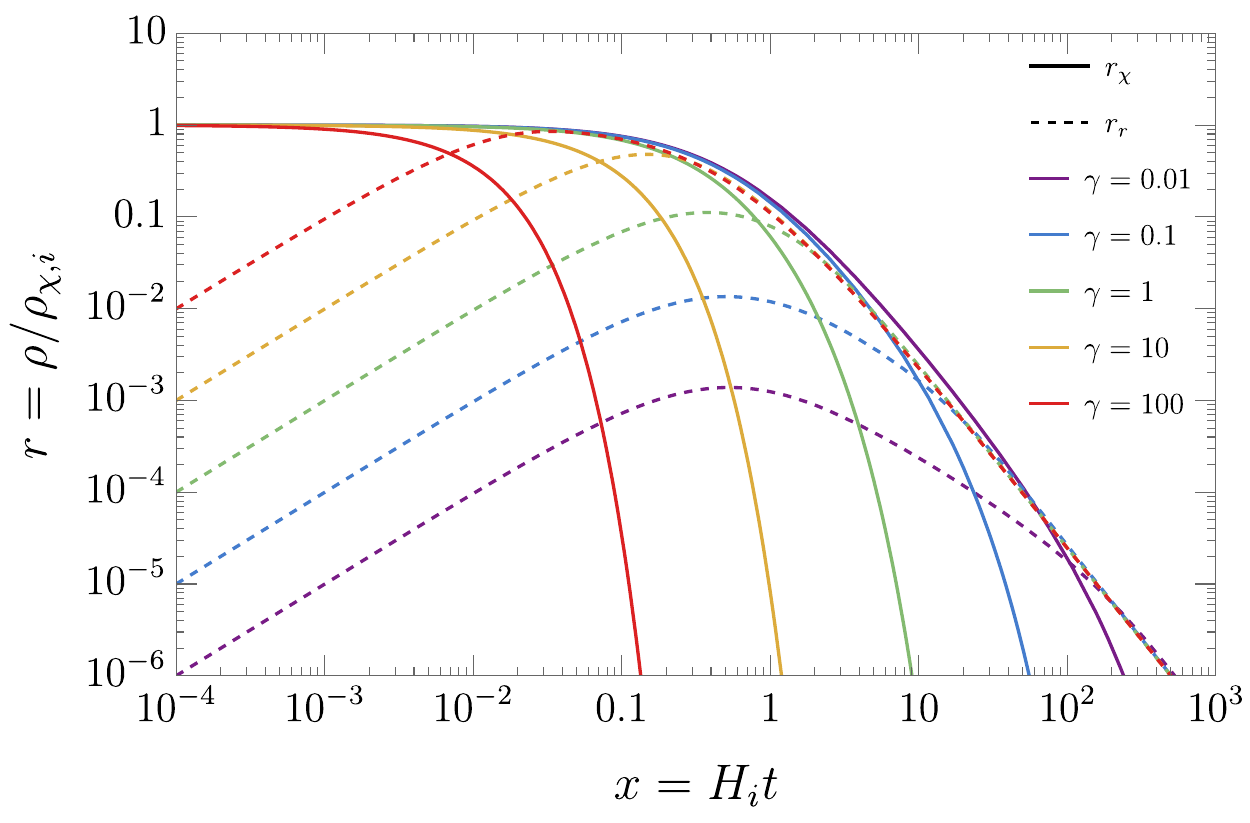}
  \caption{Thermal history of our reheaton-radiation toy model, in terms of the dimensionless quantities $x \equiv H_i t$, $r \equiv \rho / \rho_\chii$, and $\gamma \equiv \Gamma_\chi/H_i$. We plot the solutions to \Eqs{eq:chi_cont}{eq:rad_cont} for $\gamma \in \{ 0.01, 0.1, 1, 10, 100 \}$. One can easily appreciate that the reheaton-dominated era lasts for $\Delta t_\chi \sim \Gamma_\chi^{-1}$, and that the radiation temperature only increases for a time $t \lesssim \min[ \Gamma_\chi^{-1}, H_i^{-1} ]$, during which $r_r \approx \gamma \, x$.}
  \label{fig:rh_app}
\end{figure}

It is convenient to work with dimensionless quantities, which can then be appropriately rescaled. Throughout the rest of this paper we will work in terms of $x \equiv H_i t$ (with $H_i \equiv \sqrt{\rho_\chii/3 \mpl^2}$), $r \equiv \rho / \rho_\chii$, and $\gamma \equiv \Gamma_\chi/H_i$. Different thermal histories correspond then to different choices of the dimensionless parameter $\gamma$ and the only dimensionful quantity, $H_i$. In \Fig{fig:rh_app} we show the reheating history for $\gamma \in \{ 0.01, 0.1, 1, 10, 100\}$.

\subsection{Redshift}

Having solved for the thermal history of the Universe, we can determine the redshift $z_*$ between some time $t_*$ during reheating and today. This can be written in terms of the scale factor $a(t)$ as
\be\label{eq:redshift}
   1+z_* = \frac{1}{a_*} = \frac{a_\rd}{a_*} \frac{a_\vs}{a_\rd} \frac{1}{a_\vs} \ ,
\ee
where we have used the shorthand $a_X = a(t_X)$, $t_\rd$ denotes an arbitrary time during the radiation-dominated era ($\rho_r \gg \rho_\chi$), $t_\vs$ denotes a time shortly after the VS has been populated (due to portal interactions with the DS) and we have taken $a(t) = 1$ today.  The decomposition into three products is useful as some of these factors will be evaluated numerically.  The first factor is simply given by the number of $e$-folds between those two times:
\be\label{eq:ardstar}
    \frac{a_\rd}{a_*} = \exp \bl[ \int\limits_{x_*}^{x_\rd}\dd x ~ \sqrt{r_\chi(x) + r_r(x)} \br] \ .
\ee
The second term depends on the nature of the portal interaction that allows the DS to reheat the VS. For simplicity we simply assume that the VS is populated by the DS while both baths are relativistic. From energy conservation considerations, we can write
\be\label{eq:avsrd}
    \frac{a_\vs}{a_\rd} = \bl( \frac{g_*}{g_* + g_\vs} \br)^{1/4} \frac{T_\rd}{T_\vs} \ ,
\ee
where $T$ denotes the temperature of both baths, $T_X = T(t_X)$, and $g_\vs$ is the VS degrees of freedom reheated by the DS. Throughout this paper we assume that the entire SM is reheated, and therefore take $g_\vs = 106.75$. Note that $t_\rd$ is indeed arbitrary: an earlier time (and consequently hotter $T_\rd$) yields a smaller \Eq{eq:ardstar} and a larger \Eq{eq:avsrd} in the same proportion, and consequently the same \Eq{eq:redshift}. Finally, the last factor can be derived from the usual entropy considerations. Assuming no remaining DS light relics, this factor is
\be\label{eq:avs}
    \frac{1}{a_\vs} = \bl( \frac{g_* + g_\vs}{g_{s0}} \br)^{1/3} \frac{T_\vs}{T_{\gamma 0}} \ ,
\ee
where $g_{s0} = 3.94$ and $T_{\gamma 0} = 2.7255~\Kel \approx 2 \times 10^{-4}~\eV$ are today's entropy degrees of freedom and temperature, respectively.

The final exact expression is then
\be\label{eq:astarinv}
    a_*^{-1} = 1+z_* = \exp \bl[ \int\limits_{x_*}^{x_\rd}\dd x ~ \sqrt{r_\chi(x) + r_r(x)} \br] \, \bl( \frac{\bl( g_* + g_\vs \br)^{1/12}  g_*^{1/4}}{g_{s0}^{1/3}} \br) \, \frac{T_\rd}{T_{\gamma 0}} \ .
\ee

For an easier comparison to previous work, we can multiply and divide by $g_\vs^{1/3}$ and use \Eq{eq:ardstar} again to rewrite $a_*$ as \cite{Caprini:2019egz}
\ba
    a_* & = & \bxd{a_*} \ G_a \ \frac{a_* T_*}{a_\rd T_\rd} \ , \label{eq:astar_app}\\
    \text{with } \quad \bxd{a_*} & \equiv & \frac{T_{\gamma 0}}{T_*} \frac{g_{s0}^{1/3}}{g_\vs^{1/3}} \approx 8 \times 10^{-17} \ \bl( \frac{1~\TeV}{T_*} \br) \bl( \frac{106.75}{g_\vs} \br)^{1/3} \ , \label{eq:astar_bxd} \\
    \text{and } \quad G_a & \equiv & \frac{g_\vs^{1/3}}{\bl( g_* + g_\vs \br)^{1/12}  g_*^{1/4}} \ , \label{eq:Ga}
\ea
where $\bxd{a_*}$ denotes the value of $a_*$ for cPTs taking place during RD, the case commonly studied in the literature.

We can use our results to find $F_*$, the prefactor responsible for the redshift-related suppression of the GW spectra (see \Eq{eq:Fstar}), and compare it to its value $\bxd{F_*}$ for cPTs during RD \cite{Caprini:2019egz}:
\be
    F_* = a_*^4 \frac{H_*^2}{H_0^2} = \frac{T_{\gamma 0}^4 g_{s0}^{4/3}}{T_\rd^4} \frac{1}{\bl( g_* + g_\vs \br)^{1/3}  g_*} \bl( \frac{a_*}{a_\rd} \br)^4 \ \frac{\rho_{\tot,*}}{\rho_\crit} \nonumber
\ee
\ba
    \Rightarrow \quad F_* & = & \bxd{F_*} \ G_F \ \frac{a_*^4 \rho_{\tot,*}}{a_\rd^4 \rho_\rrd} \ , \label{eq:Fstar_app} \\
    \text{with } \quad \bxd{F_*} & \equiv & \frac{\rho_{\gamma,0}/2}{\rho_\crit} \, \frac{g_{s0}^{4/3}}{g_\vs^{1/3}} \approx 4 \times 10^{-5} \ \bl( \frac{106.75}{g_\vs} \br)^{1/3} \ , \label{eq:Fstar_bxd} \\
    \text{and } \quad G_F & \equiv & \bl( \frac{g_\vs}{g_* + g_\vs} \br)^{1/3} \ . \label{eq:GF}
\ea

\subsection{Analytic expressions for reheating}\label{subsec:analytic_rh}

Depending on the reheaton decay rate, one can classify the reheating history into two relatively easy to study broad classes: those with $\gamma \gg 1$, or those with $\gamma \ll 1$. For each of these cases we can find analytic expressions for various quantities of interest, such as the maximum temperature reached during reheating or the scale factor at a given temperature. We present these expressions here in a quick-and-dirty fashion. A more thorough job can be done, and is included in the explanatory notebook {\tt 01\_reheating.nb} of {\tt graphare}.

\subsubsection{Fast reheaton decays: $\gamma \gg 1$}

Before discussing the consequences of extremely fast reheaton decays, $\gamma \gg 1$, it is worth mentioning under what circumstances this can occur.  The simplest example is simply low-scale hybrid inflation~\cite{Linde:1993cn}.  In hybrid inflation, the inflaton drives inflation and a separate waterfall field abruptly ends inflation after the appropriate number of $e$-foldings.  Reheating at the end of hybrid inflation is typically extremely efficient, \eg, tachyonic reheating completes within a single oscillation~\cite{Felder:2000hj}.  Because the mass of the waterfall field (which in this context is the reheaton) is much larger than the Hubble rate, this results in $\gamma \sim m_\text{waterfall}/H \gg 1$ so that we are in the limit of a fast reheaton decay.  Alternatively, one can simply give the waterfall field a Yukawa coupling to a fermion and have it quickly decay in a more traditional manner.

If one finds low-scale inflation unappealing, an alternative manner in which to arrive at the fast reheaton decay scenario is to utilize either another phase transition or another slowly rolling scalar that controls the mass of the decay products of the reheaton.  In the early Universe the mass of the decay products is large compared to the reheaton mass kinematically forbidding the decays.  Only later will the mass of the decay products become small enough to allow for the reheaton to decay quickly.  Leaving aside any specific ways to realize fast reheaton decays, we simply focus on its phenomenological consequences.

In the $\gamma \gg 1$ limit $\Delta t_\chi \sim \Gamma_\chi^{-1} \ll H_i^{-1}$ and the $\chi$D era is very short. $\rho_\chii$ is quickly and efficiently transformed into radiation energy density, and the Universe enters into an RD era. Therefore, the maximum amount of radiation is $\rho_\rmx \approx \rho_\chii$, and the time $t_\mx$ at which this occurs is solidly within RD and before one Hubble time has passed: $t_\mx \ll H_i^{-1}$, $a_\mx \approx a_i \equiv a(t_i = 0)$ (see \Fig{fig:rh_app}).

Because an hPT can only take place while the temperature is increasing, $t_\hpt < t_\mx$ and $a_\hpt \approx a_\mx \approx a_i$. Without tuning $T_\hpt$ (or $T_c$) to be too close to $T_\mx$, the hPT will occur during $\chi$D, and $\rho_{\tot,\hpt} \approx \rho_{\chi,\hpt} \approx \rho_\chii$. On the other hand, since $T_\cpt < T_\mx$ necessarily, the cPT will always occur during RD. Of course, as is typical during RD, the energy density scales like
\be
    a^4 \rho_r = a_\mx^4 \rho_\rmx \approx a_i^4 \rho_\chii \ \Rightarrow \ a T = a_\mx T_\mx \quad \text{for $a < a_\mx \approx a_i$ (RD).} \label{eq:fastdec_rhos}
\ee

\subsubsection{Slow reheaton decays: $\gamma \ll 1$}

Here $\Delta t_\chi \sim \Gamma_\chi^{-1} \gg H_i^{-1}$ and $\chi$D is long. During most of this era the decay rate, but not the Hubble expansion rate, can be approximated as negligible for the purposes of estimating the evolution of the reheaton energy density. Since the reheaton is matter-like then $\rho_\chi \sim a^{-3}$ and $a \sim x^{2/3}$, which means that
\ba
    \frac{a_\chiR}{a_i} & \approx & \bxd{1.3} \, \gamma^{-2/3} \quad \text{($\chi$D),} \label{eq:slowdec_achir}\\
    a^3 \rho_\chi & \approx & a_i^3 \rho_\chii \quad \text{for $a_i < a < a_\chiR$ ($\chi$D)} \ , \label{eq:slowdec_rhochi_chiD} \\
    \text{and thus } \quad \rho_\chiR & \approx & \bxd{0.16} \, \gamma^2 \, \rho_\chii \ , \label{eq:slowdec_rhochiR}
\ea
with $a_\chiR$ and $\rho_\chiR$ denoting the scale factor and reheaton energy density at the time of reheaton-radiation equality.\footnote{Not to be confused with the scale factor at matter-radiation equality, $a_\eq \approx 1/3000$.} The numerical coefficients inside the square brackets can only be obtained by solving the differential equations \Eqs{eq:chi_cont}{eq:rad_cont} analytically in the $\gamma \ll 1$ limit.

The time of increasing temperature lasts for little less than one Hubble time (see \Fig{fig:rh_app}); consequently, the hPT occurs squarely during $\chi$D. As discussed before, the radiation energy density grows linearly during this time ($\rho_r \approx \rho_\chii \gamma x$), which allows us to estimate the maximum energy density in radiation as given roughly by
\ba
    \rho_\rmx & \approx & \bxd{0.14}\, \gamma \, \rho_\chii \ , \label{eq:slowdec_rhomx} \\
    \Rightarrow \quad \rho_\chiR & \approx & \gamma \, \rho_\rmx \ , \label{eq:slowdec_rhochiR2}
\ea
where once again the number in the brackets can only be obtained by solving \Eqs{eq:chi_cont}{eq:rad_cont} analytically and making use of the fact that $\dot{\rho}_r(t_\mx) = 0$ by definition. Because $t_\mx \lesssim H_i^{-1}$, then $a_\mx \approx a_i$ as well.

After one Hubble time the expansion of the Universe is felt and $\rho_\chi \sim a^{-3}$, as stated in \Eq{eq:slowdec_rhochi_chiD}. Because the Universe is in $\chi$D ($\rho_\chi \gg \rho_r$) the reheaton decays are still very much relevant to the evolution of $\rho_r$. However, the Hubble expansion wins and the temperature no longer rises with time, as clearly seen in \Fig{fig:rh_app}. Equation~\ref{eq:rad_cont} can be solved in this regime to find that $\rho_r \sim a^{-3/2}$:
\be
    a^{3/2} \rho_r \approx a_\mx^{3/2} \rho_\rmx \approx a_\chiR^{3/2} \rho_\chiR \quad \text{for $a_i \approx a_\mx < a < a_\chiR$ ($\chi$D),} \label{eq:slowdec_rhos_chiD}
\ee
where we have finally dropped the $\mathcal{O}(0.1\text{--}1)$ numerical coefficients, irrelevant to the level of precision with which we are working. It is convenient to note that during $\chi$D $\rho_\chi/\rho_r \sim a^{-3/2}$:
\be
    \frac{\rho_\chi}{\rho_r} \approx \bl( \frac{a_\mx}{a} \br)^{3/2} \frac{\rho_\chii}{\rho_\rmx} \approx \frac{\rho_r}{\gamma \rho_\rmx} \approx \frac{\rho_r}{\rho_\chiR} \quad \text{for $a_i \approx a_\mx < a < a_\chiR$ ($\chi$D).} \label{eq:slowdec_rho_ratio_chiD}
\ee
Of course, once $\chi$D ends and the Universe becomes RD, we have $\rho_r \sim a^{-4}$:
\be
    a^4 \rho_r \approx a_\chiR^4 \rho_\chiR \quad \text{for $a > a_\chiR$ (RD).} \label{eq:slowdec_rhos_RD}
\ee

\subsubsection{Summary}

The previous discussion allows us to find the important quantities $R_{\chi,*}$, $H_*$, $a_*$, $F_*$, and $d \ln T/d t$. The first four appear in the formulas for the SGWB from PTs (\Eqst{eq:Nbc}{eq:fpkbc} and \Eqst{eq:Nsw}{eq:fpksw}), while the last determines the duration of the PT, $\beta^{-1}$ (\Eq{eq:beta_exp}), which determines the peak frequency of the GW spectra. We can estimate these quantities both for hPTs and cPTs, in both the $\gamma \gg 1$ and $\gamma \ll 1$ limits.

This task is most easily accomplished by defining a handful of useful parameters with a clear physical intuition. The first is what we call the {\it reheating efficiency} parameter $\vepsrh$, which quantifies how much of the initial reheaton energy density is transformed into radiation:
\be
    \vepsrh \equiv \frac{\rho_\rmx}{\rho_\chii} \approx \min\bl[ 1, \gamma \br] \ . \label{eq:rh_eff}
\ee
It is clear that the first argument of $\min[X, Y]$ is picked when $\gamma \gg 1$, while the second when $\gamma \ll 1$, which is as we found \Eqs{eq:fastdec_rhos}{eq:slowdec_rhomx}.

Noting that in \Eqs{eq:astar_app}{eq:Fstar_app} $a_*$ and $F_*$ are defined relative to an arbitrarily chosen benchmark time during RD; we now make this choice concrete. Since for $\gamma \gg 1$ $a_\mx$ is well within RD we make this our benchmark for this limit. For $\gamma \ll 1$, assuming a sudden transition between $\chi$D and RD, we can make $a_\chiR$ our choice. It is interesting to point out that $a_\chiR < a_\mx$ for $\gamma \gg 1$, while the opposite is true for $\gamma \ll 1$. We thus choose
\be
    \ard \approx \max\bl[ a_\mx, a_\chiR \br] \ , \label{eq:ard}
\ee
where once again the first argument of $\max[X, Y]$ occurs for $\gamma \gg 1$ and the second for $\gamma \ll 1$. We can then define the relative scale factor between $t_\mx$ and $t_\rd$
\be
    \Ard \equiv \frac{a_\mx}{a_\rd} \approx \min\bl[ 1, \gamma^{2/3} \br] \ ,
\ee
as well as the redshift dilution in the radiation energy density between these two times
\be
    \Drd \equiv \frac{\rho_\rrd}{\rho_\rmx} \approx \min\bl[1 , \gamma \br] \approx A_\rd^{3/2} \ .
\ee
Of course, $\vepsrh$, $\ard$, $\Ard$, and $\Drd$, can all be written in terms of $\gamma$ and each other. But they represent quantities with distinct physical meanings, and thus it is more expedient to interpret them separately.

Based on these definitions, we can write the following expressions for hPTs (which only occur during $\chi$D and thus $a_\hpt \approx a_i \approx a_\mx$), and cPTs (which can occur during either $\chi$D or RD):
\be\label{eq:Rchi_app}
    1 + R_{\chi,*} \equiv \frac{\rho_{\tot,*}}{\rho_{r,*}} \approx
    \begin{cases}
        \frac{\rho_\chii}{\rho_\rhpt} \approx \bl( \frac{T_\mx}{T_\hpt} \br)^4 \vepsrh^{-1} & \text{hPT,}\\[2ex]
        \begin{cases}
            1 & \text{$\gamma \gg 1$, RD}\\
            \frac{\rho_\rcpt}{\rho_\chiR} \approx \bl( \frac{T_\cpt}{T_\mx} \br)^4 \Drd^{-1} > 1 & \text{$\gamma \ll 1$, $\chi$D} \\
            1 & \text{$\gamma \ll 1$, RD}
        \end{cases} & \text{cPT.}
    \end{cases}
\ee
\be\label{eq:aTratio_app}
    \frac{a_* T_*}{a_\rd T_\rd} \approx
    \begin{cases}
        \frac{T_\hpt}{T_\mx} \, \Ard \Drd^{-1/4} & \text{hPT,}\\[2ex]
        \begin{cases}
            1 & \text{$\gamma \gg 1$, RD}\\
            \bl( \frac{\rho_\chiR}{\rho_\rcpt} \br)^{2/3} \bl( \frac{\rho_\rcpt}{\rho_\chiR} \br)^{1/4} \approx \bl( \frac{T_\mx}{T_\cpt} \br)^{5/3} \, \Ard \Drd^{-1/4} < 1 & \text{$\gamma \ll 1$, $\chi$D} \\
            1 & \text{$\gamma \ll 1$, RD}
        \end{cases} & \text{cPT.}
    \end{cases}
\ee
\be\label{eq:arhoratio_app}
    \frac{a_*^4 \rho_{\tot,*}}{a_\rd^4 \rho_\rrd} \approx
    \begin{cases}
        \Ard^4 \vepsrh^{-1} \Drd^{-1} & \text{hPT,}\\[2ex]
        \begin{cases}
            1 & \text{$\gamma \gg 1$, RD}\\
            \bl( \frac{\rho_\chiR}{\rho_\rcpt} \br)^{2/3} \approx \bl( \frac{T_\mx}{T_\cpt} \br)^{8/3} \, \Ard^4 \vepsrh^{-1} \Drd^{-1} < 1 & \text{$\gamma \ll 1$, $\chi$D} \\
            1 & \text{$\gamma \ll 1$, RD}
        \end{cases} & \text{cPT.}
    \end{cases}
\ee

Putting everything together, we list
\be\label{eq:1Rchi_cases_app}
    1+ R_{\chi, *} \approx
    \begin{cases}
        \frac{\rho_\chii}{\rho_\rhpt} \approx \bl( \frac{T_\mx}{T_\hpt} \br)^4 \vepsrh^{-1} & \quad \text{hPT,}\\[2ex]
        \max\bl[ 1, \frac{\rho_\rcpt}{\gamma^2 \rho_\chii} \br] \approx \max\bl[ 1, \bl( \frac{T_\cpt}{T_\mx} \br)^4 \Drd^{-1} \br] & \quad \text{cPT;}
    \end{cases}
\ee
\be\label{eq:Hstar_cases_app}
    H_* \approx
    \begin{cases}
        H_i & \quad \text{hPT,}\\[2ex]
        H_i \sqrt{\frac{\rho_\rcpt}{\rho_\chii}} \max\bl[ 1, \sqrt{\frac{\rho_\rcpt}{\gamma^2 \rho_\chii}} \br] \approx H_r(T_\cpt) \max\bl[ 1, \bl( \frac{T_\cpt}{T_\mx} \br)^2 \Drd^{-1/2} \br] & \quad \text{cPT;}
    \end{cases}
\ee
\be\label{eq:astar_cases_app}
    a_* \approx \bxd{a_*} \, G_a \times 
    \begin{cases}
        \frac{T_\hpt}{T_\mx} \, \Ard \Drd^{-1/4} & \quad \text{hPT,}\\[2ex]
        \min\bl[1, \bl( \frac{T_\mx}{T_\cpt} \br)^{5/3} \, \Ard \Drd^{-1/4} \br] & \quad \text{cPT;}
    \end{cases}
\ee
\be\label{eq:Fstar_cases_app}
    F_* \approx \bxd{F_*} \, G_F \times
    \begin{cases}
       \Ard^4 \vepsrh^{-1} \Drd^{-1} & \text{hPT,}\\[2ex]
        \min\bl[1, \bl( \frac{\gamma^2 \rho_\chii}{\rho_\rcpt} \br)^{2/3} \br] \approx \min\bl[1, \bl( \frac{T_\mx}{T_\cpt} \br)^{8/3} \, \Ard^4 \vepsrh^{-1} \Drd^{-1} \br] & \quad \text{cPT;}
    \end{cases}
\ee
\be\label{eq:dlnTdt_cases_app}
    \frac{d \ln T}{d t} \approx
    \begin{cases}
        \frac{\Gamma_\chi}{4} \frac{\rho_\chii}{\rho_\rhpt} \approx \frac{\Gamma_\chi}{4} \bl( \frac{T_\mx}{T_\hpt} \br)^4 \vepsrh^{-1} & \quad \text{hPT,}\\[2ex]
        -\bl. H \br\vert_\cpt \approx - H_r(T_\cpt) \max\bl[ 1, \bl( \frac{T_\cpt}{T_\mx} \br)^2 \Drd^{-1/2} \br] \cr \ \approx H_i \vepsrh^{1/2} \bl( \frac{T_\cpt}{T_\mx} \br)^2 \max\bl[ 1, \bl( \frac{T_\cpt}{T_\mx} \br)^2 \Drd^{-1/2} \br] & \quad \text{cPT.}
    \end{cases}
\ee
where $H_r(T)^2 \equiv \rho_r/(3 \mpl^2) \equiv \pi^2 g_* T^4/(90 \mpl^2) = H_i^2 \rho_r/\rho_\chii$ is the usual Hubble expansion rate during RD. One can easily convince oneself that the first argument in $\min[X, Y]$ and $\max[X, Y]$ in the above expressions is picked when the cPT occurs during RD (both for $\gamma \gg 1$ and $\gamma \ll 1$), while the second argument is picked only for cPTs taking place during $\chi$D (which can only happen for $\gamma \ll 1$).

\section{First-Order Phase Transitions at Finite Temperature}
\label{app:PT}

Our analytic and numerical results for cooling and heating phase transitions have been implemented in our {\tt graphare} code as part of the {\tt PhaseTransition.wl} package, along with an explanatory notebook titled {\tt 02\_phase\_transition.nb}. Throughout this section we follow closely the notation of Refs.~\cite{Enqvist:1991xw,Cutting:2018tjt,Cutting:2020nla,Hindmarsh:2020hop}.

\subsection{Higgs model and arameters}

The key ingredient in the toy model is the (dark sector) Higgs $\Phi$, whose potential is given by\footnote{Note that our definition of $\lambda$ from the quartic term in \Eq{eq:pot_app} is a factor of $6$ larger than those in Refs.~\cite{Linde:1981zj,Enqvist:1991xw,Cutting:2020nla}.}
\be\label{eq:pot_app}
    V(\Phi) = \frac{M(T)^2}{2} \Phi^2 - \frac{\delta(T)}{3} \Phi^3 + \frac{\lambda}{4!} \Phi^4 \ ,
\ee
where $M(T)$ and $\delta(T)$ are functions of the DS temperature $T$.\footnote{We are ignoring the subleading thermal corrections to $\lambda$.} The two minima or vacua of the potential are located at $\Phis = 0$ (the {\it ``symmetric phase''}) and $\Phib = (3\delta + \sqrt{9 \delta^2 - 6 M^2 \lambda})/\lambda$ (the {\it ``broken phase''}). They are degenerate if there exists a {\it critical temperature} $T_c$ at which $M(T_c) = 2 \delta(T_c)/\sqrt{3 \lambda}$, which motivates the crucial definitions \cite{Enqvist:1991xw,Cutting:2020nla,Hindmarsh:2020hop}
\be
    M_c(T)^2 \equiv \frac{4 \delta(T)^2}{3 \lambda} \ , \quad
    \lbar(T) \equiv \frac{M(T)^2}{M_c(T)^2} \ . \label{eq:Mclbar_defs}
\ee
Clearly $M(T_c) = M_c(T_c)$ and $\lbar(T_c) = 1$. Another important quantity is the mass of the $\Phi$ field in its broken phase:
\ba
    M_b(T)^2 \equiv V^{\prime\prime}(\Phib) & = & M_c^2 ~ \frac{9 - 8 \lbar + 3 \sqrt{9 - 8 \lbar}}{4} \ , \label{eq:Mb_def} \\
    \text{with} \quad \Phib & = & \frac{2\sqrt{3} M_c}{\sqrt{\lambda}} \, \frac{3 + \sqrt{9 - 8 \lbar} }{4} \ . \label{eq:Phib_def}
\ea
Thus, the Higgs has the same mass $M_c(T_c)$ in the broken and symmetric phases at the critical temperature. In addition, it is clear that if there exists a temperature $T_0$ such that $\lbar = 0$, then $M(T_0) = 0$ and the potential minimum at the symmetric phase disappears. Analogously, if there exists a temperature $T_1$ such that $\lbar(T_1) = 9/8$, then $M_b(T_1) = 0$ and the broken phase minimum disappears. For subcritical temperatures $T \in (T_0, T_c)$ the broken phase with $\Phib \neq 0$ is the stable global minimum or {\it ``true vacuum''}, while the symmetric phase with $\Phis = 0$ is the metastable minimum or {\it ``false vacuum''}. For supercritical temperatures $T \in (T_c, T_1)$, the converse is true.

In our $\Phi$ model, inspired by the SM Higgs, the finite-temperature potential has the coefficients \cite{Linde:1981zj,Enqvist:1991xw,Quiros:1999jp}
\ba
    M(T)^2 = \mu^2 (T^2 - T_0^2) \ , &\quad& \delta(T) \equiv A T \ , \label{eq:Mdelta_defs}\\
    \text{with} \qquad \mu^2 \equiv \frac{1}{12} \sum\limits_i \, c_i N_i y_i^2  \ , &\qquad& A \equiv \frac{1}{4 \pi} \sum\limits_B \, N_B y_B^3 \ , \label{eq:muA_defs}
\ea
where $N_i$ is the number of degrees of freedom of the $i$-th particle, $y_i$ is its coupling to $\Phi$, the index $i$ runs over both bosons and fermions, the index $B$ only runs over bosons, and $c_{i=\text{boson}} = 1$ and $c_{i=\text{fermion}} = 1/2$.

This allows us to write all expressions in terms of $T_c$, $\lbar$, and the potential coefficients $\{ \mu, A, \lambda \}$:
\ba
    T = T_c \, \sqrt{\frac{1 - \Delta}{1 - \Delta \lbar}} \!\!\!\!\!\!& \Leftrightarrow &\!\!\!\!\!\! \lbar = \frac{(T/T_c)^2 - 1 + \Delta}{(T/T_c)^2 \, \Delta} \ , \label{eq:Tlbar}\\
    M_c = \mu \sqrt{\Delta} \, T \ , \quad & M = \sqrt{\lbar} \, M_c \ , & \quad \delta = \frac{\sqrt{3 \lambda}}{2} \, M_c \ , \\
    \Phib \!\!\!\!\!\!\!\!\!\!& = &\!\!\!\!\!\!\!\!\!\! \frac{2\sqrt{3} M_c}{\sqrt{\lambda}} \, \frac{3 + \sqrt{9 - 8 \lbar} }{4} \ , \\
    \text{with} \quad \Delta \!\!\!\!\!\!\!\!\!\!& \equiv &\!\!\!\!\!\!\!\!\!\! \frac{4 A^2}{3 \lambda \mu^2} \ .
\ea
Since $\lbar(T_0) = 0$ and $\lbar(T_1) = 9/8$ we find
\be
    T_0 = T_c \, \sqrt{1 - \Delta} \ , \quad T_1 = T_c \sqrt{\frac{8 - 8 \Delta}{8 - 9 \Delta}} \ .
\ee
Since $T_0$ {\it must} exist (because it corresponds to the negative Higgs mass in the potential at zero temperature), the condition for there to be a critical temperature is $\Delta \in (0, 1)$. By contrast, for $T_1$ to exist we need $\Delta \in (0, 8/9)$. Beyond this range $T_1$ does not exist and the broken minimum never disappears, even at very large temperatures. In this case the asymptotic value of $\lbar$ at large temperatures is $1/\Delta$.

From the above discussion it is clear that a cPT (which occurs for subcritical temperatures) and an hPT (occurring only for supercritical temperatures) correspond to the regimes
\be\label{eq:lbarregimes}
    \text{phase transitions: } \quad \begin{cases}
        \text{cPT regime: } & \quad T \in \bl( T_0, T_c \br) \ \Leftrightarrow \ \lbar \in \bl( 0, 1 \br) \ ,\\[2ex]
        \text{hPT regime: } & \quad T \in \bl( T_c, T_1 \br) \ \Leftrightarrow \ \lbar \in \bl( 1, \frac{9}{8} \br) \ .
    \end{cases}
\ee

Finally, at $T=0$ ($\lbar \rightarrow - \infty$) we have the usual zero-temperature spontaneous symmetry breaking at the vacuum expectation value
\be
    \angu{\Phi}_0 \equiv \angu{\Phi}_{b,T=0} = \frac{\sqrt{3} \mu \sqrt{1 - \Delta}}{\sqrt{\lambda}} \, T_c \ , \quad V_0 \equiv V(\angu{\Phi}_0) = -\frac{3 \mu^4 (1 - \Delta)^2}{8 \lambda} \, T_c^4 \ .
\ee

\subsection{Critical bubbles and bounce action}\label{app:action}

\begin{figure}[tb]
	\centering
	\includegraphics[width=0.5\linewidth]{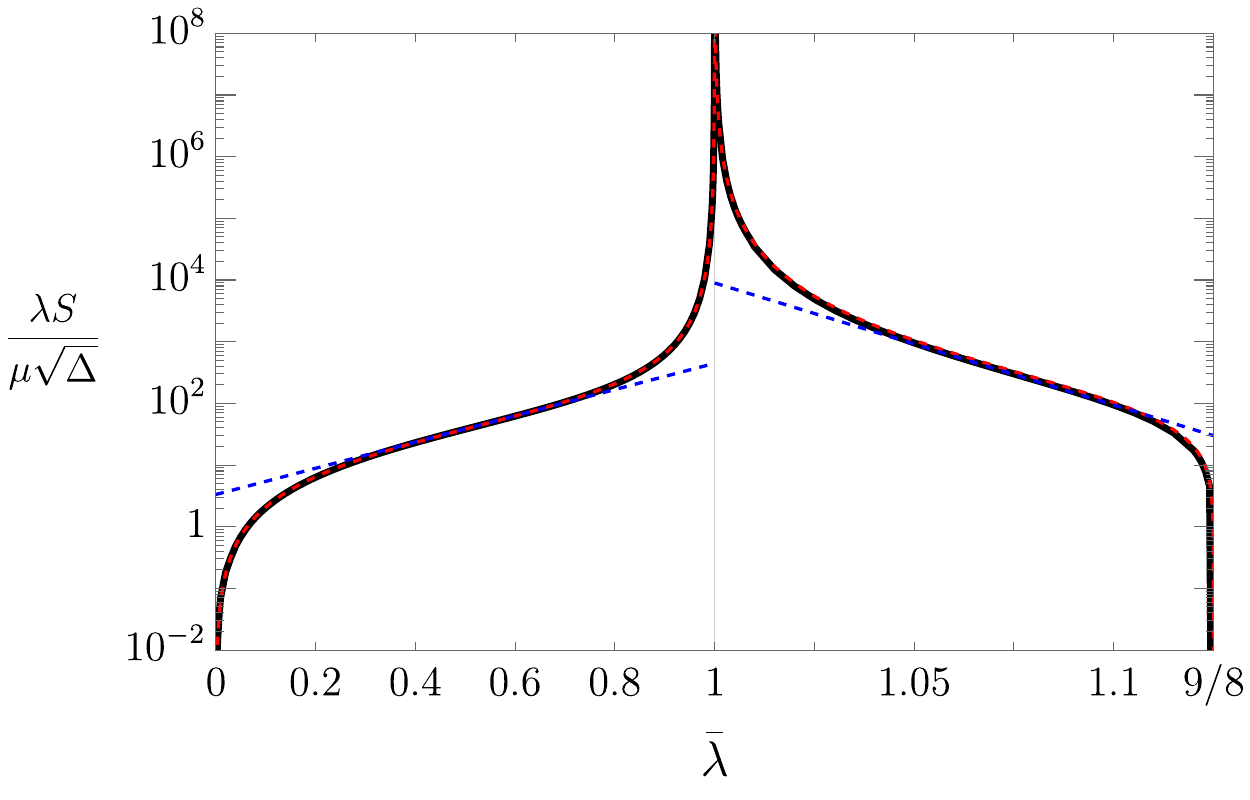}
	\caption{The action $S = S[\Phi_\bub] - S[\angu{\Phi}_\fv]$ as a function of $\lbar$. Note that it has been rescaled by $\mu \sqrt{\Delta}/\lambda = 2 A/\sqrt{3 \lambda^3}$, since $\Phib^2/(T M)$ in \Eq{eq:action} depends on just this parameter combination, the rest of it being a function of $\lbar$. The {\bf thin grey} vertical line separates the cPT (left half) and hPT (right half) regimes. The {\bf red dashed} lines are the semi-analytic expressions in \Eq{eq:semian}, while the {\bf blue dashed} lines show the estimates from \Eq{eq:estimaction}. Note that the linear scale of the abscissa for $\lbar > 1$ has been increased for easier reading.}
	\label{fig:action}
\end{figure}

A phase transition between the two minima of $V(\Phi)$ can be described by the nucleation and subsequent growth of {\it ``bubbles''} of the new, stable minimum inside a space filled with the old, metastable minimum. These bubbles are created by thermal fluctuations and they correspond to a field configuration $\Phi_\bub$ that interpolates between the new minimum at the center of the bubble, and the old minimum far from it \cite{Linde:1980tt,Linde:1981zj}. To determine the bubble configuration it is necessary to solve the Euclidean equation of motion for $\Phi$ that minimizes the Euclidean bounce action $S[\Phi]$ of the system:\footnote{Sometimes reference is made in the literature to the energy of the field configuration, which is simply $E[\Phi] = T S[\Phi]$.}
\ba
    S [\Phi] & = & \frac{\Phib^2}{T M} ~ 4\pi \int \dd\ov{r} \, \ov{r}^2 \bl[ \frac{1}{2} \bl( \frac{d \ov{\Phi}}{d \ov{r}} \br)^2 + \ov{V}(\ov{\Phi}) \br] \ , \label{eq:action}\\
    \frac{d^2 \ov{\Phi}_\bub}{d \ov{r}^2} &+& \frac{2}{\ov{r}} \frac{d \ov{\Phi}_\bub}{d \ov{r}} - \ov{V}^\prime(\ov{\Phi}_\bub) = 0 \ , \label{eq:eom} \\
    \text{with} \quad \frac{\Phib^2}{T M} & = & \frac{12 \mu \sqrt{\Delta}}{\lambda} \, \frac{\bl( 3 + \sqrt{9 - 8 \lbar} \br)^2}{16 \sqrt{\lbar}} \ , \label{eq:PS}
\ea
where we have defined the dimensionless quantities $\ov{\Phi} \equiv \Phi/\Phib$, $\ov{r} \equiv Mr$, and $\ov{V}(\ov{\Phi}) \equiv V(\Phib \ov{\Phi})/(M \Phib)^2$, and we have assumed spherical symmetry so that $d^3\mathbf{x} = 4 \pi r^2 \dd r$ and $\Phi(\mathbf{x}) = \Phi(r)$. The solutions to \Eq{eq:eom} are sometimes called {\it critical bubbles}, since these are the field configurations that will {\it not} collapse upon formation due to pressure.\footnote{Note the different meaning of the word {\it critical} here, unrelated to the critical temperature $T_c$. As we have discussed above, critical bubbles can be nucleated at both {\it sub}critical or {\it super}critical temperatures $T$. What is more, as seen in \Fig{fig:action}, the energy $E = T S$ required to nucleate a critical bubble at critical temperature is {\it infinite}!}

We have now achieved the main purpose of all of the manipulations and definitions of this section: to show that the equation of motion for $\ov{\Phi}$ depends only on $\lbar$ \cite{Cutting:2020nla,Hindmarsh:2020hop}. Indeed
\be\label{eq:dimlesspot}
    \ov{V}(\ov{\Phi}) = \frac{1}{2} \, \ov{\Phi}^2 - \frac{3 + \sqrt{9 - 8\lbar}}{4 \lbar} \, \ov{\Phi}^3 + \frac{\bl( 3 + \sqrt{9 - 8 \lbar} \br)^2}{32 \lbar} \, \ov{\Phi}^4 \ .
\ee

We find $\ov{\Phi}_\bub(\ov{r})$ by numerically solving \Eq{eq:eom} for various values of $\lbar \in (0, 9/8)$, and subsequently compute the corresponding action. Note that $S$, being an integral over infinite space, is potentially infinite. However, only action {\it differences}, taken relative to the initial metastable false vacuum value $\angu{\Phi}_\fv$, ever enter the physical quantities in which we are interested. Therefore we only compute $S[\Phi_\bub] - S[\angu{\Phi}_\fv]$, which is finite, and we denote this difference simply by $S$. For cPTs $\angu{\Phi}_\fv = \Phis$, whereas for hPTs $\angu{\Phi}_\fv = \Phib$.

\begin{figure}[tb]
	\centering
        \includegraphics[width=.475\linewidth]{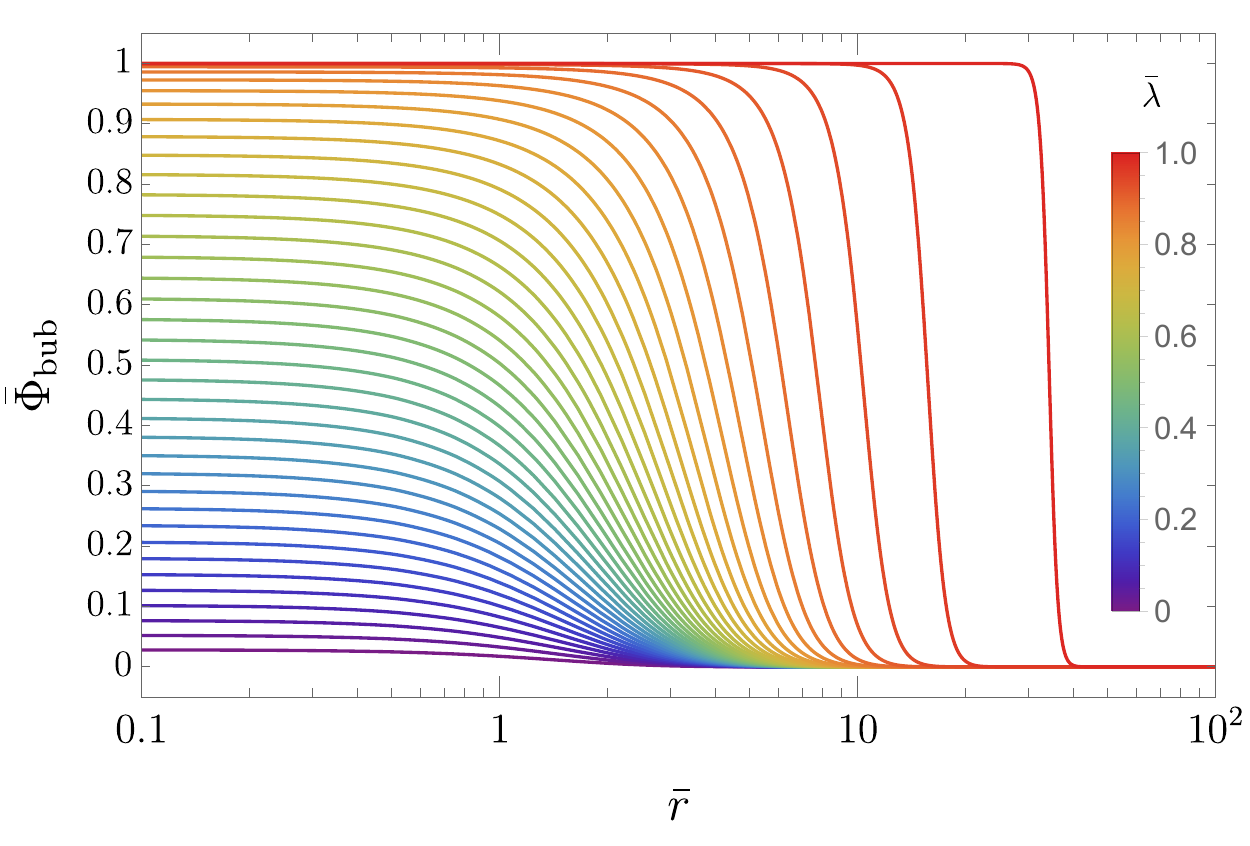}
        \includegraphics[width=.475\linewidth]{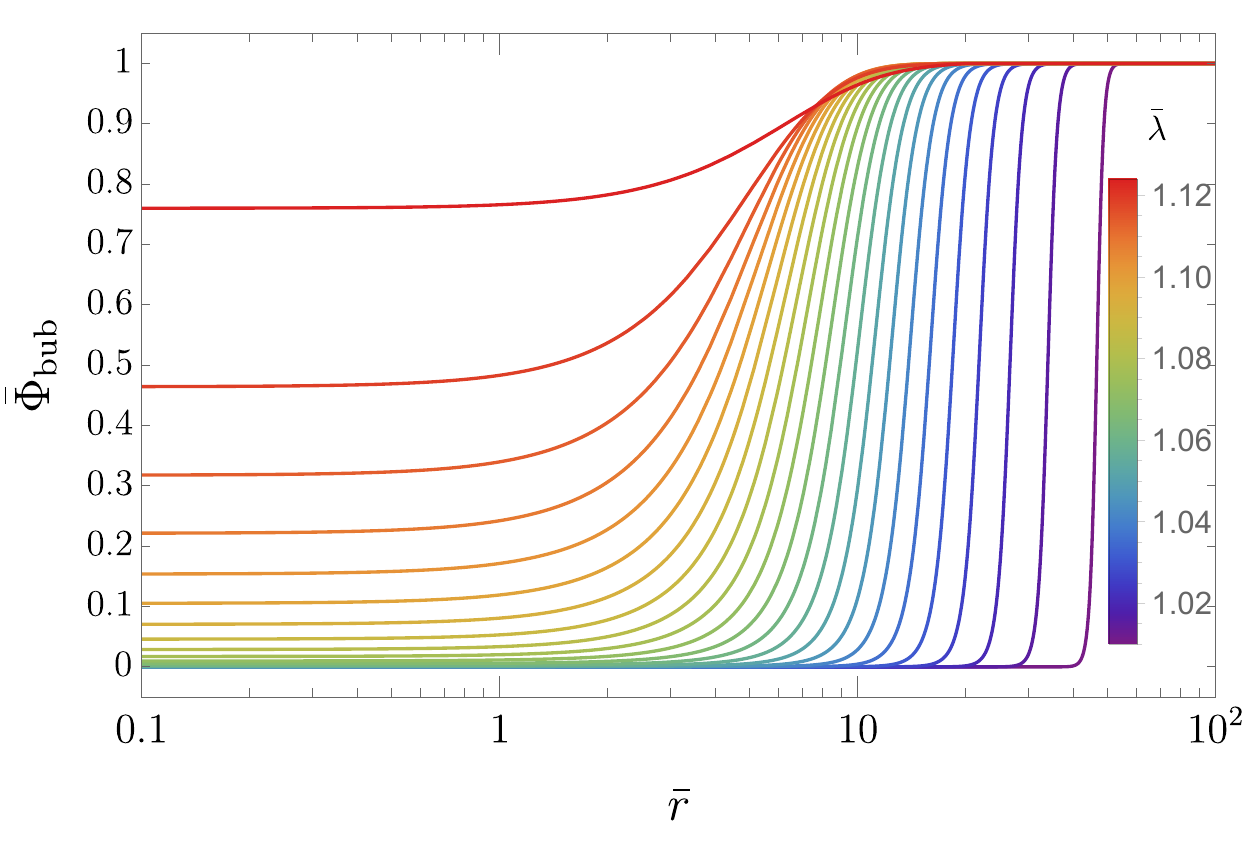}
	\caption{Solutions to \Eq{eq:eom} for different values of $\lbar$: dimensionless bubble field configurations $\ov{\Phi}_\bub = \Phi_\bub/\Phib$, as a function of the dimensionless bubble radius $\ov{r} = M r$. {\bf Left:} bubbles for subcritical temperatures $T \in (T_0, T_c)$, \ie, $\lbar \in (0, 1)$. {\bf Right:} bubbles for supercritical temperatures $T \in (T_c, T_1)$, \ie, $\lbar \in (1, 9/8)$.}
	\label{fig:bub}
\end{figure}

Our results are shown in \Figs{fig:bub}{fig:action}. In \Fig{fig:bub} we plot the bubble configurations $\ov{\Phi}(\ov{r})$ for representative values of $\lbar$; the solutions corresponding to cPTs and $\lbar \in (0, 1)$ are shown in the left plot, and those corresponding to hPTs and $\lbar \in (1, 9/8)$ are shown in the right plot. In \Fig{fig:action} we plot the corresponding values of the bubble action (difference) $S$ as a function of $\lbar$.

\subsection{Daisy resummation}\label{app:daisy}

Finally, we comment about the cubic term of \Eq{eq:pot_app} defined in \Eqs{eq:Mdelta_defs}{eq:muA_defs}. This term is of vital importance to the order of the phase transition; were it to go away, the phase transition would cease to be first order. ``Daisy'' resummation of ``ring'' diagrams effectively leads to a temperature-dependent correction to the bosonic mass stemming from its self-energy \cite{Gross:1980br,Parwani:1991gq,Carrington:1991hz,Arnold:1992rz,Quiros:1999jp}. This means that the cubic term heuristically becomes
\be
    \frac{T}{12 \pi} \sum\limits_B N_B y_B^3 \Phi^3 \rightarrow \frac{T}{12 \pi} \sum\limits_B N_B \bl( y_B^2 \Phi^2 + g_B^2 T^2 \br)^{3/2} \ ,
\ee
where $g_B T$ accounts for the model-dependent thermal mass of the $B$ boson. From now on we will simply take $g_B \sim y_B$, a conservative choice, since the thermal mass contributions are typically further suppressed by factors of $\mathcal{O}(10)$; see, for example, Ref.~\cite{Quiros:1999jp}.

\begin{figure}[tb]
	\centering
	\includegraphics[width=0.5\linewidth]{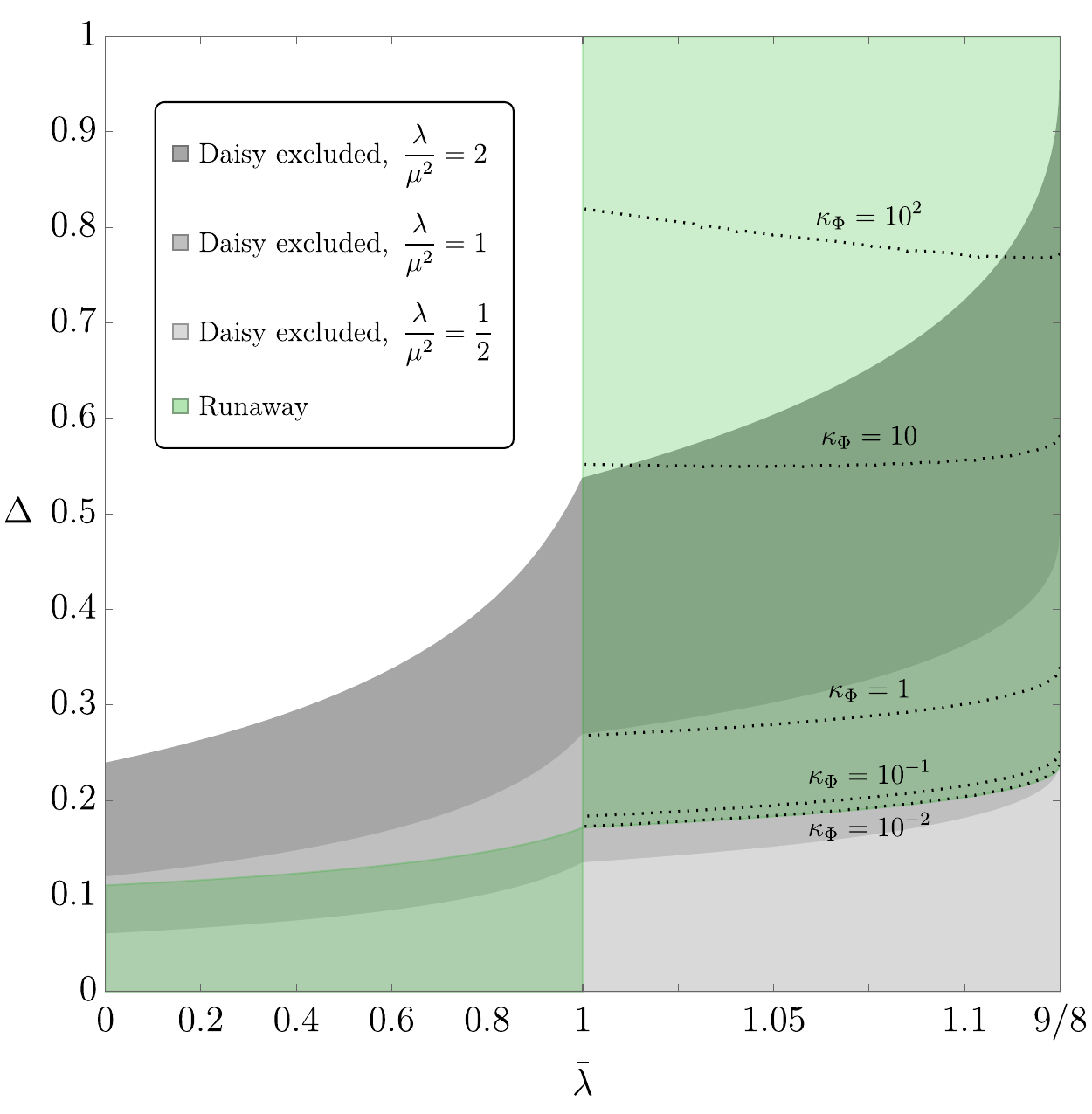}
	\caption{$\lbar$--$\Delta$ parameter space. Shaded in \textbf{grey} are the regions for which \Eq{eq:Delta_daisy} is not satisfied, for different values of $\lambda/\mu^2$. In these regions the daisy resummation corrections to the potential cubic term become relevant and the first-order phase transition ceases to be strong. We evaluate $\Delta_{\rm daisies}$ at $\lbar = 1$ to obtain the representative constraint \Eq{eq:Delta_daisy2}, which we use when presenting the sensitivity of GW detectors in our results (\eg, \Fig{fig:param}). In \textbf{green} we show the region for which \Eq{eq:runaway2} is satisfied and the bubbles are in the runaway regime. The {\bf dotted black} lines correspond to contours of the runaway efficiency $\kappa_\Phi$, see \Eq{eq:runaway2}. Note that the linear scale of the abscissa for $\lbar > 1$ has been increased for easier reading.}
    \label{fig:daisy_runaway}
\end{figure}

Evidently, for very large temperatures the $\Phi^3$ dependence vanishes. In order to ensure that the hPT is indeed strongly first order, we exclude from our results the region of parameter space that yields sizable daisy contributions. To that purpose we estimate the relative size of the self-energy corrections to be
\be\label{eq:daisies}
    \text{daisies } \approx \bl( 1 + T^2/\Phib^2 \br)^{3/2} - 1 = \bl( 1 + \frac{\lambda}{12 \mu^2 \Delta} \bl( \frac{4}{3 + \sqrt{9 - 8 \lbar}} \br)^2 \br)^{3/2} - 1 \ ,
\ee
and then conservatively demand this to be no larger than 50\%. This condition is satisfied for
\be\label{eq:Delta_daisy}
    \Delta > \Delta_{\rm daisies} \equiv \frac{\lambda}{\mu^2} \frac{4.3}{\bl( 3 + \sqrt{9 - 8 \lbar} \br)^2} \ .
\ee
Note that \Eq{eq:daisies} is function of $\lbar$ and thus of the temperature: larger temperatures yield larger daisy contributions, and thus more values of $\Delta$ violate \Eq{eq:Delta_daisy} and are thus excluded from our analysis. Since cooling and heating PTs are sensitive to different temperatures, the restrictions on the parameter $\Delta$ are different for each. However, since our 50\% threshold for the thermal corrections is an arbitrary number, and since we are working with approximate expressions (we have taken $g_B \sim y_B$), we simply evaluate this condition at $T_c$ (\ie, $\lbar = 1$). This is morally equivalent to the condition $\Phib / T_c \gg 1$ used in other parts of the literature (\eg, Ref.~\cite{Morrissey:2012db,Arakawa:2021wgz}). The final constraint is:
\be\label{eq:Delta_daisy2}
    \Delta > \Delta_{\rm daisies} \approx 0.27 \, \frac{\lambda}{\mu^2} \quad \Rightarrow \quad A \gsim 0.45 \, \lambda \ .
\ee
In \Fig{fig:daisy_runaway} we show the regions of $\lbar$--$\Delta$ parameter space for which the daisy contributions satisfy \Eq{eq:Delta_daisy}.

\subsection{Analytic expressions for the bounce action}\label{app:actionAppx}

It turns out that \Eq{eq:eom} can be solved analytically for the special cases where $\lbar \approx 1$ ($T \approx T_c$) and $\lbar \approx 0$ ($T \approx T_0$) or $\lbar \approx 9/8$ ($T \approx T_1$). In the first case the energy difference between the stable and metastable minima of the potential is very small and the {\it thin-wall approximation} can be employed. On the other hand, in the second case the stable minimum of the potential is very deep and the {\it thick-wall approximation} can be used instead. Following Ref.~\cite{Linde:1981zj} for both cases, we find that the integral in \Eq{eq:action} for the bounce action scales like\footnote{For the detailed calculation see our explanatory notebook {\tt 02\_phase\_transition.nb}.}
\be
    \frac{S}{\Phib^2/(TM)} \approx
    \begin{cases}
        \frac{8 \pi}{81 (\lbar - 1)^2} & \quad \text{thin wall, $\lbar \approx 1$;} \\[2ex]
        2.16 \, \lbar^2 & \quad \text{thick wall, $\lbar \approx 0$;} \\[2ex]
        113 \, \bl( \frac{9}{8} - \lbar \br)^{3/4} & \quad \text{thick wall, $\lbar \approx \frac{9}{8}$.}
    \end{cases}
\ee
Using \Eq{eq:PS} and the power-law behaviors listed above, we are able to find a reasonable semianalytic fit to $S(\lbar)$ for the cPT and hPT regimes ($\lbar \in (0, 1)$ and $\lbar \in (1, 9/8)$, respectively), which we show as dashed red lines in \Fig{fig:action}:
\be\label{eq:semian}
    S \approx
    \begin{cases}
        \frac{24 \, \mu \sqrt{\Delta}}{\lambda} \, \bl( \frac{3 + \sqrt{9 - 8 \lbar}}{4} \br)^2 \, \frac{\bl( 1.13 - \lbar \br)^{0.91} \lbar^{3/2}}{\bl( 1 - \lbar \br)^2} & \quad \text{cPT regime;} \\[2ex]
        \frac{20 \, \mu \sqrt{\Delta}}{\lambda} \, \bl( \frac{3 + \sqrt{9 - 8 \lbar}}{4} \br)^2 \, \frac{\bl( \frac{9}{8} - \lbar \br)^{3/4}}{\lbar^{1/2} \, \bl( 1 - \lbar \br)^2} & \quad \text{hPT regime.}
    \end{cases}
\ee
These equations, although analytic, are still too complicated to be useful for back-of-the-envelope estimates. We can do better.

While $\lbar$ is a compact variable, $S$ spans many orders of magnitude. This means we can find a linear fit of $\lbar$ to $\ln S$ to obtain an even simpler approximation that can be inverted (in order to find $\lbar(S)$ analytically). For this fit it is best to use the slope $d \ln S/d \lbar$ at the inflection point of $\ln S$, since that is where the slope is at its smallest, and thus a linear fit will have the broadest range of applicability. The inflection points and corresponding slopes for the cPT and hPT ranges are
\be\label{eq:lbardSinfl}
    \begin{cases}
        \lbar_\mathrm{infl.} \approx 0.52 \ , \quad \bl. \frac{d \ln S}{d \lbar} \br\vert_{\lbar_\mathrm{infl.}} \approx 4.9 & \quad \text{cPT regime;}\\[2ex]
        \lbar_\mathrm{infl.} \approx 1.08 \ , \quad \bl. \frac{d \ln S}{d \lbar} \br\vert_{\lbar_\mathrm{infl.}} \approx -45.6 & \quad \text{hPT regime.}
    \end{cases}
\ee
which lead to the following linear fits for $S(\lbar)$ and their inversions:
\be\label{eq:estimaction}
    S(\lbar) \sim
    \begin{cases}
        \frac{\mu \sqrt{\Delta}}{\lambda} \, \exp\bl( 1.2 + 4.9 \lbar \br) & \quad \text{cPT regime;}\\[2ex]
        \frac{\mu \sqrt{\Delta}}{\lambda} \, \exp\bl( 54.7 - 45.6 \lbar \br) & \quad \text{hPT regime.}
    \end{cases} \ \Leftrightarrow \ 
    \lbar(S) \sim
    \begin{cases}
        -0.25 + 0.21 \, \ln \bl( \frac{S \lambda}{\mu \sqrt{\Delta}} \br) & \quad \text{cPT regime,}\\[2ex]
        1.20 - 0.022 \, \ln \bl( \frac{S \lambda}{\mu \sqrt{\Delta}} \br) & \quad \text{hPT regime.}
    \end{cases}
\ee
These expressions correspond to the dashed blue lines in \Fig{fig:action}. As an illustrative benchmark, for $T \sim 1~\TeV$ and $H \sim T^2/\mpl \sim 10^{-15}~\TeV$, $S \sim 4 \ln ( T / H ) \sim 138$ and $\lbar \sim 0.74$ for a cPT and $\lbar \sim 1.09$ for an hPT, taking $\mu \sqrt{\Delta}/\lambda = 1$. Note that $\lbar$ in \Eq{eq:estimaction} depends logarithmically on $S$, so varying $H$ and $T$ has little impact on the resulting $\lbar$.

\begin{figure}[tb]
	\centering
        \includegraphics[width=0.5\linewidth]{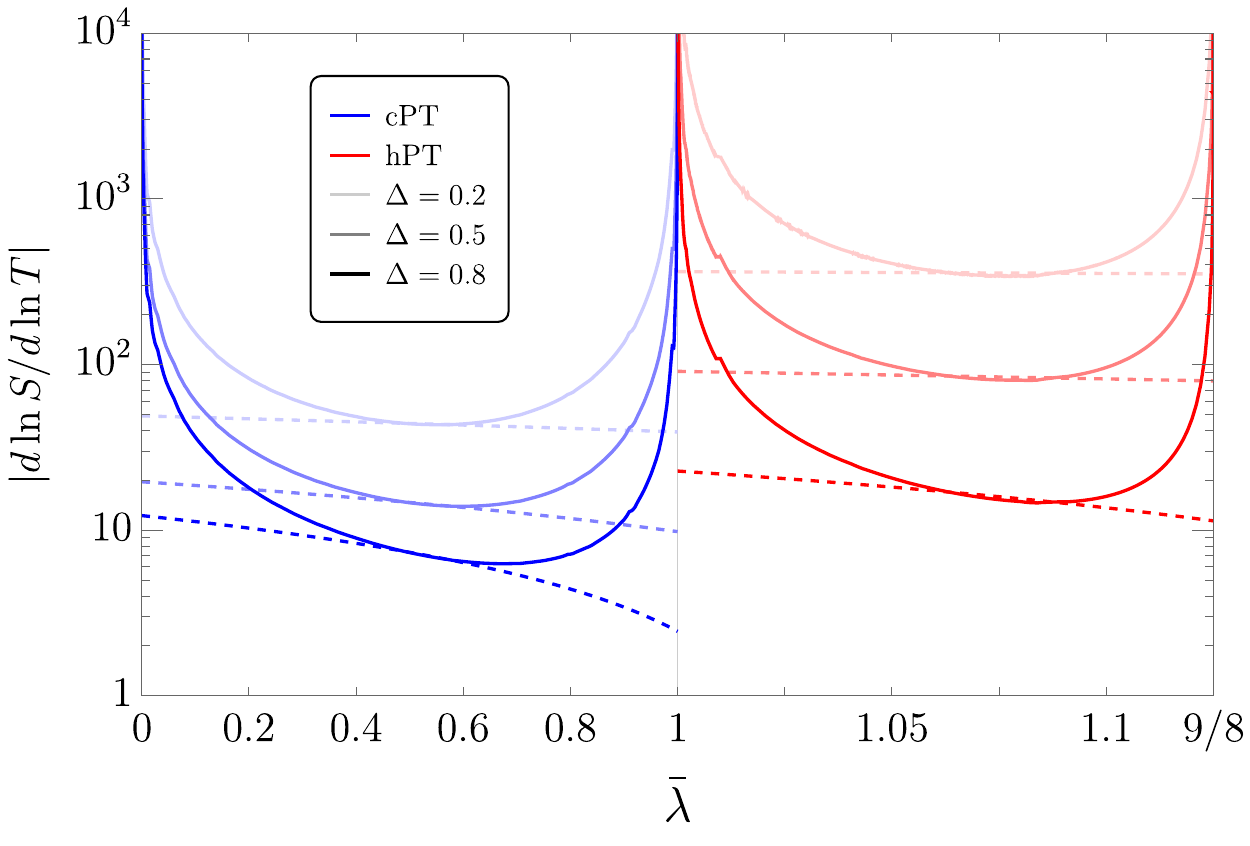}
	\caption{Plots of $\abs{d \ln S/d \ln T}$ as a function of $\lbar$ and $\Delta$, for both the cPT (\textbf{blue}) and the hPT (\textbf{red}) regimes. Also shown are the curves corresponding to the approximate expressions listed in \Eq{eq:DSDT}. Note that the linear scale of the abscissa for $\lbar > 1$ has been increased for easier reading.}
	\label{fig:DSDT}
\end{figure}

From the above approximations and \Eq{eq:Tlbar} we can obtain the rate of change of $S$ with respect to the temperature:
\be\label{eq:DSDT}
    \frac{d \ln S}{d \ln T} = \frac{d \ln S}{d \lbar} \frac{d \lbar}{d \ln T} \sim
    \begin{cases}
        \frac{9.8}{\Delta} (1 - \Delta \lbar) & \quad \text{cPT regime,}\\[2ex]
        -\frac{91}{\Delta} (1 - \Delta \lbar) & \quad \text{hPT regime.}
    \end{cases}
\ee
Contours of $\abs{d \ln S/d \ln T}$ as a function of $\Delta$ and $\lbar$, for both the cPT and hPT regimes, are shown in \Fig{fig:DSDT}, as well as their corresponding estimates from \Eq{eq:DSDT}.

\section{Bubble Nucleation and Phase Transitions}
\label{app:BN}

The Universe begins its life cold and with $\Phi$ in the broken phase minimum of the potential in \Eq{eq:pot_app}, $\Phib \neq 0$. Eventually it reheats to supercritical temperatures $T \in (T_c, T_1)$ (\ie, $\lbar \in (1, 9/8)$) at which point the stable, true vacuum is instead the symmetric phase $\Phis = 0$. Thermal fluctuations nucleate bubbles of the true vacuum, which then expand and eventually fill the entire Universe. Thus the Universe now finds itself entirely in the symmetric phase, and we say that a heating first-order phase transition (hPT) has taken place. Once reheating ends the Universe begins to cool down, eventually reaching subcritical temperatures $T \in (T_0, T_c)$ (\ie, $\lbar \in (0, 1)$). At this point the stable, true vacuum of the system is once again the broken phase $\Phib$ and the Universe may be brought to it via the nucleation and expansion of bubbles of this true vacuum, in what we call a cooling first-order phase transition (cPT).

Our treatment of bubble nucleation and expansion during both cPTs and hPTs is implemented in our {\tt graphare} code as part of the {\tt PhaseTransition.wl} package, along with an explanatory notebook titled {\tt 02\_phase\_transition.nb}.

\subsection{Bubble nucleation rate}

Having discussed the field configurations and bounce action of critical bubbles of the true vacuum in the previous section, we now focus on how they are nucleated, how they grow, and how they bring the Universe from the metastable to the stable phase.

The bubble nucleation rate per unit volume is given by \cite{Linde:1980tt,Linde:1981zj,Hindmarsh:2020hop}
\be\label{eq:GV_app}
    \GV = \mathcal{M}^4 \bl( \frac{S}{2 \pi} \br)^{3/2} \, e^{-S} \ ,
\ee
where $S$ is the finite-temperature bubble bounce action difference $S[\Phi_\bub] - S[\angu{\Phi}_\fv]$ found in \Eq{eq:action}, and $\mathcal{M}$ is a quantity with the dimensions of energy. Its precise form is model dependent and can only be computed numerically. Dimensional analysis, however, tells us that $\mathcal{M} \sim T$, and we simply take the remaining dimensionless coefficient to be $\mathcal{O}(1)$. Finally, the $S^{3/2}$ prefactor comes from the three translational zero modes that appear in the path-integral computation of the partition function, and which must be treated separately \cite{Linde:1981zj,Hindmarsh:2020hop}.

\begin{figure}[tb]
	\centering
	 \includegraphics[width=0.5\linewidth]{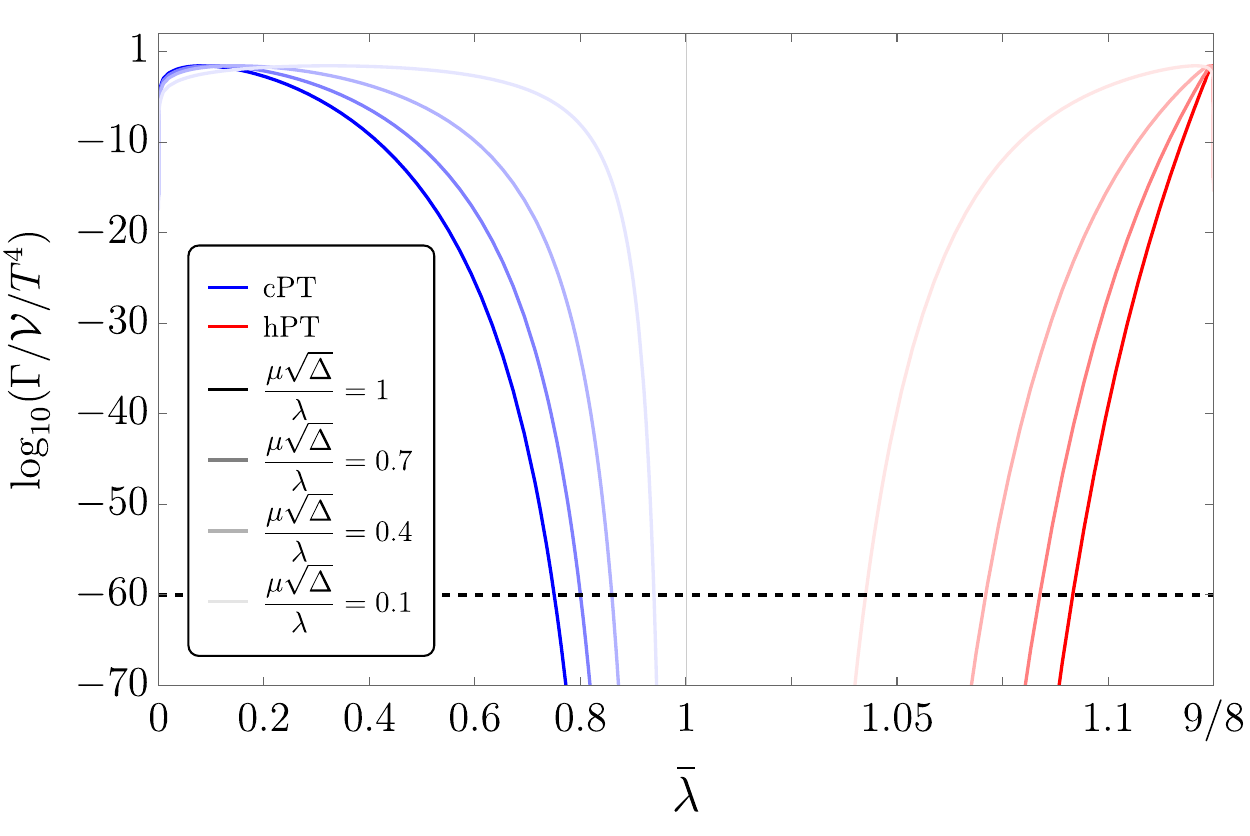}
      \caption{Bubble nucleation rate $(\GVh)/T^4$ as a function of $\lbar$, for different values of $\mu \sqrt{\Delta}/\lambda$. The {\bf blue} ({\bf red}) lines correspond to the nucleation rate during a cPT (hPT). The {\bf dashed black line} shows $(10^{-15})^4$, a typical value of $H^4$ for the $\mathcal{O}(\TeV)$ temperature scales we consider in this paper. The typical value of $\lbar$ at the nucleation time $t_n$ can then be read at the intersection of these lines. Note that the linear scale of the abscissa for $\lbar > 1$ has been increased for easier reading.}
	\label{fig:nucl_app}
\end{figure}

In \Fig{fig:nucl_app} we plot $\Gamma(\lbar)/(\mathcal{V} T^4)$ for various values of $\mu \sqrt{\Delta}/\lambda$ which, as we saw above, is the parameter combination controlling the size of the action $S$. We can see that the nucleation rate vanishes at $\lbar = 1$ and $\lbar = 0$ ($\lbar = 9/8$), corresponding to $T=T_c$ and $T = T_0$ ($T = T_1$) respectively. As long as $\mu \sqrt{\Delta}/\lambda$ is not too large, a significant number of bubbles can be nucleated within a Hubble space-time patch. Note that during reheating $T_1$ (and thus $\lbar = 9/8$) may never be reached, in which case $\Gamma/(\mathcal{V} T^4)$ will reach a maximum value at $\lbar(T_\mx)$ and then go back down to $0$ as $T$ decreases back to $T_c$.

\subsection{Bubble expansion}\label{app:expansion}

In this section we fill in the details of the bubble runaway condition described in \Sec{Sec: PT} of the main text. Recall that to test if a runaway occurs, one needs to consider the net pressure $P_\tot$ acting on the wall of a relativistic bubble, which is a sum of pressure differences (between the inside and the outside of the bubble) due to the zero-temperature vacuum energy and the plasma interactions with $\Phi$ (equal to the leading mean-field contribution to the thermal potential) \cite{Bodeker:2009qy}:
\ba
    P_\tot & = & \Delta V_0 + \Delta P_T > 0 \ , \\
    \text{with } \quad \Delta V_0 & \equiv & V_{0,\, \rm out} - V_{0,\, \rm in} = \mathrm{sign} ( 1 - \lbar ) \abs{V_0} \ , \\
    \Delta P_T & \equiv & \sum\limits_i N_i \int\limits_{m_{i, \, \rm in}^2}^{m_{i, \, \rm out}^2} \dd m_i^2 \, \int \frac{\dd^3 p}{(2\pi)^3} \frac{f_i^{\rm out}}{2 E} \approx \frac{T^2}{24} \, \sum\limits_i c_i N_i \Delta m_i^2 \ , \nonumber\\
    & = & - \mathrm{sign}\bl( 1 - \lbar \br) \frac{\mu^2}{2} T^2 \Phib^2 \ ,
\ea
where in the last equality we have used \Eq{eq:muA_defs} and the fact that
\be
    \Delta m_i^2 \equiv m_{i, \, \mathrm{out}}^2 - m_{i, \, \mathrm{in}}^2 = y_i^2 \bl( \angu{\Phi}_\fv^2 - \angu{\Phi}_\tv^2 \br) = - \mathrm{sign}\bl( 1 - \lbar \br) y_i^2 \Phib^2 \ ,
\ee
and where $\angu{\Phi}_\fv$ and $\angu{\Phi}_\tv$ denote the false and true vacuum Higgs vacuum expectation values, outside and inside the bubble respectively. Note that, of course, $\mathrm{sign}\bl( 1 - \lbar \br)$ used here and $\mathrm{sign}\bl( T_c - T \br)$ used in the main text, are equivalent.

Putting this together allows us to write the runaway condition as simply
\ba
    P_\tot & = & \mathrm{sign}\bl( 1 - \lbar \br) \bl( \abs{V_0} - \frac{\mu^2}{2} T^2 \Phib^2 \br) \nonumber \\
    & = & \frac{3 \mu^4 (1 - \Delta)^2 T_c^4}{2 \lambda} \, \mathrm{sign}\bl( 1 - \lbar \br) \, \bl( 1 - \frac{\Delta}{4} \bl( \frac{3 + \sqrt{9 - 8 \lbar}}{1 - \Delta \lbar} \br)^2 \br) > 0 \ , \label{eq:runaway1} \\
    \Rightarrow \quad \kappa_\Phi \equiv \frac{P_\tot}{\abs{V_0}} & = & \mathrm{sign}\bl( 1 - \lbar \br) \, \bl( 1 - \frac{\Delta}{4} \bl( \frac{3 + \sqrt{9 - 8 \lbar}}{1 - \Delta \lbar} \br)^2 \br) > 0 \ . \label{eq:runaway2}
\ea
where in the last line we defined the runaway efficiency $\kappa_\Phi$, which can be rewritten as $\kappa_\Phi = \mathrm{sign}( 1 - \lbar ) ( \alpha - \alpha_\infty )/\alpha$; see \Eq{eq:kappa_hpt} in the main = text. In \Fig{fig:daisy_runaway} we show the regions of $\lbar$--$\Delta$ parameter space for which the runaway condition in \Eq{eq:runaway2} is satisfied. It can be seen that, up to the different values of $\lbar$ explored by the transition, the runaway conditions for cPTs and hPTs are almost mirror images of each other, as described previously in the text. Indeed, since the (anti)friction is ultimately described in terms of the interaction strength $y_i$ of the particles with the Higgs, we see that runaway for cPTs (hPTs) requires sufficiently small (large) couplings.

Finally, a word about higher-order terms. Relatively recent developments attempt to compute the next-to-leading-order contributions to the friction coming from transition radiation of gauge bosons, finding that these scale proportionally to the Lorentz factor $\sim \gamma$ \cite{Bodeker:2017cim,Azatov:2020ufh,Gouttenoire:2021kjv,GarciaGarcia:2022yqb}.\footnote{Reference ~\cite{Hoche:2020ysm} found that these contributions scale like $\sim \gamma^2$, which would further reduce the chances of a runaway regime in a cPT. However, this result was contested in Refs.~\cite{Azatov:2020ufh,Gouttenoire:2021kjv}. We thank Filippo Sala for pointing out that contributions to the plasma friction should depend on the order parameter of the PT, $\Phib$.}

This could severely curtail the chances of a cPT entering a runaway regime. Whether such higher-order terms are present in hPTs, where the particles {\it lose} their mass upon entering the bubbles of the true vacuum, or whether they behave as a friction or antifriction, is a question left to future work. In any case, if the DS plasma does not contain such massive gauge bosons this new friction term could very well not be there \cite{Caprini:2019egz}.

\subsection{Percolation, bubble number density, and mean bubble separation}\label{app:bubbles}

The bubble nucleation rate $\GVh$ and the bubble-wall speed $v_w$ can be used to determine the phase in which the Universe finds itself at any given time. Indeed, the {\it metastable volume fraction} $h(t)$ is given by \cite{Guth:1981uk,Guth:1982pn,Enqvist:1991xw,Hindmarsh:2020hop}
\be\label{eq:h_app}
    h(t) = \exp \bl[ - \int\limits_{t_c}^{t} \! \dd t' ~ a(t')^3 \GVa{(t')} \, \frac{4 \pi}{3} v_w^3 \bl( \int\limits_{t'}^{t} \frac{\dd t''}{a(t'')} \br)^3 \br] \ ,
\ee
where $t_c$ is the time at which $T(t_c) = T_c$, $a(t)$ is the scale factor at time $t$, and we have assumed $v_w$ to be constant. The PT completes at the {\it percolation time} $t_\pt$ when the metastable volume fraction $h(t_\pt)$ has been reduced to $1/e$.

Since the bubbles can only nucleate with a probability $\GVh(t)$ in the metastable phase volume, quantified by $h(t)$, we can write the {\it bubble number density} and corresponding {\it mean bubble separation} $\ov{R}_b(t)$ as a function of time:
\ba
    n_b(t) & \equiv & a(t)^{-3} \, \int\limits_{t_c}^{t} \! \dd t' ~ a(t')^3 \GVa{(t')} \, h(t') \ , \label{eq:nb_app} \\
    \ov{R}_b (t) & \equiv & n_b(t)^{-1/3} \ . \label{eq:Rb_app}
\ea
We can then derive the mean bubble separation scale at percolation, $\ov{R}_\pt \equiv \ov{R}_b(t_\pt)$, and from there $\beta = (8 \pi)^{1/3} v_w \ov{R}_\pt^{-1}$.

As an interesting aside, we point out that one can use $\ov{R}_b (t)$ to create an alternative definition of the percolation time. Indeed, the {\it average bubble radius} is
\be\label{eq:avgR_app}
    \angu{R(t)} \equiv \frac{1}{a(t)^3 n_b(t)} \, a(t) \int\limits_{t_c}^{t} \! \dd t' ~ a(t')^3 \GVa{(t')} \, h(t') \, v_w \bl( \int\limits_{t'}^{t} \frac{\dd t''}{a(t'')} \br)
\ee
One can then {\it define} $t_\pt$ to be the time at which $\angu{R(t_\pt)} = \ov{R}_b(t_\pt)$. This seems like a natural way to determine {\it when} bubble collisions take place. We have numerically corroborated that this alternative definition yields similar results to the standard one of $h(t_\pt) = 1/e$, and thus we use the latter, more common one in our computations.

Finally, we have also numerically made sure that the impact of the expansion of the Universe is negligible, and we therefore use $a(t) = 1$ in \Eqst{eq:h_app}{eq:Rb_app} throughout this paper.

\subsection{Analytic expressions for the phase transition}\label{app:bubbleAppx}

As discussed in the main text, PTs during reheating fall into two main categories. In cPTs and most hPTs, the transition occurs via {\it exponential nucleation}, where most bubbles are produced close to the end of the PT. In hPTs for which either $T_1 > T_\mx$ or $T_1$ simply does not exist ($\Delta > 8/9$), the PT can take place via {\it simultaneous nucleation}, as long as $t_n \sim t_\mx$ or $\GVh \lsim H^4$ for all times. In this regime most of the bubbles of the new phase are produced at the time $t_\mx$, where the action $S$ reaches its minimum and the nucleation rate $\GVh$ reaches its maximum (since $T/T_c = T_\mx/T_c$ and therefore $\lbar$ is at its largest). Ignoring the expansion of the Universe by setting $a(t) = 1$ in \Eqs{eq:h_app}{eq:nb_app}, we can find approximate expressions for $h(t)$ and $n_b(t)$ in each of these regimes \cite{Cutting:2018tjt,Hindmarsh:2019phv}.

Starting with the exponential nucleation regime, we can expand $S(t)$ around $t_\pt$ in a Taylor series. To first order,
\be\label{eq:expnucl1}
    S(t) \approx S_\pt - S_1 (t_\pt - t) \ ,
\ee
with $S_\pt \equiv S(t_\pt)$, $S_1 \equiv S'(t_\pt) < 0$.

Plugging this into $\GVh$, we can then use the method of steepest descent to find analytic expressions for $h(t)$ and for $n_b(t_\pt)$, which we approximate as the asymptotic value of $n_b(t \rightarrow \infty)$ since $n_b(t)$ changes little after that time:
\ba\label{eq:expnucl2}
    h(t) & \approx & \exp\bl[ -e^{-S_1 (t - t_\pt)} \br] \ , \\
    n_b(t_\pt) & \approx & \frac{(-S_1)^3}{8\pi v_w^3} \  \\
    \Rightarrow \beta & \approx & -S_1 \ .
\ea
In the exponential nucleation regime, the condition $h(t_\pt) = 1/e$ marking the end of the PT is equivalent to
\be\label{eq:expPTcond}
    \frac{8 \pi v_w^3}{\beta^4} \GVa{(t_\pt)} = 1 \ .
\ee

Using the chain rule for $S'$ and \Eqs{eq:dlnTdt_cases_app}{eq:DSDT} we can estimate $\beta$. Taking advantage of the fact that in the exponential nucleation regime $t_c \lsim t_n \lsim t_\pt$, we note that $S_\pt \lesssim S_n$, $\rho_\chi/\rho_\rhpt \lesssim \rho_\chii/\rho_\rc$, $H_\pt \lesssim H_n \lesssim H_c < H_i$, and that $\lbar(S_\pt) \approx \lbar(S_n) \approx \lbar_{\rm infl.}$ since $\lbar$ depends only logarithmically on $S$ for a large range of values (see \Eqs{eq:lbardSinfl}{eq:estimaction}). Therefore,\footnote{These expressions are accurate to $\mathcal{O}(1)$. More precise formulas can be derived from \Eq{eq:expPTcond}, by taking $T \approx T_n$ and $S \approx S_n \approx 4 \ln ( T_n/H_n )$ everywhere except in the exponential of $\GVh$, where we use instead $e^{-S_\pt}$. From this one can obtain the logarithmic corrections $S_\pt \approx S_n + \ln(8 \pi v_w^3) - 4 \ln ( H_n^{-1} S_n \bl. \frac{d \ln S}{d \ln T} \frac{d \ln T}{d t} \br\vert_{t_n} )$. This is roughly accurate to $\mathcal{O}(10\%)$.}
\ba\label{eq:beta_exp_estim}
    \beta & \approx & - \bl. \bl( S \frac{d \ln S}{d \ln T} \frac{d \ln T}{d t} \br)\br\vert_{t_\pt} \lesssim - S_n \bl.\frac{d \ln S}{d \ln T} \br\vert_{\lbar_{\rm infl.}} \bl. \frac{d \ln T}{d t} \br\vert_{t_c} \nonumber\\
    & \sim & 4 \ln \bl( T_c/H_i \br) \times
    \begin{cases}
        91 \bl( \Delta^{-1} - 1.08 \br) \, \frac{\Gamma_\chi}{4} \bl( \frac{T_\mx}{T_c} \br)^4 \vepsrh^{-1} & \quad \text{hPT,}\\[2ex]
        9.8 \bl( \Delta^{-1} - 0.52 \br) \, H_r(T_c) \max\bl[ 1, \bl( \frac{T_c}{T_\mx} \br)^2 \Drd^{-1/2} \br] \cr \ \approx 9.8 \bl( \Delta^{-1} - 0.52 \br) \, H_i \vepsrh^{1/2} \bl( \frac{T_c}{T_\mx} \br)^2 \max\bl[ 1, \bl( \frac{T_c}{T_\mx} \br)^2 \Drd^{-1/2} \br] & \quad \text{cPT.}
    \end{cases}
\ea

For those hPTs in the simultaneous nucleation regime, the nucleation rate $\GVh$ is dominated by the minimum of the action, which takes place at the temperature the farthest from $T_c$, namely $T_\mx$. Since $T'(t_\mx) = 0$ by definition, the expansion of $S(t)$ around $t_\mx$ looks like
\be\label{eq:simnucl1}
    S(t) \approx S_\mn + \frac{1}{2} S_2^2 (t - t_\mx)^2 \ ,
\ee
with $S_\mn \equiv S(t_\mx)$ and $S_2^2 \equiv S''(t_\mx)$. Repeating the exercise, we find
\ba\label{eq:simnucl2}
    h(t) & \approx & \exp\bl[ - \frac{4 \pi}{3} n_0 v_w^3 (t - t_\mx)^2 \br] \ , \\
    n_b(t_\pt) \approx n_0 & \equiv & \frac{\sqrt{2 \pi}}{S_2} \GVa{(t_\mx)} \ , \\
    \Rightarrow \quad \beta & \approx & (8 \pi n_0)^{1/3} v_w \ .
\ea

Since at $t_\mx$ the first time derivative of $T(t)$ vanishes, then
\be\label{eq:S22}
    S_2^2 = \bl .\bl( \frac{d S}{d \ln T} \frac{d^2 \ln T}{d t^2} + \frac{d^2 S}{d \ln T^2} \bl( \frac{d \ln T}{d t} \br)^2 \br) \br \vert_{t_\mx} = \bl. S \frac{d \ln S}{d \ln T} \frac{d^2 \ln T}{d t^2} \br\vert_{t_\mx} \ .
\ee

We can then use \Eq{eq:Tlbar} and \Eq{eq:rh_eff} to find ${\lbar}_\mx \equiv \lbar(T_\mx)$ and from there $S_\mn = S(\lbar_\mx)$, $\Gamma(t_\mx)/\mathcal{V}$, and $dS/d \ln T \vert_{t_\mx}$ (using \Eqs{eq:estimaction}{eq:DSDT}). An even simpler estimate relies on noting that for simultaneous nucleation to take place $\Gamma(t_\mx)/\mathcal{V}$ cannot be too different from $H(t_\mx)^4$ (\ie, $t_n \approx t_\mx$); otherwise, $\GVh$ would be rising exponentially and the system would be in the exponential nucleation regime. Since $H_\mx \approx H_i$ we simply have $\Gamma(t_\mx)/\mathcal{V} \sim H_i^4$ and $S_\mn \sim 4 \ln ( T_\mx/H_i )$. It can be shown that $d^2 \ln T/dt^2 \sim - H_i^2 \max[ 1, \gamma ]$ at $t_\mx$, and therefore
\ba
    S_2^2 & \sim & 360 \, H_i^2 \max\bl[ 1, \gamma \br] \ln\bl( T_\mx / H_i \br) \bl( \Delta^{-1} - \lbar_\mx \br) \ , \\
    \Rightarrow \beta & \sim & H_i v_w \bl( \max\bl[ 1, \gamma \br] \ln\bl( T_\mx / H_i \br) \bl( \Delta^{-1} - \lbar_\mx \br)  \br)^{-1/6} \ .
\ea

As a final word on the topic, we would like to mention that a recent study \cite{Cutting:2018tjt} showed that the GW spectrum coming from PTs is essentially the same regardless of whether the PT took place in a simultaneous or exponential regime, although others \cite{Weir:2016tov,Jinno:2017ixd} claim there are $\mathcal{O}(1)$ differences.

\section{Additional Results}\label{app:additional}
\label{app:AR}

\begin{figure}[tb]
	\centering
        \includegraphics[width=.475\linewidth]{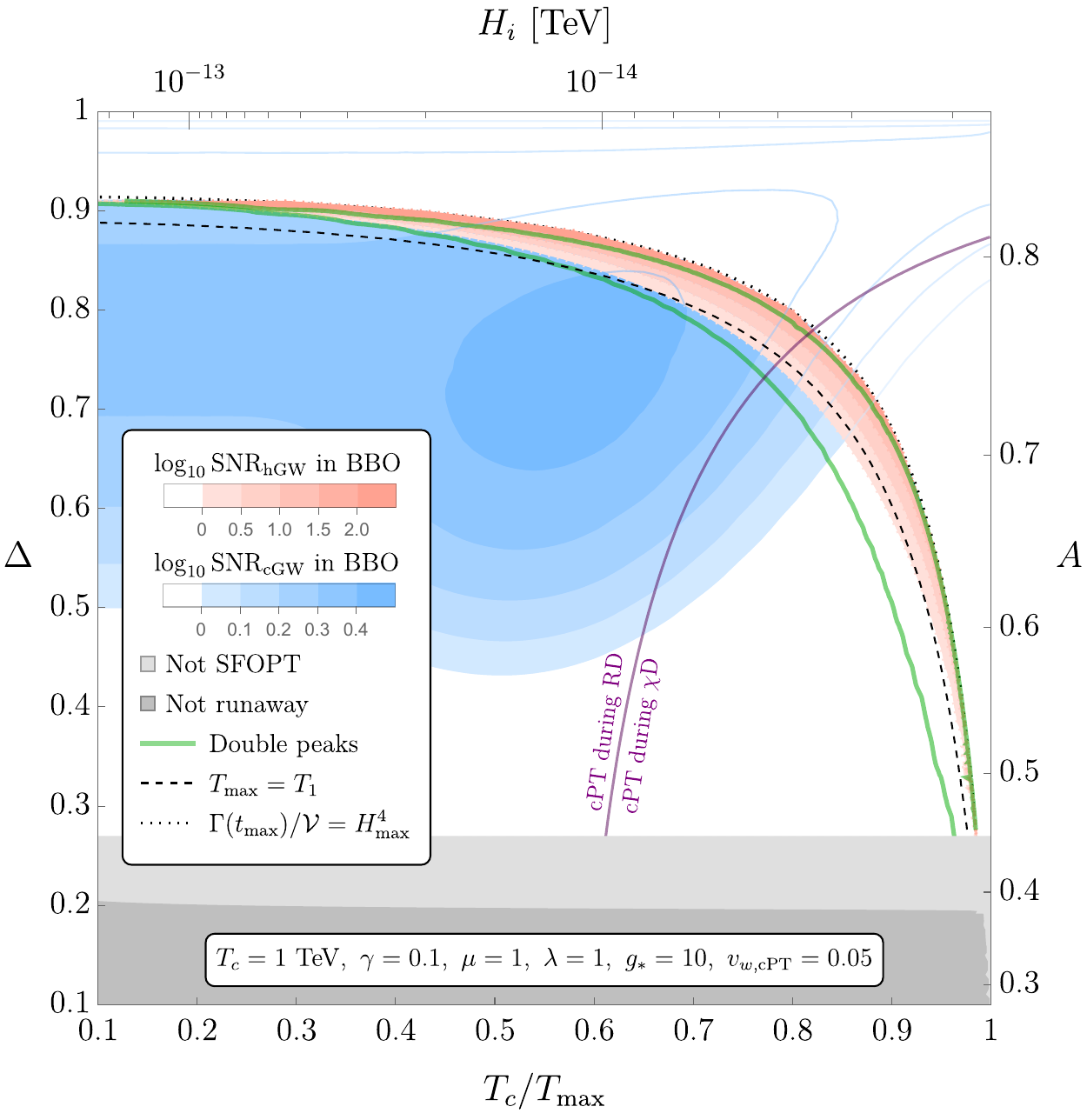}
        ~~~~\includegraphics[width=.475\linewidth]{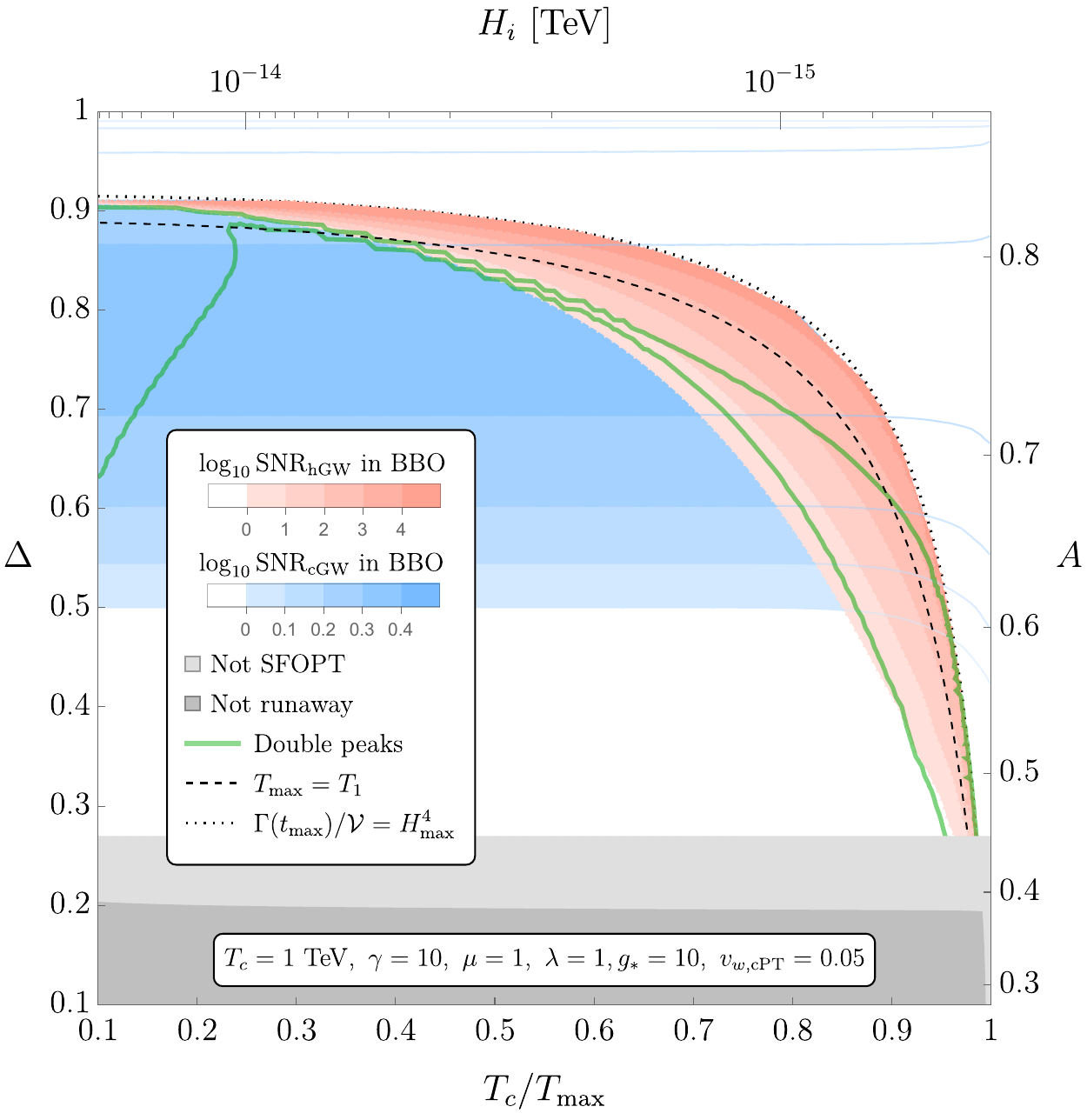}
	\caption{SNR contours for 1-year observation time of the SGWB spectra generated during reheating by hPT bubble collisions ({\bf red}) and cPT sound waves ({\bf blue}), for the upcoming BBO detector \cite{Crowder:2005nr,Corbin:2005ny,Harry:2006fi}, as a function of $T_c/T_\mx$ and $\Delta$, with $\gamma = 0.1$ ({\bf left}) and $\gamma = 10$ ({\bf right}). The corresponding values of $H_i$ and $A$ are also shown. The notation is the same as in \Fig{fig:param}, namely, the points inside the {\bf green} contour have double peaks (from both the hGWs and the cGWs), the region above the {\bf dashed line} has $T_\mx < T_1$, the region above the {\bf dotted line} has $\GVh < H^4$ at $t_\mx$, the {\bf dark grey region} has no runaway hPT bubbles, and the {\bf light grey region} has daisy contributions to the thermal potential that prevent a SFOPT. In the left panel, $\gamma = 0.1$ and we have added a {\bf purple line} separating the parameter regions in which the cPT takes place during RD and during $\chi$D; for $\gamma = 10$ in the right panel and $\gamma = 1$ in \Fig{fig:param}, the cPT always takes place during RD. For this plot we chose $T_c = 1~\TeV$, $g_* = 10$, $\{ \mu, \lambda \} = \{ 1, 1 \}$, and $v_{w,\cpt} = 0.05$.}
	\label{fig:param2}
\end{figure}

\begin{figure}[tb]
	\centering
        \includegraphics[width=.475\linewidth]{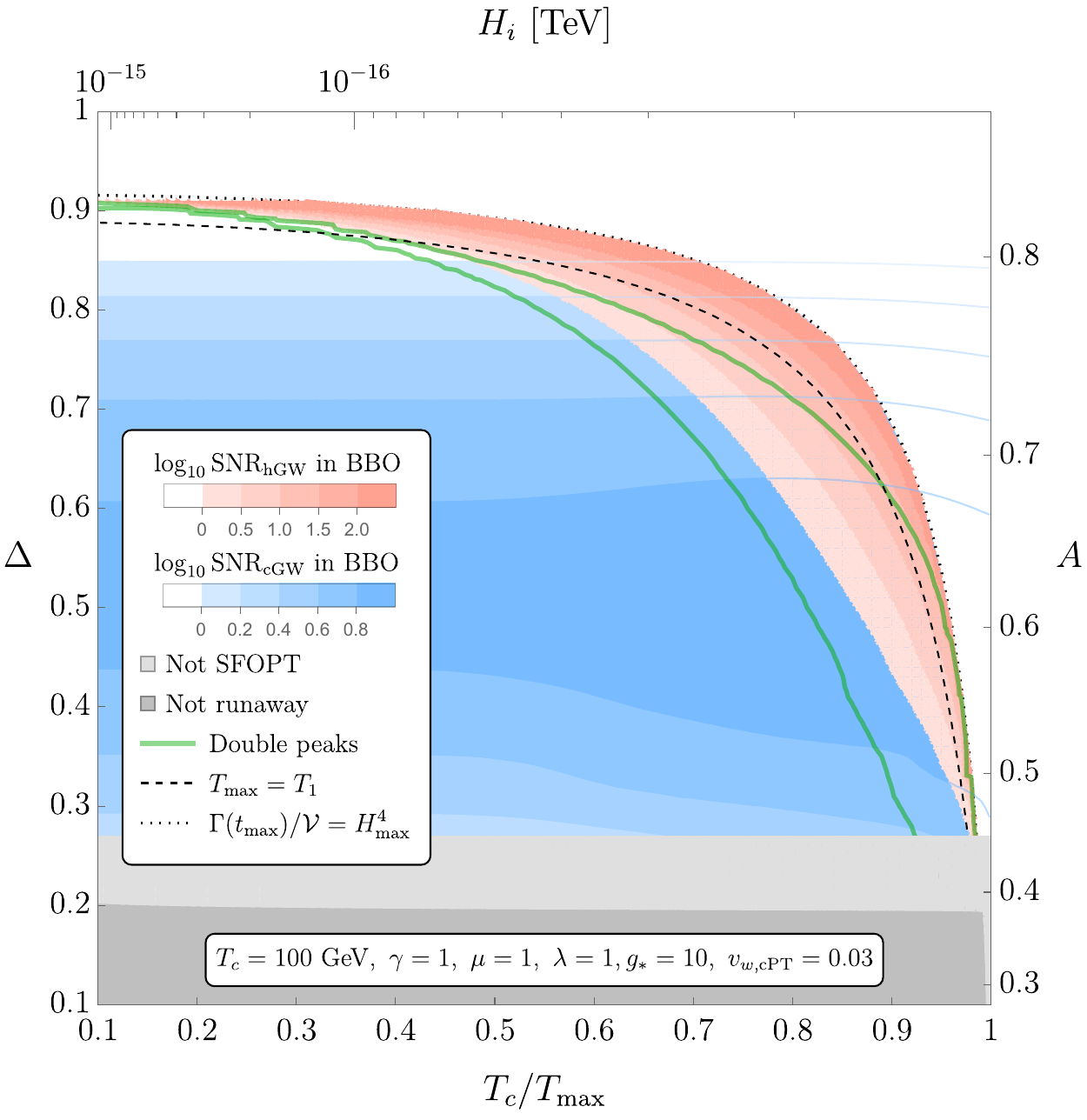}
        ~~~~\includegraphics[width=.475\linewidth]{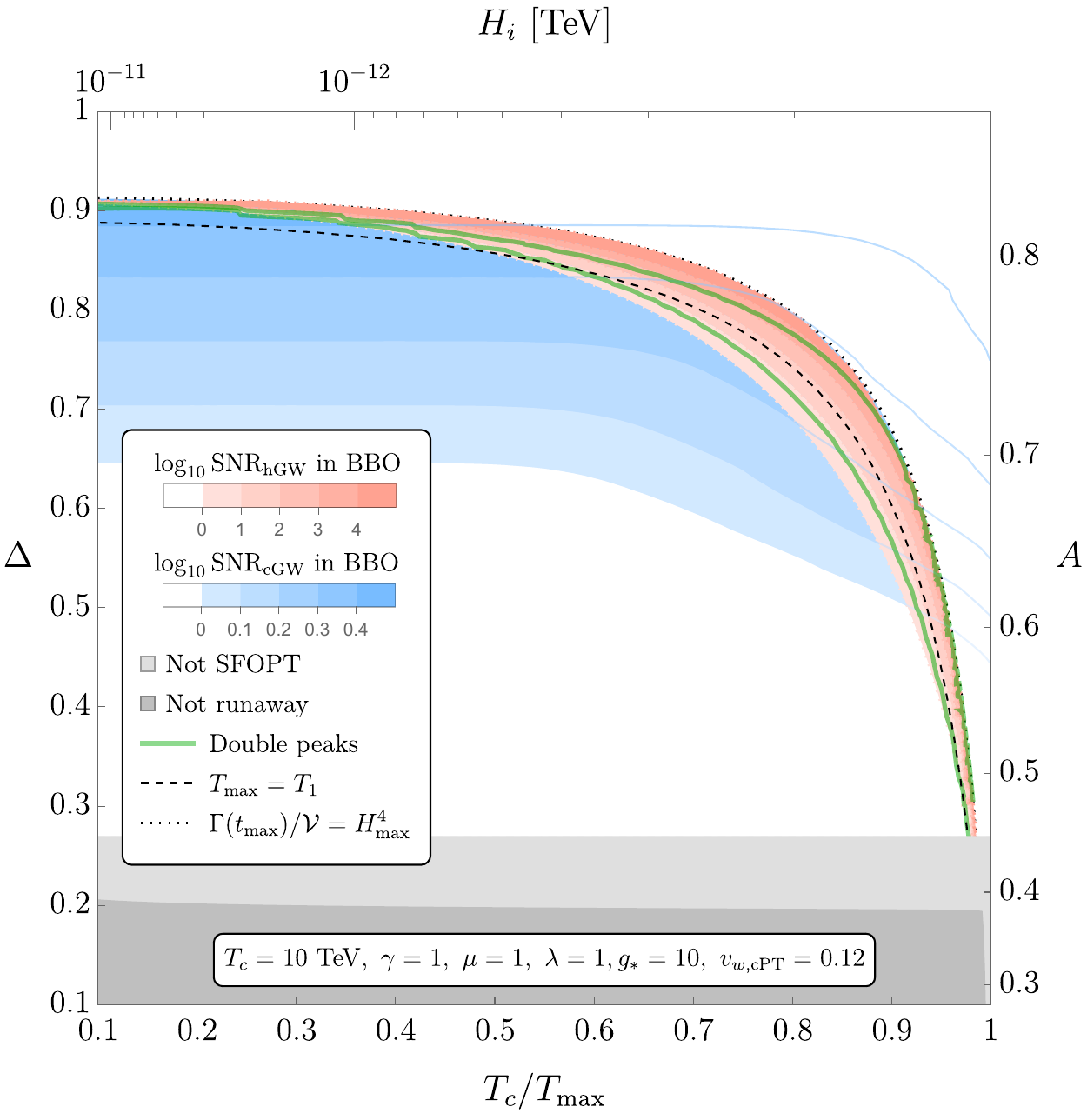}
	\caption{SNR contours for 1-year observation time of the SGWB spectra generated during reheating by hPT bubble collisions ({\bf red}) and cPT sound waves ({\bf blue}), for the upcoming BBO detector \cite{Crowder:2005nr,Corbin:2005ny,Harry:2006fi}, as a function of $T_c/T_\mx$ and $\Delta$, with $T_c = 100~\GeV$ and $v_{w,\cpt} = 0.03$ ({\bf left}), and $T_c = 10~\TeV$ and $v_{w,\cpt} = 0.12$ ({\bf right}). Note that we tune $v_{w,\cpt}$ so that the cGWs and hGWs have similar amplitudes. The corresponding values of $H_i$ and $A$ are also shown. The notation is the same as in \Fig{fig:param}, namely: the points inside the {\bf green} contour have double peaks  (from both the hGWs and the cGWs), the region above the {\bf dashed line} has $T_\mx < T_1$, the region above the {\bf dotted line} has $\GVh < H^4$ at $t_\mx$, the {\bf dark grey region} has no runaway hPT bubbles, and the {\bf light grey region} has daisy contributions to the thermal potential that prevent a SFOPT. For this plot we chose $\gamma = 1$, $g_* = 10$, and $\{ \mu, \lambda \} = \{ 1, 1 \}$.}
	\label{fig:param3}
\end{figure}

In this appendix, we show more plots of the BBO SNR of the SGWB from cPTs (blue contours) and hPTs (red contours) as a function of the $\Delta$--$T_c/T_\mx$ parameter space, for choices of $\gamma = \Gamma_\chi/H_i$ and $T_c$ different from the one presented in the main text (namely, $\gamma = 1$ and $T_c = 1~\TeV$). We do this for $\gamma \in \{ 0.1, \, 10 \}$ in \Fig{fig:param2} and for $T_c \in \{ 100~\GeV, \, 10~\TeV \}$ in \Fig{fig:param3}. The notations and legends in these plots are the same as in \Fig{fig:param}, and their qualitative features are the same, and have been described in detail in \Sec{Sec: prospects}. The characteristic crescent shape of the hPT contours, determined by the BBO sensitivity at low $T_c/T_\mx$ (and consequently more redshift) and by the increasing difficulty in nucleating bubbles at large $\Delta$, is clearly seen in all the figures. The crescent is thinner for smaller values of $\gamma$, because the era of reheaton domination becomes longer and thus the relative redshifting between the hGWs and cGWs increases. The cPT contours depend mostly on $\Delta$. Their insensitivity to low values of $T_c/T_\mx$, caused by the cPT taking place well after reheating during RD, is also evident, while the increasing SNR values as $T_c/T_\mx$ approaches 1 can also be seen.

\begin{figure}[tb]
	\centering
        \includegraphics[width=.475\linewidth]{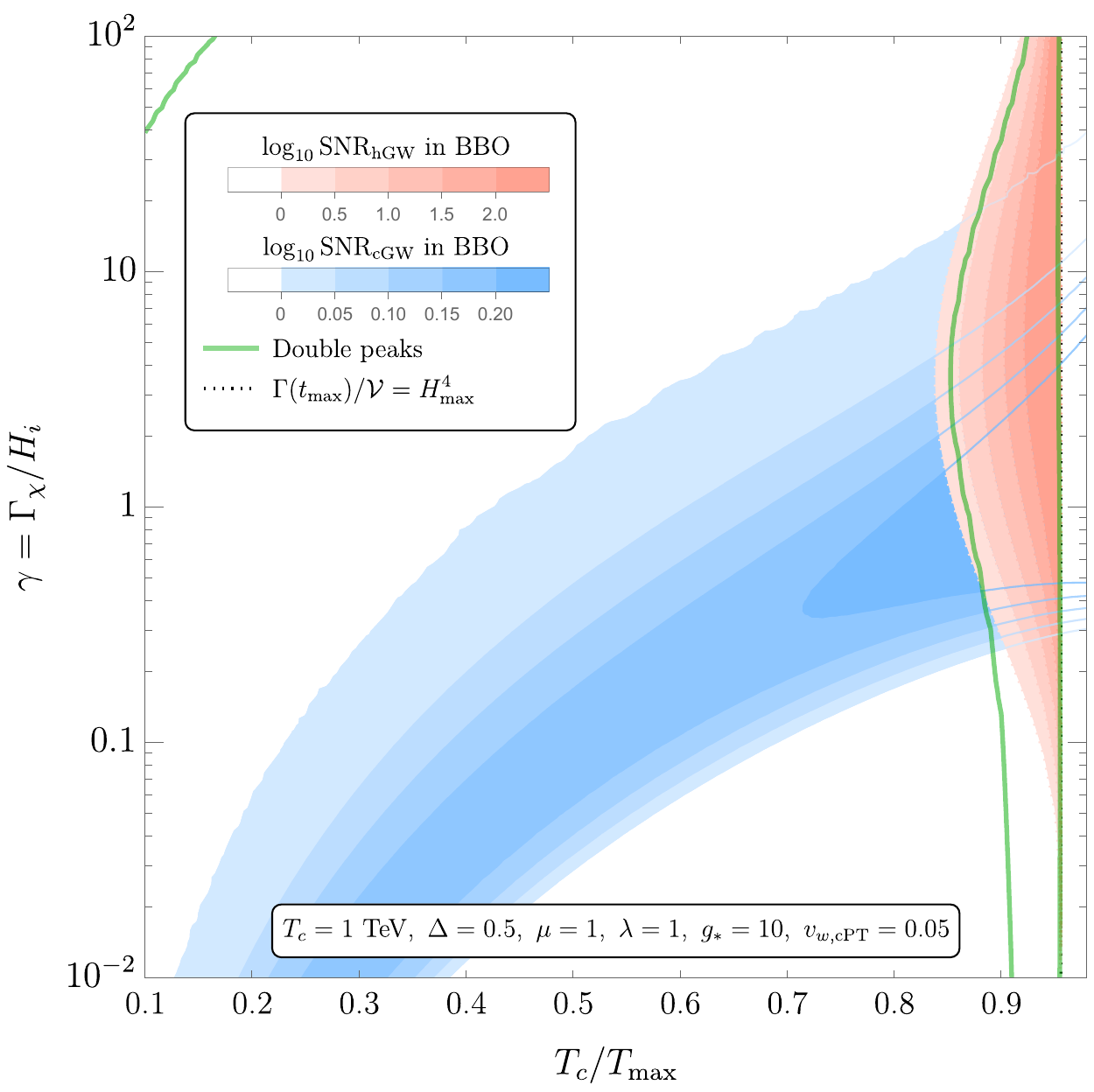}
        ~~~~\includegraphics[width=.475\linewidth]{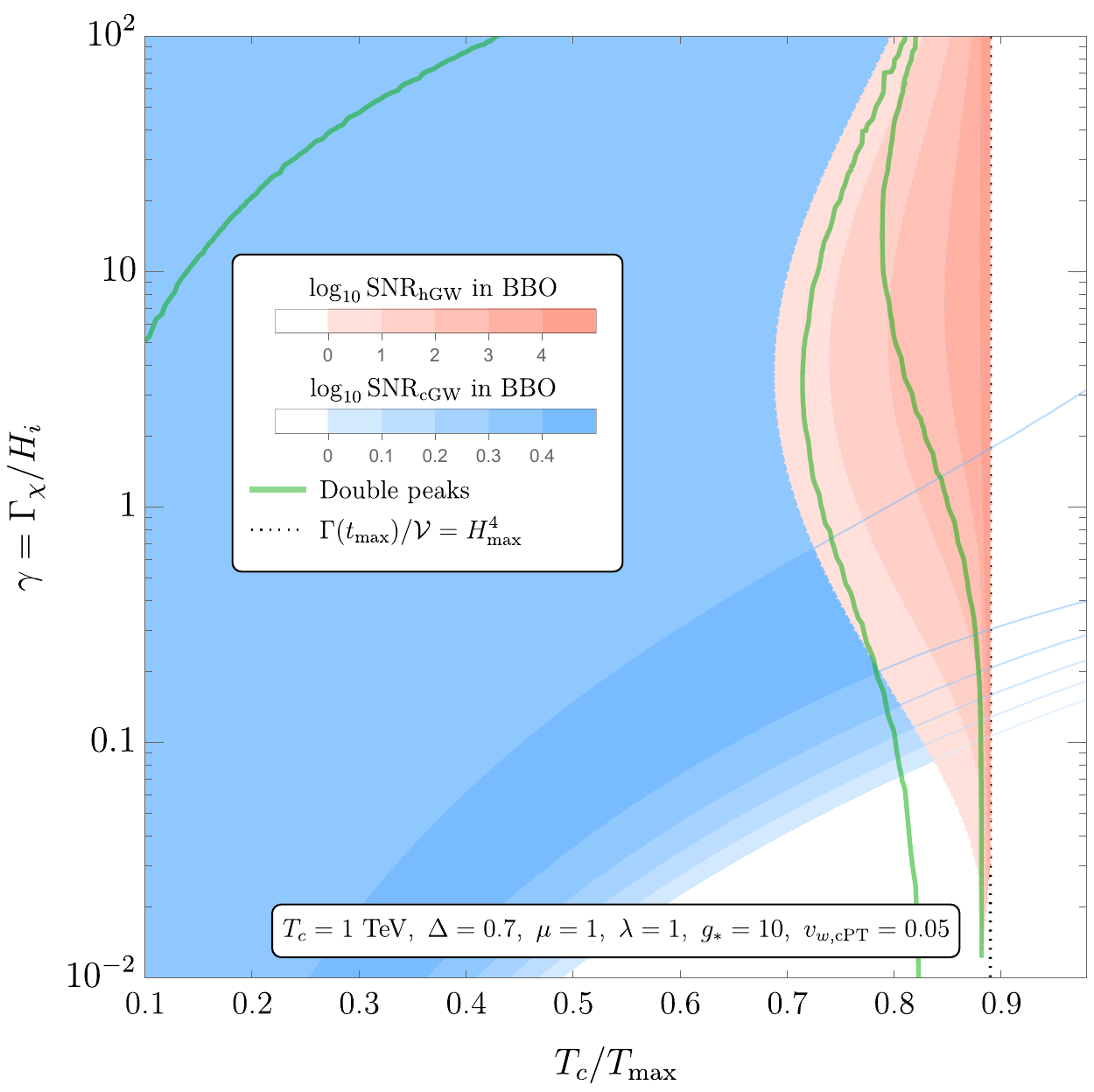}
	\caption{SNR contours for 1-year observation time of the SGWB spectra generated during reheating by hPT bubble collisions ({\bf red}) and by cPT plasma sound waves ({\bf blue}), for the upcoming BBO detector \cite{Crowder:2005nr,Corbin:2005ny,Harry:2006fi}, as a function of $T_c/T_\mx$ and $\gamma$, with $\Delta = 0.5$ ({\bf left}), and $\Delta = 0.7$ ({\bf right}). The corresponding values of $H_i$ are also shown. The regions between the {\bf green contour} near the dotted line and above the {\bf green contour} at the left-upper corner have double peaks (from both the hGWs and the cGWs), as in \Fig{fig:param}. The region to the right of the {\bf dotted line} has $\GVh < H^4$ at $t_\mx$. For this plot we chose $T_c = 1$ TeV, $g_* = 10$, $\{ \mu, \lambda \} = \{ 1, 1 \}$, and $v_{w,\textrm{cPT}}=0.05$.}
	\label{fig:param4}
\end{figure}

Furthermore, in \Fig{fig:param4} we show the $\gamma$--$T_c/T_\mx$ slice of the parameter space for two example values of $\Delta$ ($0.5$ and $0.7$). We can see that there is a sharp drop in the GW signature for a sufficiently large value of $T_c/T_\mx$, which corresponds to the PT never taking place because no bubbles are nucleated within a Hubble patch (\ie, the intersection of the dotted lines in \Fig{fig:param} with the value of $\Delta$ under consideration). At lower values of $T_c/T_\mx$ the hGW SNR also drops because the hGWs become invisible to BBO due to redshifting (the left-hand boundary of the crescent shape in Figs. \ref{fig:param}, \ref{fig:param2}, and \ref{fig:param3}). In addition, the hGW SNR contours vanish for values of $\gamma$ that are very small (due to the already discussed large relative redshift between hPT and cPT caused by a long $\chi$D era) or very large (due to the corresponding increase in $\beta_\hpt$, which means both that the hGWs are quieter because the PT lasts for a shorter $\beta^{-1}_\hpt$ time and that the peak frequency $f_\hpt$ of the hGW spectrum can move outside the BBO sensitivity). The shape of the SNR contours of the cGWs can similarly be understood, in terms of the duration of the $\chi$D era and the duration $\beta^{-1}_\cpt$ of the PT, with the aid of Figs. \ref{fig:param}, \ref{fig:param2}, and \ref{fig:param3} and the discussions of \Sec{Sec: prospects} and the above paragraph.


\end{document}